\documentclass[12pt]{article}
\usepackage{fullpage}
\usepackage{epsf}
\usepackage{epsfig}
\usepackage{amsthm}
\usepackage{amsfonts}
\usepackage{color}
\usepackage{amsmath}
\usepackage{setspace}
\usepackage{natbib}
\usepackage{url}
\usepackage{algorithmic}
\usepackage[boxed]{algorithm}
\usepackage[thinlines]{easymat}

\begin{document}

\newcommand{\bm}[1]{\mbox{\boldmath $#1$}}
\newcommand{\mb}[1]{\mathbf{#1}}
\newcommand{\bE}[0]{\mathbb{E}}
\newcommand{\bP}[0]{\mathbb{P}}
\newcommand{\ve}[0]{\varepsilon}
\newcommand{\Var}[0]{\mathbb{V}\mathrm{ar}}
\newcommand{\Corr}[0]{\mathbb{C}\mathrm{orr}}
\newcommand{\Cov}[0]{\mathbb{C}\mathrm{ov}}
\newcommand{\mN}[0]{\mathcal{N}}
\newcommand{\iidsim}[0]{\stackrel{\mathrm{iid}}{\sim}}
\newcommand{\NA}[0]{{\tt NA}}
\newcommand{\argmax}{\operatornamewithlimits{argmax}}
\newcommand{\acv}{auto-covariance function}
\newcommand{\acf}{auto-correlation function}
\newcommand{\iid}{independent and identically distributed}
\newcommand{\backshift}{\ensuremath{\mathcal{B}}}
\newcommand{\innov}{\ensuremath{\varepsilon_t}}
\newcommand{\ARMA}[1]{ARMA\ensuremath{(#1)}}
\newcommand{\AR}[1]{AR\ensuremath{(#1)}}
\newcommand{\MA}[1]{MA\ensuremath{(#1)}}
\newcommand{\sdf}{spectral density function}
\newcommand{\pdf}{probability density function}
\newcommand{\FI}[1]{FI\ensuremath{(#1)}}
\newcommand{\FID}[1]{\FAR{0,#1,0}}
\newcommand{\FAR}[1]{ARFIMA\ensuremath{(#1)}}
\newcommand{\dint}{\ensuremath{(-\frac{1}{2},\frac{1}{2})}}
\newcommand{\MH}{Metropolis--Hastings}
\newcommand{\Sstandard}[2]{\ensuremath{\mathcal{S}_{#1,#2}}}
\newtheorem{Theorem}{Theorem}
\newtheorem{Corollary}{Corollary}
\newtheorem{Lemma}{Lemma}

\title{\vspace{-1cm} Efficient Bayesian inference for ARFIMA processes}
\author{
  Timothy Graves\thanks{URS Corporation, London, UK}
  \and
  Robert B.~Gramacy\thanks{Corresponding author:
    The University of Chicago Booth School of Business; 5807 S. Woodlawn Avenue, Chicago, IL 60637; 
    {\tt rbgramacy@chicagobooth.edu}}\\
  \and
  Christian Franzke\thanks{Meteorologisches Institut, University of Hamburg, Germany}
  \and
  Nicholas Watkins\thanks{Max Planck Institute for the Physics of Complex
  Systems, Dresden, Germany; Centre for Complexity and Design, Open
  University, Milton Keynes, UK; Centre for the Analysis of Time Series,
  London School of Economics and Political Science, London, UK; and Centre for
  Fusion Space and Astrophysics, University of Warwick, Coventry, UK.} }

\date{}

\maketitle

\vspace{-0.5cm}

\begin{abstract}
In forecasting problems it is important to know whether or not recent events
represent a regime change (low long-term predictive potential), or rather a
local  manifestation of longer term effects (potentially higher predictive
potential). Mathematically, a key question is about whether the underlying
stochastic process exhibits ``memory", and if so whether the memory is ``long"
in a precise sense. Being able to detect or rule out such effects can have a
profound impact on speculative investment (e.g., in financial markets) and
inform public policy (e.g., characterising the size and timescales of the
earth system's response to the anthropogenic $CO_2$ perturbation). Most
previous work on inference of long memory effects is frequentist in nature.
Here we provide a systematic treatment of Bayesian inference for long memory
processes via the Autoregressive Fractional Integrated Moving Average (ARFIMA)
model. In particular, we provide a new approximate likelihood for efficient
parameter inference, and show how nuisance parameters (e.g., short memory
effects) can be integrated over in order to focus on long memory parameters
and hypothesis testing more directly than ever before. We illustrate our new
methodology on both synthetic and observational data, with favorable
comparison to the standard estimators.\\

\end{abstract}

%\doublespacing

\section{Introduction}
\label{sec:intro}

In this paper we are concerned with Bayesian analysis of specific types of
stochastic processes capable of possessing `long memory', or
``long-range dependence" (LRD) \citep{Beran_1994,Palma_2007,Beran_2013}. Long
memory is the notion of there being correlation between the present and {\em
all} points in the past. A standard definition is that a (finite variance,
stationary) process has {\em long memory} if its autocorrelation function
(ACF) has  power-law decay: $\rho(\cdot)$ such that $\rho(k) \sim c_\rho\,
k^{2d-1}$ as $k \rightarrow \infty$, for some non-zero constant $c_\rho$, and
where $0<d<\frac{1}{2}$. The parameter $d$ is the memory parameter; if $d=0$
the process does not exhibit long memory, while if $-\frac{1}{2}<d<0$ the
process is said to have {\em negative} memory.

%The asymptotic power law form of the ACF  corresponds to an absence of a
%characteristic decay timescale, in striking contrast to many standard
%(stationary) stochastic processes  where   the effect of each data point
%decays so fast that it rapidly becomes indistinguishable from noise.
% An example of the latter is the exponential ACF where the e-folding scale sets
% a characteristic correlation time. 
%The study of processes that \emph{do}
%possess long memory is important because they exhibit unusual properties,
%because many familiar mathematical results fail to hold, and because of the
%numerous examples of datasets where LRD is seen.

The study of long memory originated in the 1950s in the field of hydrology,
where studies of the levels of the river Nile \citep{Hurst_1951} demonstrated
anomalously fast growth of the rescaled range of the time series.
%, the so-called  ``Hurst effect". 
After protracted debates\footnote{For a detailed
exposition of this period of mathematical history, see \citet{Graves_etal_2014b}.} about whether this was a transient (finite time) effect,
the mathematical pioneer Beno\^{\i}t B.~Mandelbrot showed that if
one retained the assumption of stationarity, novel mathematics would then be
essential to sufficiently explain the Hurst effect. In doing so he rigorously
defined  \citep{Mandelbrot_1968a,Mandelbrot_1968b} the concept of long memory.
% which he dubbed the ``Joseph effect". 
%Such processes can be characterised as possessing fluctuations over most
%frequencies, in particular for low frequencies (or equivalently long time
%periods). Because of this presence of low frequencies, one might observe long
%periods of `high' values followed by long periods of `low' values. In
%classical time series analysis this might indicate non-stationarity; however
%such behaviour is not unusual for long memory processes, which can be seen as
%forming an intermediate ``boundary layer" between short range dependence and
%nonstationarity.

%Since then, many other fields have found applications for long memory
%processes: computer network traffic, econometrics, astrophysics and geophysics
%\citep[see, e.g.,][]{Taqqu_col}. As a particular example of its relevance to
%climatology; the existence, or not, of long memory in weather systems has
%potential ramifications for the understanding of climate
%behaviour, particularly 
%quantification of the anthropogenic warming trend \citep{Franzke:2012}.

Most research into long memory and its properties has been based on classical
statistical methods, spanning parametric, semi-parametric and non-parametric
modeling \citep[see][for a review]{Beran_2013}.  Very few Bayesian methods
have been studied, most probably due to computational difficulties. The
earliest works are parametric and include \citet{Koop_1997}
\citet{Pai_1998}, and \citet{Hsu_2003}.
If computational challenges could be mitigated, the Bayesian paradigm would
offer advantages over classical methods including flexibility in
specification of priors (i.e., physical expertise could be used to elicit an
informative prior). It would offer the ability to marginalise out aspects of a
model apparatus and data, such as short memory or seasonal effects and missing
observations, so that statements about long memory effects can be made
unconditionally.

Towards easing the computational burden, we focus on the ARFIMA class of
processes \citep{Granger_Joyeux_1980,Hosking_1981} as the basis of developing a systematic and unifying Bayesian
framework for modeling a variety of common time series phenomena, with
particular emphasis on detecting potential long memory effects. 
ARFIMA has become very popular  in statistics and econometrics because it is generalisable
and its connection to the ARMA family (and to fractional Gaussian noise) is relatively transparent.
A key property of ARFIMA is its ability to simultaneously yet separately model
long and short memory.
%
%
%\citet{Hsu_2003} argued that neither of these two methods are very efficient
%and developed a faster sampling method. This method essentially approximates
%an ARFIMA model by an appropriately chosen ARMA model, which is then corrected
%using importance sampling. The advantages of this are that ARMA processes are
%generally easier to deal with.
%
% Bayesian inference concerning long memory processes is not restricted to the
% parametric setting. 
Both \citet{Liseo_2001} and \citet{Holan_2009} argued,
echoing a sentiment in the classical literature, that full parametric long
memory models (like ARFIMA) are `too hard' to work with. Furthermore, often
$d$ is the only object of real interest, and consideration of a single class
of models, such as ARFIMA, is too restrictive. They therefore developed
methods which have similarities to classical periodograms.

We think ARFIMA deserves another look---that many of the above drawbacks,
to ARFIMA in particular and Bayesian computation more generally,
can be addressed with a careful treatment.
We provide a new
approximate likelihood for ARFIMA processes that can be computed quickly for
repeated evaluation on large time series, and which underpins an
efficient MCMC scheme for Bayesian inference.  Our sampling scheme can be best
described as a modernisation of a blocked MCMC scheme proposed by
\cite{Pai_1998}---adapting it to the approximate
likelihood and extending it to handle a richer form of (known) short memory
effects.  We then further extend the analysis to the case where the short
memory form is unknown, which requires transdimensional MCMC.  This aspect is
similar to the work of
\citet{Ehlers_2008} who considered the simpler 
ARIMA model class, and to \citet{Holan_2009} who worked with a nonparametric
long memory process. Our contribution has aspects in common with
\citet{Egrioglu_2010} who presented a more limited method
focused on model selection rather than averaging. The advantage of averaging
is that the unknown form of short memory effects can
be integrated out, focusing on long-memory
without conditioning on nuisance parameters.

The aim of this paper is to introduce an efficient Bayesian algorithm for the inference of the parameters of the \FAR{p,d,q} model, with particular emphasis on the LRD parameter $d$. Our Bayesian inference algorithm has been designed in a flexible fashion so that, for instance, the innovations can come from a wide class of different distributions; e.g., $\alpha$-stable or $t$-distribution. The remainder of the paper is organised as follows. Section
\ref{sec:preamble} summarises of ARFIMA required for our purposes.
Section \ref{sec:lik} discusses the important numerical calculation of
likelihoods, representing a hybrid between earlier classical statistical
methods, and our new contributions towards a full Bayesian approach. Section
\ref{sec:long} describes our proposed Bayesian framework and methodology
method in detail, focusing on long-memory only. Then, in Section
\ref{sec:short}, we consider extensions for additional short memory. Empirical
illustration and comparison of all methods is provided in Section
\ref{sec:empirical}. The paper concludes with a discussion in Section
\ref{sec:discuss} focused on potential for further extension.

\section{Time series definitions and the ARFIMA model}  
\label{sec:preamble}
Following \citep{Brockwell_1991}  a {\em time series} will mean a set of
univariate real-valued observations $\{x_t\}$, each   recorded at a specified
time $t\in\mathbb{Z}$, and sampled at discrete,  regular, intervals.  A {\em
process} will   refer to a corresponding set of random variables $\{X_t\}$.
The process $\{X_t\}$ is   \emph{strictly} stationary if the joint
distributions   $(X_{t_1},\ldots,X_{t_k})^\top$ and
$(X_{t_1+h},\ldots,X_{t_k+h})^\top$ are the same for all positive integers
$k$, and for all $t_1,\ldots,t_k,h \in \mathbb{Z}$. It is
\emph{weakly} stationary if: (1) $\bE X_t =\mu < \infty$ for all $t \in
\mathbb{Z}$; (2) $\bE|X_t|^2 < \infty$ for all $t \in \mathbb{Z}$; and (3)
$\Cov(X_r,X_s) = \Cov(X_{r+t},X_{s+t})$ for all $r,s,t \in \mathbb{Z}$.  A process $\{X_t\}$ is Gaussian if the distribution of
$(X_{t_1},\ldots,X_{t_k})^\top$ is multivariate normal (MVN) for all positive
integers $k$, and for all $t_1,\ldots,t_k \in \mathbb{Z}$. 
Throughout, stationary Gaussian processes will be assumed for convenience, where `strong' and `weak' are equivalent and consequently those qualifiers will be dropped.

From the above, we see that the covariance depends only
on the temporal difference   which motivates defining an autocovariance {\em ACV}
$\gamma(\cdot)$ of a weakly stationary process as
$\gamma(k)=\mathrm{Cov}(X_t,X_{t+k})$, where $k$ is referred to as the  (time) `lag'.
The (normalised) autocorrelation function {\em ACF} $\rho(\cdot)$ is defined as:
$\rho(k)=\frac{\gamma(k)}{\gamma(0)}$. %Note that the ACV is an even
%function, i.e.\ $\gamma(k)=\gamma(-k)$, so one only need be interested in the
%behaviour of the ACV for $k\geq0$. 
%A Gaussian process $\{X_t\}$ is said to
%be {\em white noise} if it is stationary, has zero mean, and $\gamma(k) = 0$
%for all $k>0$, i.e., implying that they are \iid{} (iid).

Another useful time domain 
tool is the  `backshift' operator \backshift{}, where
$\backshift{}X_t=X_{t-1}$, and powers of $\backshift{}$ are defined
iteratively: $
%\label{eqn: backshift_iterate}
\backshift{}^kX_t = \backshift{}^{k-1}(\backshift{}X_t) =
\backshift{}^{k-1}X_{t-1} = \cdots = X_{t-k}$. 
\label{defn: causality}
A stationary process $\{X_t\}$ is said to be \emph{causal} if there exists a
sequence of coefficients $\{\psi_k\}$, with finite total mean square  $\sum_{k=0}^{\infty}\psi_k^2 <
\infty$ such that  for all $t$, a given member of the process can be expanded
as a power series in the backshift operator acting on the `innovations',
$\{\innov{}\}$:
\begin{equation}
\label{eqn:causaldefn1} %cn
X_t = \Psi(\backshift{})\innov{}, \quad \hbox{where} \; \Psi(z) = \sum_{k=0}^{\infty}\psi_k z^k.
\end{equation}
The innovations are  a white (i.e. stationary, zero mean, iid) noise process   with
variance $\sigma^2$. Causality specifies that for every $t$, $X_t$ can only depend
  on the past and present values of the innovations $\{\innov{}\}$. Furthermore Wold's theorem shows that any purely non-deterministic stationary process has a unique causal representation
(referred to as the Wold expansion).

%For
%most practical studies this is a natural assumption due to the unidirectional
%nature of time---today's weather cannot be determined by tomorrow. 

% Wold's decomposition theorem, which can help clarify the relationship between stationarity and causality, says that 
% any discrete time stationary process $\{X_t\}$ can be uniquely written in the form:
% \begin{equation*}
% %\label{equation: Wold decomposition theorem}
% X_t = \sum_{k=0}^{\infty}\psi_k\innov{}_{-k} + D_t,
% \end{equation*}
% where $\{D_t\}$ is a purely deterministic process \citep[see,
% e.g.,][\S5.2.1.5]{Madsen_2008}. 
A stationary process $\{X_t\}$ is said to be \emph{invertible} if there exists
a sequence of coefficients $\{\pi_k\}$ such that $\sum_{k=0}^{\infty}\pi_k^2 <
\infty$, allowing innovations to be written as a power series
\begin{equation}
\label{eqn:invertibilitydefn1} %cn
\innov{} = \Pi(\backshift{})X_t, \quad \hbox{where} \; \Pi(z) = \sum_{k=0}^{\infty}\pi_k z^k.
\end{equation}
The expansion in 
\eqref{eqn:invertibilitydefn1} has many uses, but
an additional reason for assuming invertibility is that it is closely related to
identifiability---it is possible for two different processes to have the same
ACF, however this cannot happen for two \emph{invertible} ones. Therefore
in what follows we restrict ourselves to models that are  causal (and
hence stationary) and in  addition invertible.

A process $\{X_t\}$ is said to be an {\em auto-regressive process of order
$p$}, \AR{p}, if for all $t$:
\begin{equation}
\label{eqn:ARdefn1} %cn
\Phi(\backshift{})X_t = \innov{},
%% \end{equation}
%% \begin{equation}
%% \label{eqn: AR poly defns} %cn
\quad \hbox{where} \quad \Phi(z) = 1 + \sum_{k=1}^{p}\phi_k z^k, \quad \hbox{and} \quad  (\phi_1,\ldots,\phi_p) \in \mathbb{R}^p.
\end{equation}
\AR{p} processes are invertible, stationary and causal if and only if
$\Phi(z)\neq0$ for all $z \in \mathbb{C}$ such that $|z|\leq 1$. From
\eqref{eqn:invertibilitydefn1} invertibility is equivalent to the
process having an \AR{\infty} representation. Similarly, $\{X_t\}$
is said to be a {\em moving average process of order $q$}, \MA{q}, if 
\begin{equation}
\label{eqn:MAdefn1} %cn
X_t = \Theta(\backshift{})\innov{},
%\end{equation}
%\begin{equation}
%\label{eqn: MA poly defns} %cn
\quad \hbox{where} \quad \Theta(z) = 1 + \sum_{k=1}^{q}\theta_k z^k,
\quad \hbox{and} \quad (\theta_1,\ldots,\theta_p) \in \mathbb{R}^q,
\end{equation}
for all
$t$.\footnote{Many authors define $\Phi(z) =
1 - \sum\phi_k z^k$.  Our version emphasises connections between $\Phi$ and (\ref{eqn:ARdefn1}--\ref{eqn:MAdefn1}).}
\MA{q} processes are stationary and causal, and are invertible if and only if
$\Theta(z)\neq0$ for all $z \in \mathbb{C}$ such that $|z|\leq 1$.
%Note that \eqref{eqn:causaldefn1} shows that causality is
%equivalent to the process having an \MA{\infty} representation.

A natural extension of the AR and MA classes arises by combining them \citep{Box_1970}.
The process $\{X_t\}$ is said to be an {\em auto-regressive moving average
(ARMA) process} process of orders $p$ and $q$, \ARMA{p,q}, if for all $t$:
\begin{equation}
\label{eqn:ARMAtimedefn} %cn
\Phi(\backshift{})X_t = \Theta(\backshift{})\innov{}.
\end{equation}
% where $\Phi$ and $\Theta$ are as given in \eqref{eqn:ARdefn1} and
% \eqref{eqn:MAdefn1} respectively.  
%$\{X_t\}$ is said to be an \ARMA{p,q}
%process with mean $\mu$ if $\{X_t-\mu\}$ is an \ARMA{p,q} process. The
%conditions required to ensure that an ARMA process is both causal and
%invertible are inherited from AR and MA processes.  These are equivalent to
%the simple condition that neither $\Phi$ nor $\Theta$ can have any roots on,
%or within, the unit circle. Secondly, 
Although there is no simple closed form
for the ACV of an ARMA process with arbitrary $p$ and $q$, so long as the
process is causal and invertible, then $|\rho(k)| \leq Cr^k$, for $k>0$, i.e.,
it decays exponentially fast.  In other words, although correlation between
nearby points may be high, dependence between distant points is negligible.
%This type of behaviour is typically referred to qualitatively as `short range
%dependence', or `short memory'. Our main focus in this paper is on processes
%that have a correlation structure that \emph{cannot} (entirely) be described
%in this way.

Before turning to `long memory', we require one further result. Under some
extra conditions, % \citep[see e.g.][\S5.4]{Madsen_2008}, 
stationary processes
with ACV $\gamma(\cdot)$ possess a spectral density function (SDF) $f(\cdot)$
defined such that: $\gamma(k) =
\int_{-\pi}^{\pi}e^{ik\lambda}f(\lambda)\,d\lambda$, $\forall k \in
\mathbb{Z}$. This can be inverted to obtain an explicit expression for the
SDF \citep[e.g.][\S4.3]{Brockwell_1991}:
$f(\lambda)=\frac{1}{2\pi}\sum_{k=-\infty}^{\infty}\gamma(k)e^{-ik\lambda}$,
where $-\pi \leq \lambda \leq \pi$.\footnote{Since ACV of
a stationary process is an even function of lag, the above equation implies that the associated
SDF is an even function. One therefore only needs to be interested positive arguments:  $0
\leq \lambda \leq \pi$.} Finally, the SDF of an ARMA process is
\begin{equation}
\label{eqn:ARMAsdf1}
f(\lambda) = \frac{\sigma^2}{2\pi}\frac{|\Theta(e^{-i\lambda})|^2}{|\Phi(e^{-i\lambda})|^2}, \qquad 0 \leq \lambda \leq \pi.
\end{equation}

%\section{Long memory}
%\label{sec:tslongmem}

%As noted above, our paper is concerned with `long memory', where the ACF has power-law decay:
%$\rho(\cdot)$ such
%that $\rho(k) \sim c_\rho\, k^{2d-1}$ as $k
%\rightarrow
%\infty$, for some non-zero constant $c_\rho$, and $0<d<\frac{1}{2}$. 
%The
%parameter $d$ is the memory parameter; if $d=0$ the process does not exhibit
%long memory.
%% Moreover, it is specific about what strength power-law is permitted. 
%% Clearly such processes do not have short memory because this decay cannot be
%% geometrically bounded.
%
%There exist stationary processes that are neither short nor long memory. A
%stationary process is said to have \emph{negative} memory if it has an ACF
%where the same power-law decay holds, but and where $-\frac{1}{2}<d<0$. 
% Some
% authors do not use the term `negative memory', but instead maintain a strict
% dichotomy between short and long memory.   
The restriction $|d|<\frac{1}{2}$
is necessary to ensure stationarity; clearly if $|d|\geq \frac{1}{2}$ the
ACF would not decay. 
%A process with such $d$ can be said to have
%`non-stationary long memory'. 
The continuity between stationary and
non-stationary processes around $|d|=\frac{1}{2}$ is similar to that which
occurs for \AR{1} process with $|\phi_1|\rightarrow 1$ (such processes are
stationary for $|\phi_1|<1$, but the case $|\phi_1|=1$ is the non-stationary
random-walk). 
%A key fact about stationary long (and negative) memory processes
%is that, qualitatively speaking, they are `less stationary' than stationary
%short memory models, and in some sense provide an intermediate type of
%process.

There are a number of alternative definitions of LRD, one of which is
 particularly useful, as it considers the frequency domain:  %During the 1960s,
 %it was noticed that many economic indices had spectra that appeared to `blow
% up' at the origin
%\citep[e.g.][]{Adelman_1965,Granger_1966}. These spectral densities could be
%well-approximated by a function with a pole at the origin---a concept which
%was not widely leveraged by time series analysts at that time, although a key
%point in Mandelbrot's work. 
A stationary process 
has long memory when its SDF follows $ f(\lambda)
\sim c_f \lambda ^{-2d}$, as $\lambda \rightarrow 0^+$ for some positive
constant $c_f$, and where $0<d<\frac{1}{2}$. Similarly, it is said to have
\emph{negative} memory if that relationship holds for $-\frac{1}{2}<d<0$.
%These definitions are clearly related to the original ones since both demand
%power-law behaviours. But they are not entirely equivalent without extra
%conditions (essentially ultimate monotonicity of the \acv{} (ACV)). For a more
%detailed discussion see \citet{Cox_1984}, \citet{Taqqu_2003},
%\citet{Palma_2007} or \citet{Beran_2010}. Throughout, we assume sufficient
%regularity to ensure these definitions are
%all equivalent.

%\subsection{ARFIMA processes}
%\label{sec:arfima}

The simplest way of {\em creating}   a process which exhibits long
memory is through the SDF. Consider  $f(\lambda) =
|1-e^{i\lambda}|^{-2d}$, where $0<|d|<\frac{1}{2}$. By simple algebraic
manipulation, this is equivalently $ f(\lambda) = \left(2\sin
\frac{\lambda}{2} \right)^{-2d}$, from which we deduce that $f(\lambda) \sim
\lambda^{-2d}$ as $\lambda \rightarrow 0^+$. Therefore, assuming stationarity,
the process which has this
SDF (or any scalar multiple of it) is a long memory process. 
More generally, a process having spectral density 
\begin{equation}
\label{eqn:unknowingdefnofFI(d)1} %cn
f(\lambda) =\frac{\sigma^2}{2\pi} 
\left|1-e^{i\lambda}\right|^{-2d}, \qquad 0 < \lambda \leq \pi.
\end{equation}
is called {\em fractionally integrated} with memory parameter $d$, \FI{d} \citep{Barnes_1966,Adenstedt_1974}. The full trichotomy of
negative, short, and long memory is determined solely by $d$. 
%% When $d<0$ or $d>0$, the process has negative or long memory respectively. 
When $d=0$,
the SDF is flat, yielding white noise. 

In practice this
model is of limited appeal to time series analysts because the entire memory structure
determined by just one parameter, $d$. 
One often therefore generalises by taking any short memory SDF
$f^*(\cdot)$, and defining a new SDF: $f(\lambda) = f^*(\lambda)
\left|1-e^{i\lambda}\right|^{-2d}$, $0 \leq \lambda \leq \pi$. An obvious
class of short memory processes to use this way is ARMA. Taking $f^*$ from
\eqref{eqn:ARMAsdf1} yields so-called
auto-regressive fractionally integrated moving average process with parameter
$d$, and orders $p$ and $q$ (\FAR{p,d,q}), having SDF:
\begin{equation}
\label{eqn:spectraldefnofARFIMA(p,d,q)1} %cn
f(\lambda) = \frac{\sigma^2}{2\pi}\frac{|\Theta(e^{-i\lambda})|^2}{|\Phi(e^{-i\lambda})|^2}|1-e^{i\lambda}|^{-2d}, \qquad 0 \leq \lambda \leq \pi.
\end{equation}
Choosing $p=q=0$ recovers 
\FI{d} $ \equiv $ \FID{d}.

%% Skipped bit about other short memory models EXP -> FEXP (Bloomfield),
%% and non-parametric approaches

Practical utility from the perspective of (Bayesian) inference demands
finding a representation in the temporal domain. To obtain this, consider the
operator $(1-\mathcal{B})^d$ for real $d>-1$, which is formally defined using
the generalised form of the binomial expansion \citep[Eq.~13.2.2]{Brockwell_1991}:
\begin{align}
(1-\mathcal{B})^d & =: \sum_{k=0}^{\infty} \pi_k^{(d)} \mathcal{B}^k, 
%\label{eqn: fractional differencing defn} \\
&& \mathrm{where} &
\pi_k^{(d)} & = (-1)^k\frac{1}{\Gamma(k+1)}\frac{\Gamma(d+1)}{\Gamma(d-k+1)}. 
\label{eqn:FIpidefn1}
\end{align}
From this observation, one can show that $X_t=(1-\mathcal{B})^{-d}Z_t$, where
$\{Z_t\}$ is an ARMA process, has SDF
\eqref{eqn:spectraldefnofARFIMA(p,d,q)1}. The operator $(1-\mathcal{B})^d$ is
called the `fractional differencing' operator since it allows a degree of
differencing between zeroth and first order. The process $\{X_t\}$ is  
fractionally `inverse-differenced', i.e.\ it is an `integrated' process. 
The
operator is used to redefine both the \FID{d} and more general \FAR{p,d,q}
processes in the time domain.  
A process $\{X_t\}$ is an \FID{d} process if for all $t$:
%\begin{equation}
%\label{eqn:FIeqn}
$(1-\mathcal{B})^d X_t=\innov{}$.
%\end{equation}
Likewise, a process $\{X_t\}$ is an \FAR{p,d,q} process if for all $t$:
%\begin{equation}
%\label{eqn:ARFIMAeqn}
$\Phi(\mathcal{B})(1-\mathcal{B})^d X_t=\Theta(\mathcal{B})\innov{}$,
%\end{equation}
where $\Phi$ and $\Theta$ are given in \eqref{eqn:ARdefn1} and
\eqref{eqn:MAdefn1} respectively. 
%$\{X_t\}$ is said to be an
%\FAR{p,d,q} process with mean $\mu$ if $\{X_t-\mu\}$ is an \FAR{p,d,q}
%process.
%An important result \citep[e.g.][\S3.2.1]{Palma_2007} is that the conditions
%required to ensure that an ARFIMA process is stationary, causal and invertible
%are inherited from ARMA processes when
%$|d|<\frac{1}{2}$. 
% Note that $d$ outside this range still has a useful
% interpretation: for example, an \FAR{p,d,q} process that has
% $d\in(\frac{1}{2},\frac{3}{2})$ will be non-stationary but its first
% difference will be a stationary \FAR{p,d-1,q} process.

%% Skipped confirmation of LM via definition 2.22 (LM 4/4)
%% Not sure if this next bit is really necessary, but it seems to add color
%% by bringing us full circle.

Finally, to connect back to our first definition of long memory, consider the
ACV of the \FID{d} process. By using the definition of spectral density to
directly integrate \eqref{eqn:unknowingdefnofFI(d)1}, and an alternative
expression for $\pi_k^{(d)}$ in \eqref{eqn:FIpidefn1}
\begin{equation}
\pi_k^{(d)} = \frac{1}{\Gamma(k+1)}\frac{\Gamma(k-d)}{\Gamma(-d)},
\label{eqn:altpi1}
\end{equation}
%it is straightforward
%\citep[using an identity from][]{Gradshteyn_2000} to
one can obtain the
following representation of the ACV of the \FID{d} process:
\begin{align}
\gamma_d(k;\sigma) &= \sigma^2\frac{\Gamma(1-2d)}{\Gamma(1-d)\Gamma(d)}\frac{\Gamma(k+d)}{\Gamma(1+k-d)}. \label{eqn:gammaddefn}
\intertext{Because the parameter $\sigma^2$ is just a scalar multiplier, we
may simplify notation by defining $\gamma_d(k) = \gamma_d(k;
\sigma)/\sigma^2$, whereby $\gamma_d(\cdot) \equiv \gamma_d(\cdot;1)$. Then
the ACF is:}
\rho_d(k) &= \frac{\Gamma(1-d)}{\Gamma(d)}\frac{\Gamma(k+d)}{\Gamma(1+k-d)},
\end{align}
from which Stirling's approximation gives $\rho_d(k) \sim
\frac{\Gamma(1-d)}{\Gamma(d)} k^{2d-1}$, confirming a power-law relationship
for the ACF. Finally, note that \eqref{eqn:altpi1} can be used to
represent \FID{d} as an \AR{\infty} process, as $X_{t} +
\sum_{k=1}^{\infty}\pi_k^{(d)}X_{t-k} = \innov{}$.  And noting that
$\psi_k^{(d)} = \pi_k^{(-d)}$, leads to the following \MA{\infty} analog: $X_t
= \sum_{k=0}^{\infty}\frac{1}{\Gamma(k+1)}\frac{\Gamma(k+d)}{\Gamma(d)}
\innov{}_{-k}$.

\section{Likelihood evaluation for Bayesian inference}
\label{sec:lik}

For now we restrict our attention to (a Bayesian) analysis of an \FID{d}
process, having no short-ranged ARMA components, placing emphasis
squarely on the memory parameter $d$. We present two alternative
likelihoods, `exact' and `approximate'.  The exact one is not
original, but is presented here to highlight some important
(particularly computational) issues that prevent effective
use in a Bayesian context where MCMC inference requires thousands of
evaluations. The approximate one represents a novel contribution.

\subsection{Exact likelihood calculation}
\label{sec:exactlik}

For Gaussian processes, all information is contained in the covariance
structure, so inference about memory behaviour only can proceed through
the covariance matrix $\Sigma$ given $\sigma$ and $d$:
$\Sigma(\sigma,d)_{(i,j)} = \sigma^2\gamma_d(i-j)$,
where $\gamma_d(\cdot) = \sigma^2\frac{\Gamma(1-2d)}{\Gamma(1-d)\Gamma(d)}\frac{\Gamma(k+d)}{\Gamma(1+k-d)}$. Therefore,
 the vector $\mathbf{X}=(X_1,\ldots,X_n)^\top$ is
MVN with mean $\mu\mathbf{1}_n$  and covariance $\Sigma(\sigma,d)$, so the
likelihood is:
\[ %\begin{equation}
%\label{eqn:likelihood}
 L(\mathbf{x} | \mu,\sigma,d)=(2\pi)^{-\frac{n}{2}} \{\det [\Sigma(\sigma,d)] \}^{-\frac{1}{2}}\exp\left[-\frac{1}{2}(\mathbf{x}-\mu \mathbf{1}_n)^t \Sigma(\sigma,d)^{-1}(\mathbf{x}-\mu \mathbf{1}_n)\right].
\] %\end{equation}
To simplify the development below, write $\sigma^2\Sigma_d$ as a shorthand for
$\Sigma(\sigma,d)$, whereby we have $\det [\Sigma(\sigma,d)] = \sigma^{2n}
\det (\Sigma_d)$.  Also, denote the quadratic term as:
$Q(\mathbf{x}|\mu,d) = (\mathbf{x}-\mu \mathbf{1}_n)^t
\Sigma_d^{-1}(\mathbf{x}-\mu \mathbf{1}_n)$, so the log-likelihood can be
re-written as
\begin{equation}
\label{eqn:log-likelihoodwithQ}
\ell(\mathbf{x} |  \mu,\sigma,d) = -n\log \sigma - \frac{1}{2}\log[\det (\Sigma_d)]  - \frac{1}{2\sigma^2}Q(\mathbf{x}|\mu,d).
\end{equation}
Numerical evaluation requires computing 
the determinant and inverse of a dense, symmetric positive-definite
$n\times n$ matrix, an $\mathcal{O}(n^3)$ operation---too
slow for the large $n$ typically encountered in long memory 
contexts.\footnote{$\Sigma_d$ is often also poorly conditioned, complicating decomposition
\citep[appendix A]{Chen_2006}.} 
% \footnote{More precisely, one must compute the determinant and the
%term $Q(\mathbf{x}|\mu,d)$.} 
% \citet{Sowell_1992} was the first to attempt these calculations
Simplifications arise upon recognising that $\Sigma_d$ is symmetric {\em Toeplitz}, being
expressible by just $n$ scalars $c_0,\ldots,c_{n-1}$, i.e.,
$\Sigma(d, \sigma)_{i,j} = c_{i-j}$ for $i \geq j$.   The Durbin--Levinson
algorithm \citep[\S4.1.2]{Palma_2007} exploits this form, yielding an
$\mathcal{O}(n^2)$ cost. % \citep{Ammar_1998}. 
However even that remains too large in practice for most 
applications.

\subsection{Approximate likelihood calculation}
\label{sec:alik}

Here we develop an efficient scheme for evaluating the
(log) likelihood, via approximation. Throughout, suppose that we have
observed the vector $\mathbf{x}=(x_1,\ldots,x_n)^\top$ as a
realisation of a stationary, causal and invertible \FID{d} process
$\{X_t\}$ with mean $\mu \in \mathbb{R}$. The innovations will be
assumed to be independent, and taken from a zero-mean {\em location-scale
probability density} $f(\cdot;0,\sigma,\bm{\lambda})$, which means the density
can be written as 
$f(x; \delta,\sigma,\bm{\lambda}) 
\equiv \frac{1}{\sigma}f\left(\frac{x-\delta}{\sigma} ; 0,1,\bm{\lambda}\right)$.
The parameters $\delta$ and $\sigma$ are called the `location' and `scale'
parameters respectively. The $m$--dimensional $\bm{\lambda}$ is a 
`shape' parameter (if it exists, i.e. $m>0$).  An common example is
the Gaussian $\mathcal{N}(\mu,\sigma^2)$, where $\delta \equiv \mu$ and there
is $\bm{\lambda}$. We classify the four parameters
$\mu$, $\sigma$, $\bm{\lambda}$, and $d$, into three
distinct classes: (1) the mean of process, $\mu$; (2) innovation distribution
parameters, $\bm{\upsilon}=(\sigma,\bm{\lambda})$; and (3) memory
structure, $d$. Together, $\bm{\psi} =
(\mu,\bm{\upsilon},\bm{\omega})$, where $\bm{\omega}$ will later encompass the
short-range parameters $p$ and $q$. 
% The parameter spaces follow  obvious
% notation: $\bm{\upsilon} \in \Upsilon$, $\bm{\omega}
% \in \Omega$ and $\bm{\psi} \in \Psi = \mathbb{R}\times \Upsilon \times \Omega$.

Our proposed likelihood approximation uses a truncated \AR{\infty} approximation (cf. \cite{Haslett_1989}).
We first re-write the \AR{\infty} approximation of \FID{d} to
incorporate the unknown parameter $\mu$, and drop the $(d)$
superscript for convenience: $X_{t}-\mu = \innov{} -
\sum_{k=1}^{\infty}\pi_k(X_{t-k}-\mu)$. Then we truncate this \AR{\infty}
representation to obtain an \AR{P} one, with $P$ large
enough to retain low frequency effects, e.g., $P=n$. 
% suggest $P=n$ as using a shorter truncation would compromise accuracy
% (albeit minimally), whereas any longer truncation would result in the
% latent variables exceeding the observed data series and introduce
% unnecessary complication. 
We denote: $\Pi_P = \sum_{k=0}^P\pi_k$ and,
with $\pi_0=1$, rearrange terms to obtain the following
modified model:
\begin{equation}
\label{eqn:ARmodelassumptionconvenienform}
X_{t} = \innov{} + \Pi_P\mu - \sum_{k=1}^{P}\pi_k X_{t-k}.
\end{equation}

It is now possible to write down a \emph{conditional} likelihood. For
convenience the notation $\mathbf{x}_k=(x_1,\ldots,x_k)^\top$ for
$k=1,\ldots,n$ will be used (and $\mathbf{x}_0$ is interpreted as appropriate
where necessary). Denote the unobserved $P$--vector of random variables
$(x_{1-P},\ldots,x_{-1},x_0)^\top$  by $\mathbf{x}_A$ (in the Bayesian
context these will be `auxiliary', hence `$A$').
Consider the likelihood $L(\mathbf{x}|\bm{\psi})$ as a joint density which can
be factorised as a product of conditionals. Writing $g_t(x_t |
\mathbf{x}_{t-1},\bm{\psi})$ for the density of $X_t$ conditional on
$\mathbf{x}_{t-1}$, we obtain $L(\mathbf{x}|\bm{\psi}) =
\prod_{t=1}^{n}g_t(x_t|\mathbf{x}_{t-1},\bm{\psi})$.

This is still of little use because the $g_t$ may have a complicated
form. However by further
conditioning on $\mathbf{x}_A$, and writing $h_t(x_t |\mathbf{x}_A,
\mathbf{x}_{t-1},\bm{\psi})$ for the density of $X_t$ conditional on
$\mathbf{x}_{t-1}$ \emph{and} $\mathbf{x}_A$, we obtain:
$L(\mathbf{x}|\bm{\psi},\mathbf{x}_A) =
\prod_{t=1}^{n}h_t(x_t|\mathbf{x}_{A},\mathbf{x}_{t-1},\bm{\psi})$. Returning
to \eqref{eqn:ARmodelassumptionconvenienform} observe that, conditional on
both the observed and \emph{un}observed past values, $X_t$ is simply
distributed according to the innovations' density $f$ with a suitable change in
location: $ X_t|\mathbf{x}_{t-1},\mathbf{x}_{A} \sim f\left(\cdot;
\left[\Pi_P\mu - \sum_{k=1}^{P}\pi_k x_{t-k}\right]
,\sigma,\bm{\lambda}\right)$. Then using location-scale representation:
\begin{align}
h_t(x_t|\mathbf{x}_{A},\mathbf{x}_{t-1},\bm{\psi}) &\approx f\left(x_t; \left[\Pi_P\mu - \sum_{k=1}^{P}\pi_k x_{t-k}\right],\sigma,\bm{\lambda}\right) \label{eqn:defnofcvec} \\
& \equiv \frac{1}{\sigma} f\left(\frac{c_t-\Pi_P\mu}{\sigma};0,1,\bm{\lambda}\right),
\quad \hbox{where} \quad 
c_t = \sum_{k=0}^{P}\pi_k x_{t-k},\qquad t=1,\ldots,n. \nonumber
\end{align}
Therefore, $L(\mathbf{x}|\bm{\psi},\mathbf{x}_A) \approx \sigma^{-n}\prod_{t=1}^{n}  f\left(\frac{c_t-\Pi_P\mu}{\sigma};\bm{\lambda}\right)$,
or equivalently:
\begin{equation}
\label{eqn:formofapproxloglikelihood1}
\ell(\mathbf{x}|\bm{\psi},\mathbf{x}_A) \approx -n\log\sigma + \sum_{t=1}^{n}  \log\left\{f\left(\frac{c_t-\Pi_P\mu}{\sigma};\bm{\lambda}\right)\right\}.
\end{equation}

Evaluating this expression efficiently depends upon
efficient calculation of $\bm{c}=(c_1,\ldots,c_n)^t$ and 
$\log f(\cdot)$. From \eqref{eqn:defnofcvec},
$\bm{c}$ is a convolution of the augmented data,
$(\mathbf{x},\mathbf{x}_A)$, and coefficients depending on 
$d$, which can be evaluated quickly in \textsf{R}
via \texttt{convolve} via  FFT. Consequently,
evaluation of the \emph{conditional} likelihood in the Gaussian case costs
only $\mathcal{O}(n\log n)$---a clear improvement over the `exact' method.
Obtaining the \emph{un}conditional likelihood requires marginalisation over
$\mathbf{x}_A$, which is analytically infeasible. However this conditional
form will suffice in the context of our Bayesian inferential scheme, presented
below.

\section{A Bayesian approach to long memory inference}
\label{sec:long}

We are now ready to consider Bayesian inference for \FID{d} processes. 
% The development will be divided into three parts. Throughout we assume familiarity
% with simulations-based inference techniques like Markov chain Monte Carlo
% (MCMC), Metropolis--Hastings (MH), Gibbs sampling, and combinations thereof.
%
Our method can be succinctly described as a modernisation of
the blocked MCMC method of \citet{Pai_1998}.  Isolating parameters by
blocking provides significant scope for modularisation which helps
accommodate our extensions for short memory. Pairing with
efficient likelihood evaluations allows much
longer time series to be entertained than ever before. Our description
begins with appropriate specification of priors which are more general
than previous choices, yet still encourages tractable inference.  We
then provide the relevant updating calculations for all parameters,
including those for auxiliary parameters $\mathbf{x}_A$.

%\subsubsection*{Priors}

We follow earlier work \citep{Koop_1997,Pai_1998} and assume {\em a priori}
independence for components of $\psi$.  Each component will
leverage familiar prior forms with diffuse versions as limiting cases.
Specifically, we use a diffuse Gaussian prior on $\mu$: $\mu \sim
\mathcal{N}(\mu_0,\sigma^2_{0})$, with $\sigma_0$ large.
The improper flat prior is obtained as the limiting distribution when
$\sigma_0 \rightarrow \infty$: $ p_\mu(\mu) \propto 1$. We place a
gamma prior on the precision $\tau=\sigma^{-2}$ implying 
a {\em Root-Inverse Gamma} distribution $\mathcal{R}(\alpha_0,\beta_0)$ for $\sigma$, with density $f(\sigma) =
\frac{2}{\Gamma(\alpha)}{\beta_0}^{\alpha_0} \sigma^{-(2\alpha_0 +1)}
\exp\left(-\frac{\beta_0}{y^2}\right)$, $\sigma>0$. 
A diffuse/improper prior is obtained as the
limiting distribution when $\alpha_0,\beta_0 \rightarrow 0$:
$p_\sigma(\sigma) \propto \sigma^{-1}$.  Finally, we specify
$d\sim\mathcal{U}\left(-\frac{1}{2},\frac{1}{2}\right)$.

%\subsubsection*{Posterior updating of parameters}

{\bf Updating $\mu$:}  Following
\citet{Pai_1998}, we use a symmetric random walk (RW) MH update with proposals
$\xi_\mu
\sim \mathcal{N}(\mu,\sigma_\mu^2)$, for some $\sigma_\mu^2$. The acceptance
ratio is
\begin{equation}
A_\mu(\mu,\xi_\mu) = 
\sum_{t=1}^{n} \log\left\{\!f\left(\frac{c_t-\Pi_P\xi_\mu}{\sigma};\bm{\lambda}\right)\!\right\} - \sum_{t=1}^{n}  
\log\left\{\!f\left(\frac{c_t-\Pi_P\mu}{\sigma};\bm{\lambda}\right)\!\right\}
 + \log\left[\frac{p_\mu(\xi_\mu)} {p_\mu(\mu)} \right]
\label{eqn:formofapproxloglikelihood}
\end{equation}
under the approximate likelihood. With the exact
likelihood, recall \eqref{eqn:log-likelihoodwithQ} to obtain:
\begin{align*}
A_\mu(\mu,\xi_\mu) &= \frac{1}{2\sigma^2}\left[ Q(\mathbf{x}|\mu,d) - Q(\mathbf{x}|\xi_\mu,d) \right] + \log\left[\frac{p_\mu(\xi_\mu)} {p_\mu(\mu)} \right].
\end{align*}

{\bf Updating $\sigma$:} 
We diverge from \citet{Pai_1998} here, who suggest independent MH with
moment-matched inverse gamma proposals, finding poor performance under
poor moment estimates.
We instead prefer a Random Walk (RW) Metropolis-Hastings (MH) approach, which we conduct in log space since the
domain is $\mathbb{R}^+$. Specifically, set: $\log\xi_\sigma =
\log\sigma + \upsilon$, where $\upsilon \sim \mathcal{N}(0,\sigma^2_\sigma)$
for some $\sigma^2_\sigma$. $\xi_\sigma|\sigma$ is log-normal and we obtain:
$\frac{q(\sigma;\xi_\sigma)}{q(\xi_\sigma;\sigma)} = \frac{\xi_\sigma}{\sigma}$.
Recalling \eqref{eqn:formofapproxloglikelihood} the MH acceptance ratio 
under the approximate likelihood is
\begin{align*}
A_\sigma(\sigma,\xi_\sigma)  &=  \sum_{t=1}^{n} \log\left\{f\left(\frac{c_t-\Pi_P\mu}{\xi_\sigma};\bm{\lambda}\right)\right\} - \sum_{t=1}^{n}  \log\left\{f\left(\frac{c_t-\Pi_P\mu}{\sigma};\bm{\lambda}\right)\right\} \\
&\;\;\; + \log\left[\frac{p_\sigma(\xi_\sigma)} {p_\sigma(\sigma)} \right] + (n-1)\log\left[\frac{\sigma}{\xi_\sigma} \right]. 
\intertext{When using the exact likelihood, \eqref{eqn:log-likelihoodwithQ} gives}
A_\sigma(\sigma,\xi_\sigma)  &=  \frac{1}{2} \left(\frac{1}{\sigma^2} - \frac{1}{\xi_\sigma^2}\right)Q(\mathbf{x}|\mu,d)  + \log\left[\frac{p_\sigma(\xi_\sigma)} {p_\sigma(\sigma)} \right] + (n-1)\log\left[\frac{\sigma}{\xi_\sigma} \right].
\end{align*}

The MH algorithm, applied alternately in a Metropolis-within-Gibbs fashion to
the parameters $\mu$ and $\sigma$, works well. However {\em actual} Gibbs
sampling is an efficient alternative in this two-parameter case (i.e., for
known $d$).  Since inference for $d$ is a primary goal,
we have relegated a derivation of the resulting updates to Appendix
\ref{sec:gibbs}.

{\bf Update of $d$:} Updating the memory parameter $d$ is far less
straightforward than either $\mu$ or $\sigma$. Regardless of the innovations'
distribution, the conditional posterior
$\pi_{d|\psi_{-d}}(d|\bm{\psi}_{-d},\mathbf{x})$ is not amenable to Gibbs sampling. 
We use RW proposals from truncated Gaussian
$\xi_d \sim \mathcal{N}^{(a,b)}(\mu,\sigma^2)$, with density
\begin{equation}
f(x;\mu,\sigma,a,b) = \frac{1}{\sigma}\frac{\phi[(x-\mu)/\sigma]}{\Phi[(b-\mu)/\sigma]-\Phi[(a -\mu)/\sigma]},\qquad a<x<b.
\label{eqn:univariatetruncatedgaussian}
\end{equation}
In particular, we use $\xi_d|d \sim \mathcal{N}^{(-1/2,1/2)}(d,\sigma^2_d)$
via rejection sampling from $\mathcal{N}(d,\sigma^2_d)$ until
$\xi_d \in$ \dint{}. Although this may seem inefficient, it is
perfectly acceptable: as an example, if $\sigma_d=0.5$ the expected number of
required variates is still less than 2, regardless of $d$. More refined
methods of directly sampling from truncated normal distributions exist---see
for example \citet{Robert_1995}---but we find little added benefit in our
context.

A useful cancellation in $q(d;\xi_d)/q(\xi_d;d)$ obtained from
\eqref{eqn:univariatetruncatedgaussian} yields
\begin{equation*}
%\begin{split}
A_d  =  \ell(\mathbf{x}|\xi_d,\bm{\psi}_{-d}) - \ell(\mathbf{x}|d,\bm{\psi}_{-d}) + \log\left[\frac{p_d(\xi_d)} {p_d(d)} \right] + \log\left\{\frac{\Phi[(\frac{1}{2}-d)/\sigma_d]-\Phi[(-\frac{1}{2} -d)/\sigma_d]}{\Phi[(\frac{1}{2}-\xi_d)/\sigma_d]-\Phi[(-\frac{1}{2} -\xi_d)/\sigma_d]}\right\}.
%\end{split}
\end{equation*}
Denote $\xi_{c_t} = \sum_{k=0}^{P}\xi_{\pi_k} x_{t-k}$ for
$t=1,\ldots,n$, where $\{\xi_{\pi_k}\}$ are the proposed coefficients
$\{\pi_k^{(\xi_d)} \}; \pi_k^{(d)} = \frac{1}{\Gamma(k+1)}\frac{\Gamma(k-d)}{\Gamma(-d)}$. Denote $\xi_{\Pi_P} = \sum_{k=0}^P\xi_{\pi_k}$. Then in the approximate case:
\begin{align}
\label{eqn: A_d approx}
A_d &= \sum_{t=1}^{n} \log\left\{ f\left(\frac{\xi_{c_t}-\xi_{\Pi_P}\mu}{\sigma};\bm{\lambda}\right)\right\} - \sum_{t=1}^{n} \log\left\{ f\left(\frac{c_t-\Pi_P\mu}{\sigma};\bm{\lambda}\right)\right\} \nonumber \\ 
&\;\;\; + \log\left[\frac{p_d(\xi_d)} {p_d(d)} \right] + \log\left\{\frac{\Phi[(\frac{1}{2}-d)/\sigma_d]-\Phi[(-\frac{1}{2} -d)/\sigma_d]}{\Phi[(\frac{1}{2}-\xi_d)/\sigma_d]-\Phi[(-\frac{1}{2} -\xi_d)/\sigma_d]}\right\}.
\intertext{In the exact likelihood case, from \eqref{eqn:log-likelihoodwithQ} we obtain:}
A_d &= \frac{1}{2}\log[\det (\Sigma_d)] - \frac{1}{2}\log[\det (\Sigma_{\xi_d})] + \frac{1}{2\sigma^2}\left[Q(\mathbf{x}|\mu,d) -Q(\mathbf{x}|\mu,\xi_d) \right] \nonumber \\
&\;\;\; + \log\left[\frac{p_d(\xi_d)} {p_d(d)} \right] + \log\left\{\frac{\Phi[(\frac{1}{2}-d)/\sigma_d]-\Phi[(-\frac{1}{2} -d)/\sigma_d]}{\Phi[(\frac{1}{2}-\xi_d)/\sigma_d]-\Phi[(-\frac{1}{2} -\xi_d)/\sigma_d]}\right\}.
\end{align}

{\bf Optional update of $\mathbf{x}_A$:}
When using the approximate likelihood method, one must account for the
auxiliary variables $\mathbf{x}_A$, a $P$--vector (where $P=n$ is
sensible). 
We find that, in practice, it is not necessary to update all the auxiliary
parameters at each iteration. In fact the method can be shown to work
perfectly well, empirically, if we \emph{never} update them, provided they are
given a sensible initial value (such as the sample mean of the observed data
$\bar{x}$).  This is not an uncommon tactic in the long memory (big-$n$)
context \cite[e.g.,][]{Beran_1994a}; for further discussion refer to
\citet[][Appendix C]{graves:phd}.

For a full MH approach, we recommend an independence sampler to `backward
project' the observed time series. Specifically, first relabel the observed
data: $ y_{-i} = x_{i+1}$, $i=0,\ldots n-1$. Then use the vector
$(y_{-(n-1)},\ldots,y_{-1},y_0)^t$ to generate a new vector of length $n$,
$(Y_1,\ldots,Y_n)^t$ where $Y_t$ via
\eqref{eqn:ARmodelassumptionconvenienform}:
$Y_{t} = \innov{} + \Pi_P\mu - \sum_{k=1}^{n}\pi_k Y_{t-k}$,
where the coefficients $\{\pi\}$ are determined by the current value of the
memory parameter(s). Then take the proposed $\mathbf{x}_A$, denoted
$\bm{\xi}_{\mathbf{x}_A}$, as the reverse sequence: $\xi_{x_{-i}} = y_{i+1}$,
$i = 0,\ldots,n-1$. Since this is an independence sampler, calculation of the
acceptance probability is straightforward. It is only necessary to evaluate
the proposal density $q(\bm{\xi}_{\mathbf{x}_A}|\mathbf{x},\bm{\psi})$. But
this is easy using the results from section \ref{sec:alik}. For simplicity, we
prefer uniform prior for $\mb{x}_A$.

Besides simplicity, justification for this approach lies primarily in is
preservation of the auto-correlation structure---this is clear since the
ACF is symmetric in time. The proposed vector has a low acceptance rate,
and the potential remedies (e.g., multiple-try methods) seem unnecessarily
complicated given the success of the simpler method.

\section{Extensions to accommodate short memory}
\label{sec:short}

Simple \FID{d} are mathematically convenient but have limited practical
applicability because the entire memory structure is determined by just one
parameter, $d$. Although $d$ is often of primary interest, it may
be unrealistic to assume no short memory effects. This issue is often
implicitly acknowledged since semi-parametric estimation methods, such as
those used as comparators in Section
\ref{sec:emplm}, are motivated by a desire to circumvent the problem of
specifying precisely (and inferring) the form of short memory (i.e., the
values of $p$ and $q$ in an ARIMA model). Full parametric Bayesian modelling
of \FAR{p,d,q} processes represents an essentially untried alternative,
primarily due to computational challenges.  Related, more discrete,
alternatives show potential. \citet{Pai_1998} considered all four
models with $p,q \leq 1$, whereas \citet{Koop_1997}~considered 
sixteen  with $p,q \leq 3$.

Such approaches, especially ones allowing larger $p,q$, can be computationally
burdensome as much effort is spent modelling unsuitable processes towards a goal
(inferring $p,q$) which is not of primary interest ($d$ is). To develop an
efficient, fully-parametric, Bayesian method of inference that properly
accounts for varying models, and to marginalise out these nuisance quantities,
we use reversible-jump (RJ) MCMC \citep{Green_1995}. We extend the
parameter space 
to include the set of models ($p$ and $q$), with chains
moving \emph{between} and within models, and focus on the marginal posterior
distribution of $d$ obtained by (Monte Carlo) integration over all models and
parameters therein.  RJ methods have previously been applied to both
auto-regressive models
\citep{Vermaak_2004}, and full ARMA models
\citep{Ehlers_2006,Ehlers_2008}. In the long memory context,
\citet{Holan_2009} applied RJ to FEXP processes. However for
ARFIMA, the only related work we are aware of is by
\citet{Egrioglu_2010} who demonstrated a promising if limited alternative.

Below we show how the likelihood may be calculated with extra short-memory
components when $p$ and $q$ are known, and subsequently how Bayesian inference
can be applied in this case. Then, the more general case of  unknown $p$ and
$q$ via RJ is described.

\subsection{Likelihood derivation and inference for known short memory}
\label{sec:pdqlik}

Recall that short memory components of an ARFIMA process are defined by the AR
and MA polynomials, $\Phi$ and $\Theta$ respectively, (see Section  
\ref{sec:preamble}). Here, we distinguish between the
polynomial, $\Phi$, and the vector of its coefficients,
$\bm{\phi}=(\phi_1,\ldots,\phi_p)$. When the polynomial degree is required
explicitly, bracketed superscripts will be used; $\Phi^{(p)}$,
$\bm{\phi}^{(p)}$, $\Theta^{(p)}$, $\bm{\theta}^{(p)}$, respectively. 

We combine the short memory parameters $\bm{\phi}$ and $\bm{\theta}$ with $d$
to create a single `memory' parameter, $\bm{\omega} =
(\bm{\phi},\bm{\theta},d)$. For a given unit-variance \FAR{p,d,q} process,
we denote its ACV by $\gamma_{\bm{\omega}}(\cdot)$,  with $\gamma_d(\cdot)$ and
$\gamma_{\bm{\phi},\bm{\theta}}(\cdot)$ those of the relevant
unit-variance \FID{d} and \ARMA{p,q} processes respectively. The SDF of the
unit-variance \FAR{p,d,q} process is written as $f_{\bm{\omega}}(\cdot)$, and
its covariance matrix is $\Sigma_{\bm{\omega}}$. Therefore, in the general
Gaussian \FAR{p,d,q} case, we can update the likelihood in
\eqref{eqn:log-likelihoodwithQ} to obtain
%\begin{equation*}
$
\ell(\mathbf{x} |  \mu,\sigma,\bm{\omega}) = -n\log \sigma - \frac{1}{2}\log[\det (\Sigma_{\bm{\omega}})]  - \frac{1}{2\sigma^2}Q(\mathbf{x}|\mu,\bm{\omega})$.
% \end{equation*}

An `exact' likelihood evaluation requires an explicit calculation
of the ACV $\gamma_{\bm{\omega}}(\cdot)$, however there is no
simple closed form for arbitrary ARFIMA processes.
Fortunately, our proposed
approximate likelihood method of section \ref{sec:alik} can be ported over
directly. Given the coefficients $\{\pi_k^{(d)}\}$ and polynomials $\Phi$ and
$\Theta$, it is trivial to calculate the $\{\pi_k^{(\bm{\omega})}\}$
coefficients required by again applying the numerical methods of
\citet[\S3.3]{Brockwell_1991}.
  
% Bayesian posterior sampling requires MCMC. 
 To focus the exposition, % and fix ideas, 
 consider the simple, yet useful,
\FAR{1,d,0} model where the full memory parameter is
$\bm{\omega}=(d,\phi_1)$. Because the parameter spaces of $d$ and $\phi_1$ are
independent, it is simplest to update each of these parameters separately; $d$
with the methods of section
\ref{sec:long} and $\phi_1$ similarly: $\xi_{\phi_1} | \phi_1 \sim
\mathcal{N}^{(-1,1)}(\phi_1,\sigma^2_{\phi_1})$, for some $\sigma^2_{\phi_1}$.
In practice however, the posteriors of $d$ and $\phi_1$ typically exhibit
significant correlation so independent proposals are inefficient. One solution
would be to reparametrise to some $d^*$ and orthogonal $\phi_2^*$, but the
interpretation of $d^*$ would not be clear. An alternative to explicit
reparametrisation is to update the parameters jointly, but in such a way that
proposals are aligned with the correlation structure. This will ensure a
reasonable acceptance rate and mixing.

To propose parameters in the manner described above, a two-dimensional,
suitably truncated Gaussian random walk, with covariance matrix aligned with
the posterior covariance, is required.  To make proposals of this sort, and
indeed for arbitrary $\bm{\omega}$ in larger $p$ and $q$ cases, requires
sampling from a {\em hypercuboid}-truncated MVN
$\mathcal{N}_r^{(\mathbf{a},\mathbf{b})}(\bm{\omega,\Sigma_{\bm{\omega}}})$,
where $(\mathbf{a},\mathbf{b})$ describe the coordinates of the hypercube. We
find that rejection sampling based unconstrained similarly parameterised MVNs
samples [e.g., using {\tt mvtnorm} \citep{R_mvtnorm}] works well, because in
the RW setup the mode of the distribution always lies inside the hypercuboid.
% Out-of-the box methods designed to work well for all/other cases, e.g., the
% {\tt tmvtnorm} package for {\sf R} \citep{tmvtnorm} provide slightly better
% results when $r$ is large. 
Returning to the specific \FAR{1,d,0} case, clearly
$r=2$, $\mathbf{b}=(0.5,1)$ and $\mathbf{a}=-\mathbf{b}$, is appropriate.
Calculation of the MH acceptance ratio
$A_{\bm{\omega}}(\bm{\omega},\bm{\xi}_{\bm{\omega}})$ is trivial; it simply
requires numerical evaluation of
$\Phi_r(\cdot;\cdot,\bm{\Sigma}_{\bm{\omega}})$, e.g., via {\tt mvtnorm},
since the ratios of hypercuboid normalisation terms would cancel.  We find
that initial $\phi^{[0]}$ chosen uniformly  in $\mathcal{C}_1$, i.e.\ the
interval $(-1,1)$, and $d^{[0]}$ are systematically from
$\{-0.4,-0.2,0,0.2,0.4\}$ work well. Any choice of prior for $\bm{\omega}$ can
be made, although we prefer flat (proper) priors.

The only technical difficulty is the choice of proposal covariance matrix
$\Sigma_{\bm{\omega}}$. Ideally, it would be aligned with the posterior
covariance---however this is not \emph{a priori} known. 
We find that running a `pilot' chain
with independent proposals via
$\mathcal{N}_r^{(\mathbf{a},\mathbf{b})}(\bm{\omega},\sigma_{\bm{\omega}}^2
\bm{I}_r)$ can help choose a $\Sigma_{\bm{\omega}}$.
A rescaled version of
the sample covariance matrix from the pilot posterior chain, following
\citet{Roberts_2001}, works well [see Section \ref{sec:empsm}].  

\subsection{Unknown short memory form}
\label{sec:rj}

We now expand the parameter space to include models $M \in
\mathcal{M}$, the set of ARFIMA models with 
$p$ and $q$ short memory parameters, indexing the size
of the parameter space $\Psi^{(M)}$. For our `transdimensional moves',
we only consider adjacent models, on which we will be more specific later. For
now, note that the choice of bijective function mapping between models
spaces (whose Jacobian term appears in the acceptance ratio), is crucial to
the success of the sampler. To illustrate, consider transforming from
$\Phi^{(p+1)}\in \mathcal{C}_{p+1}$ down to $\Phi^{(p)}\in
\mathcal{C}_{p}$. This turns out to be a non-trivial problem
however because, for $p>1$, $\mathcal{C}_p$ has a very complicated shape. The
most natural map would be: $(\phi_1,\ldots,\phi_{p},\phi_{p+1}) \longmapsto
(\phi_1,\ldots,\phi_{p})$. However there is no guarantee that the image will
lie in $\mathcal{C}_{p}$. Even if the model dimension is fixed, difficulties are still
encountered; a natural proposal method would be to update each component of
$\bm{\phi}$ separately but, because of the awkward shape of $\mathcal{C}_p$,
the `allowable' values for each component are a complicated function of the
others. % Note that this problem was avoided in section
% \ref{sec:pdqlik} because only the simple case $p=1$ was considered.
%
Nontrivial proposals are required. 

A potential approach is to reparametrise in
terms of the inverse roots (poles) of $\Phi$, as advocated by
\citet{Ehlers_2006,Ehlers_2008}: By writing $\Phi(z) =
\prod_{i=1}^{p}(1-\alpha_i z)$, we have that $\bm{\phi}^{(p)} \in
\mathcal{C}_p \Longleftrightarrow  |\alpha_i | <1$ for all $i$. This looks
attractive because it transforms $\mathcal{C}_p$ into $D^p = D \times \cdots
\times D$ ($p$ times) where $D$ is the open unit disc, which is easy to
sample from. But this method has serious drawbacks when we consider
the RJ step. To decrease dimension, the natural map would be to  remove
one of the roots from the polynomial. But because it is assumed that $\Phi$
has real coefficients (otherwise the model has no realistic interpretation),
any complex $\alpha_i$ must appear as conjugate pairs. There is then no
obvious way to remove a root; a contrived method might be to remove the
conjugate pair and replace it with a real root with the same modulus, however
it is unclear how this new polynomial is related to the original, and
to other aspects of the process, like ACV.

\subsubsection*{Reparametrisation of $\Phi$ and $\Theta$}

We therefore propose reparameterising $\Phi$ (and $\Theta$) using the
bijection between $\mathcal{C}_p$ and $(-1,1)^p$ advocated by various authors,
e.g., \citet{Marriott_1995} and \citet{Vermaak_2004}.  To our knowledge, these
methods have not previously been deployed towards integrating out short memory
components in Bayesian analysis of ARFIMA processes.

\citet{Monahan_1984} defined a mapping
% $\bm{\phi}^{(p)}=(\phi_1^{(p)},\ldots,\phi_p^{(p)}) \longleftrightarrow \bm{\varphi}^{(p)}=(\varphi_1^{(p)},\ldots,\varphi_p^{(p)})$ 
$\bm{\phi}^{(p)} \longleftrightarrow \bm{\varphi}^{(p)}$
recursively as
follows:
\begin{equation}
\label{eqn:reparameterisationforwardrecursion}
\phi_i^{(k-1)} = \frac{\phi_i^{(k)}-\phi_k^{(k)}\phi_{k-i}^{(k)}}{1-\left(\phi_k^{(k)}\right)^2}, \qquad k=p,\ldots,2, \qquad i=1,\ldots,k-1.
\end{equation}
Then set $\varphi_k^{(p)} = \phi_k^{(k)}$ for $k=1,\ldots,p$.
The reverse recursion is given by:
\begin{equation*}
\phi_i^{(k)} = \left\{\begin{array}{lcll}
\varphi_k^{(p)} & \mathrm{for} & i=k & k=1,\ldots,p \\
\phi_i^{(k-1)}+ \varphi_k^{(p)}\phi_{k-i}^{(k-1)} & \mathrm{for} & i=1,\ldots,k-1 & k=2,\ldots,p
\end{array} \right. .
\end{equation*}
Note that $\phi_p^{(p)}=\varphi_p^{(p)}$. Moreover, if $p=1$, the two
parametrisations are the same, i.e.\ $\phi_1=\varphi_1$ (consequently the
brief study of \FAR{1,d,0} in section \ref{sec:pdqlik} fits in this framework).
The
equivalent reparametrised form for $\bm{\theta}$ is $\bm{\vartheta}$. The full
memory parameter $\bm{\omega}$ is reparametrised as $\bar{\Omega} = (-1/2,1/2)
\times (\hbox{the image of } \mathcal{C}_{p,q})$.  However recall that in
practice, $\mathcal{C}_{p,q}$ will be assumed equivalent to
$\mathcal{C}_p\times \mathcal{C}_q$, so the parameter space is effectively:
$\bar{\Omega} = (-1/2,1/2) \times(-1,1)^{p+q}$.

Besides mathematical convenience, this bijection has a very useful property \citep[cf.]{Kay_1981}
which helps motivate its use in defining RJ maps .  In Appendix \ref{sec:bijection} we show that the ACFs
of the $\bm{\phi}^{(p)}$ and $\bm{\phi}^{(p-1)}$ are identical. In other words, if
$d=q=0$, using this parametrisation for $\bm{\varphi}$ when moving between
different values of $p$ allows one to automatically choose processes that have
very closely matching ACFs at low lags. In the MCMC context this is useful
because it allows the chain to propose models that have a similar correlation
structure to the current one. Although this property is nice, it may be of
limited value for full ARFIMA models, since the proof of the main result
[Theorem \ref{theorem:firstmatching}] 
does not easily lend itself to the
inclusion of either a MA or long memory component. Nevertheless,
our empirical results similarly indicate a
`near-match' for a full \FAR{p,d,q} model.

\subsubsection*{Application of RJ MCMC to \FAR{p,d,q} processes}

We now use this reparametrisation to efficiently propose new parameter
values. Firstly, it is necessary to propose a new memory parameter
$\bm{\varpi}$ whilst keeping the model fixed. Attempts at updating each
component individually suffer from the same problems of excessive posterior
correlation that were encountered in section \ref{sec:pdqlik}.
Therefore the simultaneous update of the entire $r=(p+q+1)$-dimensional
parameter $\bm{\varpi}$ is performed using the hypercuboid-truncated Gaussian
distribution from definition $\bm{\xi}_{\bm{\varpi}}|\bm{\varpi} \sim \mathcal{N}^{\mathcal{H}_r}_r(\bm{\varpi},\Sigma_{\bm{\varpi}})$, where $\mathcal{H}_r$
defines the $r$-dimensional rectangle.
The covariance matrix $\Sigma_{\bm{\varpi}}$ is discussed in some detail below. The
choice of prior $p_{\bm{\varpi}}(\cdot)$ is arbitrary. \citet{Pai_1998} used a
uniform prior for $\bm{\omega}$ which has an explicit expression in the
$\bm{\varpi}$ parameterisation \citep{Monahan_1984}. However, their
expression is unnecessarily complicated since a uniform prior over
$\Omega$ holds no special interpretation. We therefore prefer
uniform prior over $\bar{\Omega}$:
%% \begin{equation}
%% \label{eqn: prior of varpi}
$p_{\bm{\varpi}}(\bm{\varpi}) \propto 1$, $\bm{\varpi} \in \bar{\Omega}$.
%% \end{equation}

Now consider the `between-models' transition. We must first choose a model
prior $p_\mathcal{M}(\cdot)$. A variety of priors are possible; the simplest
option would be to have a uniform prior over $\mathcal{M}$, but this would of
course be improper. We may in practice want to restrict the possible values of
$p,q$ to $0\leq p \leq P$ and $0\leq q \leq Q$ for some $P$,$Q$ (say 5), which
would render the uniform prior proper. However even in this formulation, a lot
of prior weight is being put onto complicated models which, in the interests
of parsimony, might be undesired. We prefer a truncated joint
Poisson distribution with parameter $\lambda$:
%% \begin{equation}
%% \label{eqn: model prior Poisson}
$p_{\mathcal{M}}(p,q) \propto \frac{\lambda^{p+q}}{p!q!}\mathbb{I}(p\leq P, q\leq Q)$.
%% \end{equation}

Now, denote the probability of jumping from model $M_{p,q}$ to model
$M_{p',q'}$ by $U_{(p,q),(p',q')}$.  $U$ could allocate non-zero probability for
every model pair, but for convenience we severely restrict the possible jumps
(whilst retaining irreducibility) using a two-dimensional bounded birth and
death process. Consider the subgraph of $\mathbb{Z}^2$: $G = \{(p,q) : 0\leq p
\leq P,\: 0\leq q \leq Q\}$, and allocate uniform non-zero probability
only to neighboring values, i.e., if and only if $|p-p'| + |q-q'|=1$.
Each point in
the `body' of $G$ has four neighbours; each point on the `line boundaries' has
three; and each of the four `corner points' has only two
neighbours. Therefore the model transition probabilities $U_{(p,q),(p',q')}$
are either 1/4, 1/3, 1/2, or 0.

Now suppose the current $(p+q+3)$-dimensional parameter is
$\bm{\psi}^{(p,q)}$, given by $\bm{\psi}^{(p,q)} =
(\mu,\sigma,d,\bm{\varphi}^{(p)},\bm{\vartheta}^{(q)})$, using a slight abuse
of notation. Because the mathematical detail of
the AR and MA components are almost identical, we consider 
only the case of de/increasing $p$ by 1 here; all of the following
remains valid if $p$ is replaced by $q$, and $\bm{\varphi}$ replaced by
$\bm{\vartheta}$. We therefore seek to propose a parameter $\bm{\xi}^{(p+1,q)}
= (\xi_\mu,\xi_{\sigma},\xi_d,\bm{\xi}_{\bm{\varphi}}^{(p+1)},
\bm{\xi}_{\bm{\vartheta}}^{(q)})$, that is somehow based on $\bm{\psi}^{(p,q)}$. We
further simplify by regarding the other three parameters ($\mu$, $\sigma$, and
$d$) as having the same interpretation in every model, choosing
$\xi_\mu = \mu$, $\xi_{\sigma} = \sigma$ and $\xi_d = d$. For simplicity we
also set $\bm{\xi}_{\bm{\vartheta}}^{(q)} =  \bm{\vartheta}^{(q)}$. Now
consider the map $\bm{\varphi}^{(p)}\rightarrow
\bm{\xi}_{\bm{\varphi}}^{(p+1)}$. To specify a bijection we 
`dimension-match' by adding in a random scalar $u$. The most obvious map is to
specify $u$ so that its support is the interval $(-1,1)$ and then set:
$\bm{\xi}_{\bm{\varphi}}^{(p+1)} =
\left(\bm{\varphi}^{(p)},u\right)$. The corresponding map for decreasing the
dimension is $\bm{\varphi}^{(p+1)}\rightarrow \bm{\xi}_{\bm{\varphi}}^{(p)}$ is
$\bm{\xi}_{\bm{\varphi}}^{(p)} =
\left(\varphi_1^{(p+1)},\ldots,\varphi_p^{(p+1)}\right)$. In other words, we
either add, or remove the final parameter, whilst keeping all others fixed
with the identity map, so the Jacobian is unity. The proposal 
$q(u|\bm{\psi}^{(p,q)})$ can be made in many ways---we prefer the simple
$\mathcal{U}(-1,1)$.  With these choices the RJ acceptance ratio
is
\begin{equation*}
A =  \ell_{(p',q')}(\mathbf{x}|\bm{\xi}^{(p',q')}) - \ell_{(p,q)}(\mathbf{x}|\bm{\psi}^{(p,q)})  + \log \left\{\frac{p_{\mathcal{M}}(p',q')}{p_{\mathcal{M}}(p,q)} \frac{U_{(p',q'),(p,q)}}{U_{(p,q),(p',q')}} \right\},
\end{equation*}
which applies to both increasing and decreasing dimensional moves.

\noindent
{\bf Construction of $\Sigma_{\bm{\varpi}}$:} Much of the efficiency of the
above scheme, including within- and between-model moves, depends on the choice
of $\Sigma_{\bm{\varpi}} \equiv \Sigma^{(p,q)}$, the within-model move RW
proposal covariance matrix.  We first seek  an appropriate
$\Sigma^{(1,1)}$, as in section \ref{sec:pdqlik}, with a pilot tuning scheme.
That matrix is shown on the left below, where we've `blocked it out'
\begin{align}
\Sigma^{(1,1)} &= \left(
\begin{MAT}(@,40pt,20pt){c|c;c}
\aligntop
\sigma^2_d & \sigma_{d,\varphi_1} & \sigma_{d,\vartheta_1} \\|
 & \sigma^2_{\varphi_1} & \sigma_{\varphi_1,\vartheta_1} \\;
\alignbottom
 &  & \sigma^2_{\vartheta_1}  \\
\end{MAT}
\right), &
\label{eqn:blockformforSigma^(p,q)}
\Sigma^{(p,q)} &= \left(
\begin{MAT}(@,40pt,20pt){c|c;c}
\aligntop
\sigma^2_d & \Sigma_{d,\bm{\varphi}^{(p)}} & \Sigma_{d,\bm{\vartheta}^{(q)}} \\|
  & \Sigma_{\bm{\varphi}^{(p)},\bm{\varphi}^{(p)}}  & \Sigma_{\bm{\varphi}^{(p)},\bm{\vartheta}^{(q)}} \\;
\alignbottom
&  & \Sigma_{\bm{\vartheta}^{(q)},\bm{\vartheta}^{(q)}} \\
\end{MAT}
\right),
\end{align}
(where each block is a scalar) so that we can extend this idea to the $(p,q)$
 case in the obvious way---on the right above---where
 $\Sigma_{\bm{\varphi}^{(p)},\bm{\varphi}^{(p)}}$ is a $p\times p$ matrix,
 $\Sigma_{\bm{\vartheta}^{(q)},\bm{\vartheta}^{(q)}}$ is a $q\times q$ matrix,
 etc. If either (or both) $p,q=0$ then the relevant blocks are simply omitted.
 To specify the various sub-matrices, we propose $\varphi_2,\ldots,\varphi_p$
 with equal variances, and \emph{independently} of $d,\varphi_1,\vartheta_1$,
 (and similarly for $\vartheta_2,\ldots,\vartheta_q$). In the context of
 (\ref{eqn:blockformforSigma^(p,q)}), the following hold:
\begin{align*}
\Sigma_{d,\bm{\varphi}^{(p)}} &= \left(
\begin{MAT}(@,40pt,20pt){c.c}
\sigma_{d,\varphi_1} & \mathbf{0}  \\
\end{MAT}
\right)
,&
\Sigma_{d,\bm{\vartheta}^{(q)}} &= \left(
\begin{MAT}(@,40pt,20pt){c.c}
\sigma_{d,\vartheta_1} & \mathbf{0}  \\
\end{MAT}
\right), \\
\Sigma_{\bm{\varphi}^{(p)},\bm{\varphi}^{(p)}} &= \left(
\begin{MAT}(@,40pt,20pt){c.c}
\aligntop
\sigma^2_{\varphi_1} & \mathbf{0}  \\.
\alignbottom
\mathbf{0}  & \sigma^2_{\bm{\varphi}}I_{p-1}   \\
\end{MAT}
\right)
, &
\Sigma_{\bm{\vartheta}^{(q)},\bm{\vartheta}^{(q)}} &= \left(
\begin{MAT}(@,40pt,20pt){c.c}
\aligntop
\sigma^2_{\vartheta_1} & \mathbf{0}  \\.
\alignbottom
\mathbf{0}  & \sigma^2_{\bm{\vartheta}}I_{q-1}   \\
\end{MAT}
\right), \\
%\end{align*} 
%\begin{equation*}
\Sigma_{\bm{\varphi}^{(p)},\bm{\vartheta}^{(q)}} &= \left(
\begin{MAT}(@,40pt,20pt){c.c}
\aligntop
\sigma_{\varphi_1,\vartheta_1} & \mathbf{0}  \\.
\alignbottom
\mathbf{0}  & \mathbf{O}   \\
\end{MAT}
\right),
%\end{equation*}
\end{align*}
where the dotted lines indicate further blocking, $\mathbf{0}$ is
a row-vector of zeros, and $\mathbf{O}$ is a zero matrix. 
This choice of $\Sigma_{\bm{\varpi}}$ is conceptually
simple, computationally easy and preserves the positive-definiteness as
required; this is shown by a simple relabeling of the rows/columns, and then
repeated application of theorems
\ref{theorem:positivedefinitenessofSigmaoneway} and
\ref{theorem:positivedefinitenessofSigmaotherway} which appear in
\ref{sec:bijection}.

\section{Empirical illustration and comparison}
\label{sec:empirical}

Here we provide empirical illustrations for the methods above: for
classical and Bayesian analysis of long memory models, and extensions
for short memory. To ensure consistency throughout,
the location and scale parameters will always be chosen as $\mu_I=0$ and
$\sigma_I=1$. Furthermore, unless stated otherwise, the simulated
series will be of length $n=2^{10}=1024$. This is a reasonable size
for many applications; it is equivalent to 85 years' monthly observations.
When using the approximate likelihood method we set $P=n$. Unless
otherwise stated the priors used will be those simple defaults suggested in
the previous sections.

\subsection{Long memory}
\label{sec:emplm}

We begin by demonstrating that the approximate likelihood of section
\ref{sec:alik} is accurate.  We then conduct a Monte
Carlo study varying length of the input, $n$. Finally, we
compare the Bayesian point-estimates and with
common non/semi-parametric alternatives.

Standard MCMC diagnostics were used throughout to ensure, and tune for, good
mixing. Because $d$ is the parameter of primary interest, the initial values
$d^{[0]}$ will be chosen to systematically cover its parameter space, usually
starting five chains at the regularly-spaced points $\{-0.4, -0.2, 0, 0.2,
0.4\}$. Initial values for other parameters are not varied: $\mu$ will start
at the sample mean $\bar{x}$; $\sigma$ at the sample standard deviation of the
observed series $\mathbf{x}$. When using the approximate likelihood method,
setting each of $\mathbf{x}_A$ to $\bar{x}$ turns out to be a sufficiently
good strategy.  For other MCMC particulars, see
\citet[][\S4.3.3]{graves:phd}.

\subsubsection*{Efficacy of approximate likelihood method}
%\label{section: Efficacy of Approximate method}

Start with the `null case', i.e., how does the algorithm perform when the data
are not from a long memory process? One hundred independent \FID{0}, or
Gaussian white noise, processes are simulated, from which  marginal posterior
means, standard deviations, and credibility interval endpoints are extracted.
Table \ref{table:test1.a} shows averages over the runs.

\begin{table}[ht!]
\centering
\caption{Posterior summary statistics for \FID{0} process. Average of 100 runs.}
\label{table:test1.a}
\vspace{3 mm}
\begin{tabular}{crc|rr} 
 & mean & std & \multicolumn{2}{c}{95\% CI} \\ \hline
$d$        & $0.006$    & $0.025$ & $-0.042$ & $0.055$ \\
$\mu$      & $-0.004$   & $0.035$ & $-0.073$ & $0.063$ \\
$\sigma$ & $1.002$    & $0.022$ & $0.956$  & $1.041$
\end{tabular}
\end{table}
The average estimate for each of the
three parameters is less than a quarter of a standard deviation away from the
truth. Credibility intervals are nearly symmetric about the
estimate and the marginal posteriors are, to a good approximation, locally
Gaussian (not shown). Upon, applying a proxy `credible-interval-based hypothesis
test' one would conclude in ninety-eight of the cases that $d=0$ could not
be ruled out.
A similar analysis for $\mu$ and $\sigma$
shows that hypotheses $\mu=0$ and $\sigma=1$ would each have been accepted
ninety-six times. These results indicate that the 95\% credibility intervals
are approximately correctly sized.

\begin{figure}[ht!]
\begin{center}
\begin{tabular}{cc}
\includegraphics[width=15pc]{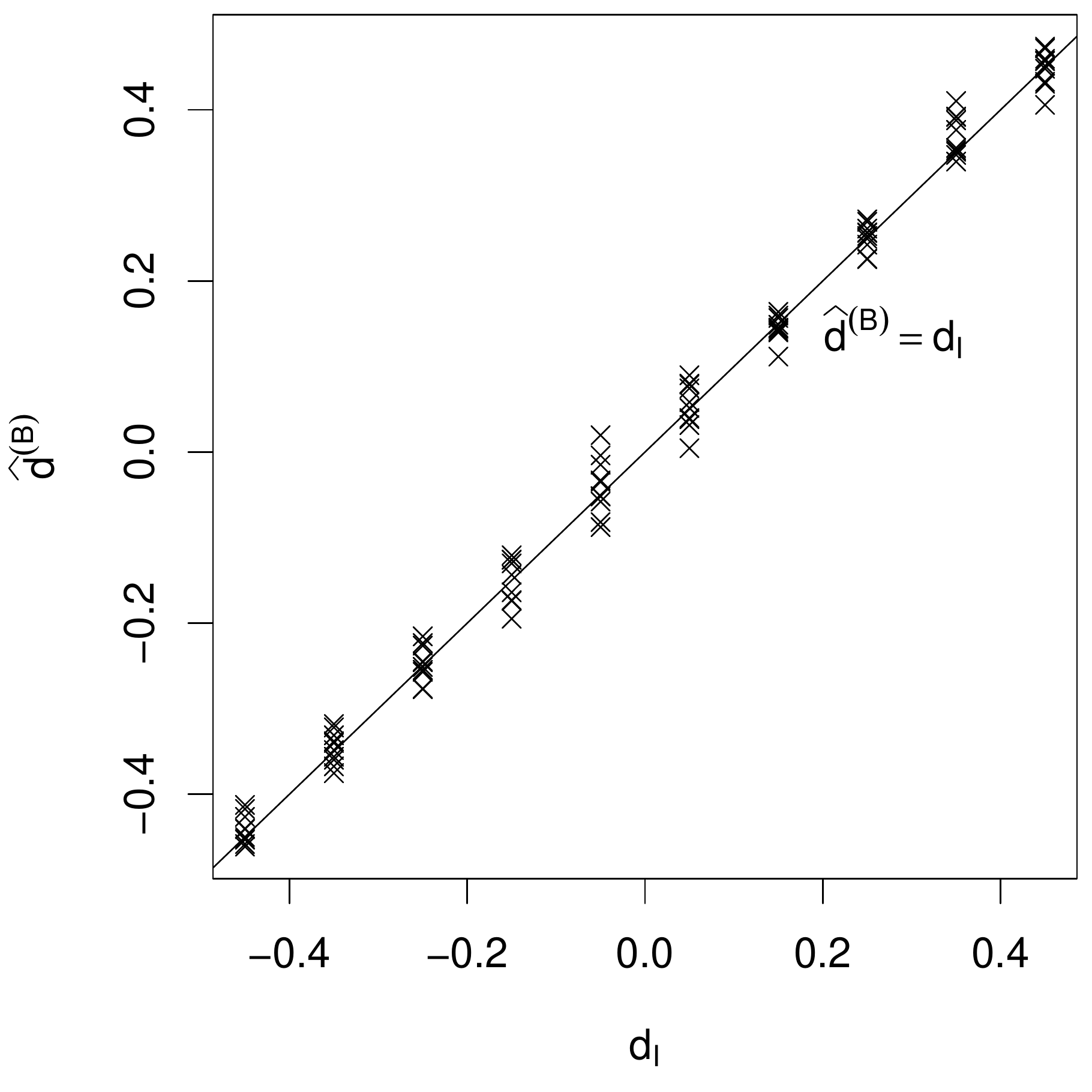} & \includegraphics[width=15pc]{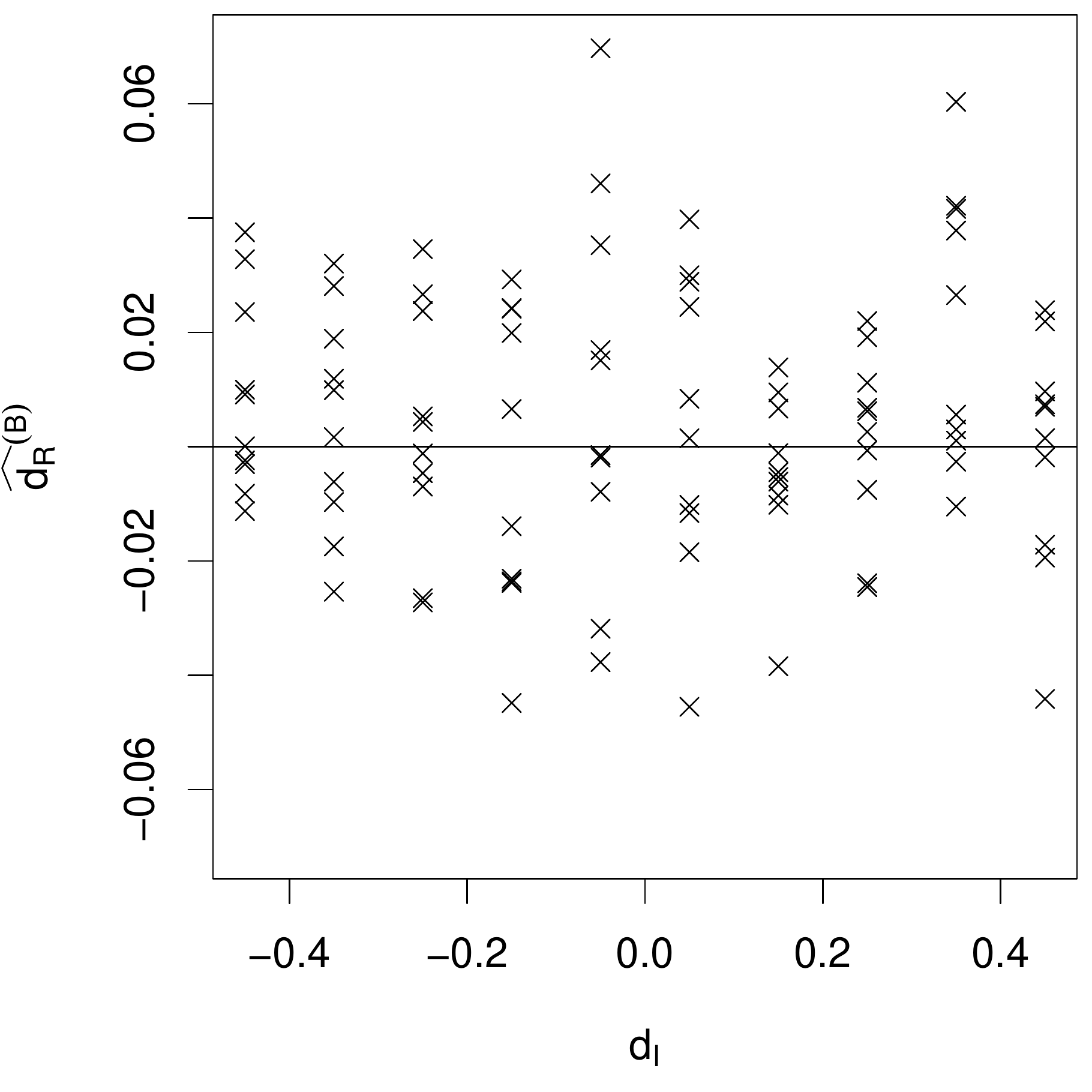} \\
$\quad\quad\quad$ (a) & $\quad\quad\quad$ (b)
\end{tabular}
\caption{Posterior outputs; (a)
$\widehat{d}^{(B)}$ against $d_I$, (b) $\widehat{d_R}^{(B)}$ against $d_I$.}
\label{fig:1_b_dd_1and1_b_dd_3}
\end{center}
\end{figure}

Next, consider the more interesting case of $d_I\neq 0$. We repeat the above
experiment except that ten processes are generated with $d_I$ set to each of
$\{-0.45,-0.35,\ldots,0.45\}$, giving 100 series total.  Figure
\ref{fig:1_b_dd_1and1_b_dd_3} shows a graphical analog of results from this
experiment.  The plot axes involve a Bayesian residual estimate of $d$,
$\widehat{d_R}^{(B)}$, defined as $\widehat{d_R}^{(B)} = \widehat{d}^{(B)} -
d_I$, where $\widehat{d}^{(B)}$ is the Bayesian estimate of $d$.
From the figure is clear that the estimator for $d$ is performing
well. Plot (a) shows how `tight' the estimates of $d$ are around the input
value---recall that the parameter space for $d$ is the whole interval
\dint{}. Moreover, plot (b) indicates that there is no significant change of
posterior bias or variance as $d_I$ is varied. 
% These remarks are further
% corroborated by directly studying the posterior standard deviations (not shown
% here for the sake of brevity).

\begin{figure}[ht!]
\begin{center}
\begin{tabular}{ccc}
\includegraphics[width=12pc,trim=25 0 0 0]{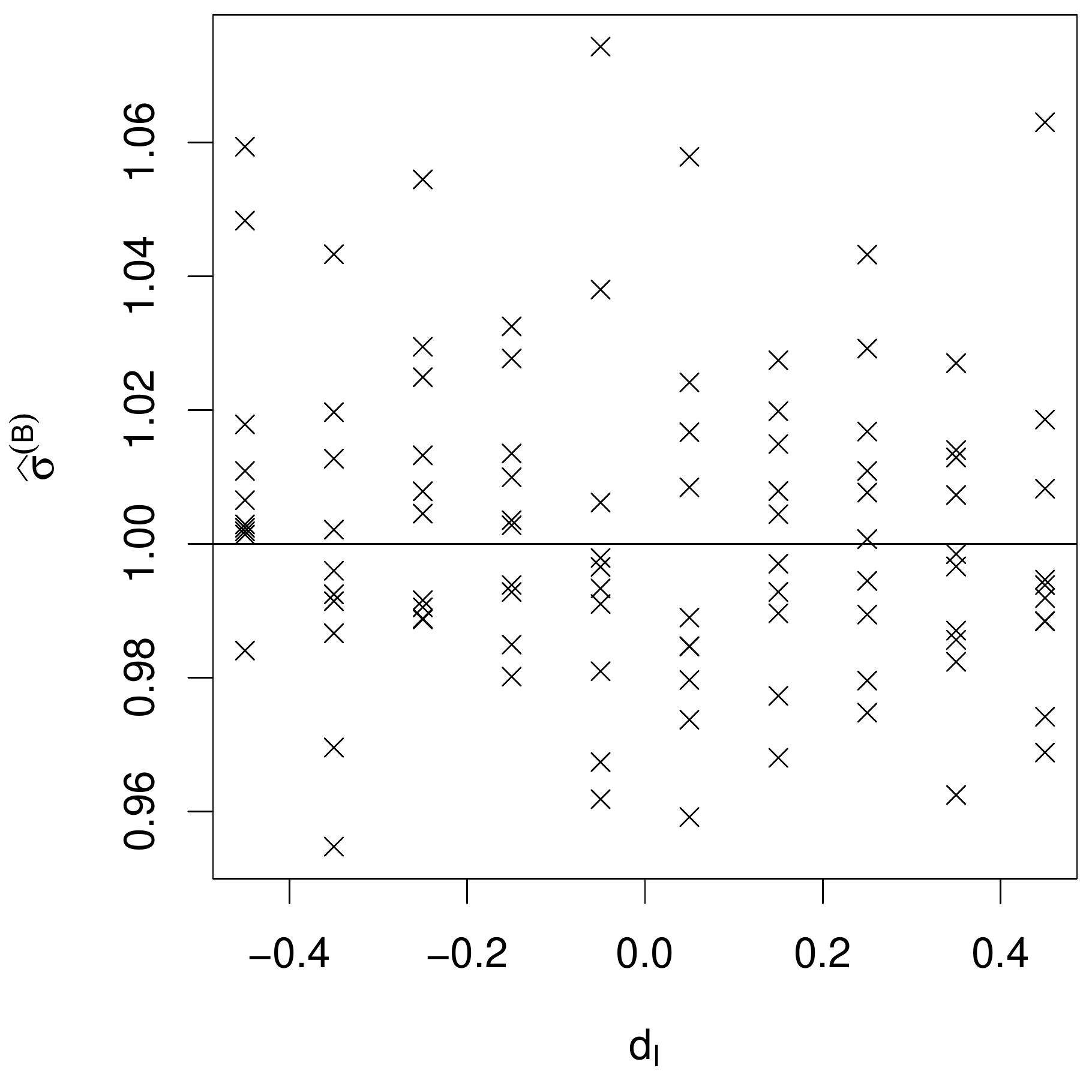} \ & 
\includegraphics[width=12pc,trim=25 0 0 0]{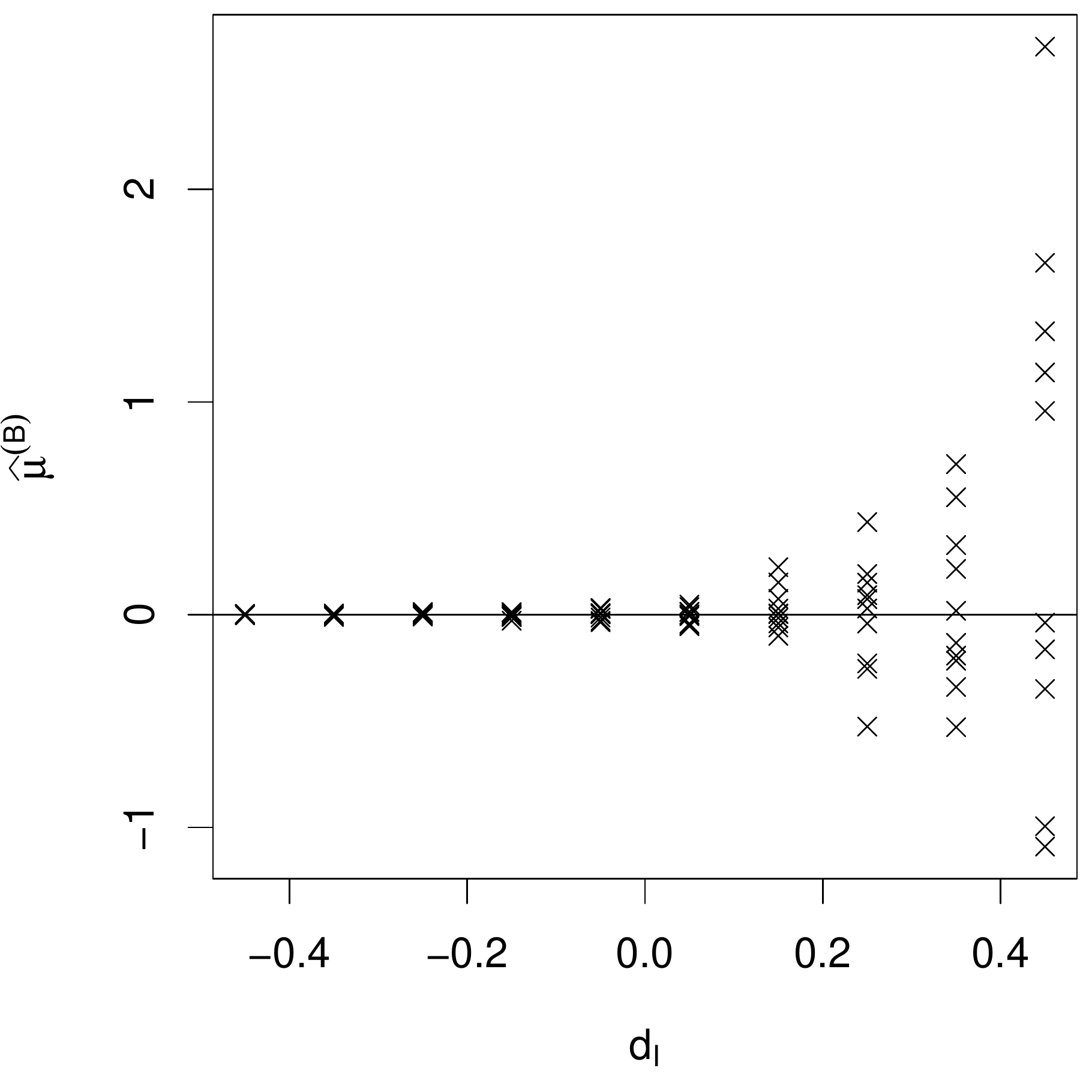} \ &
\includegraphics[width=12pc,trim=25 0 0 0]{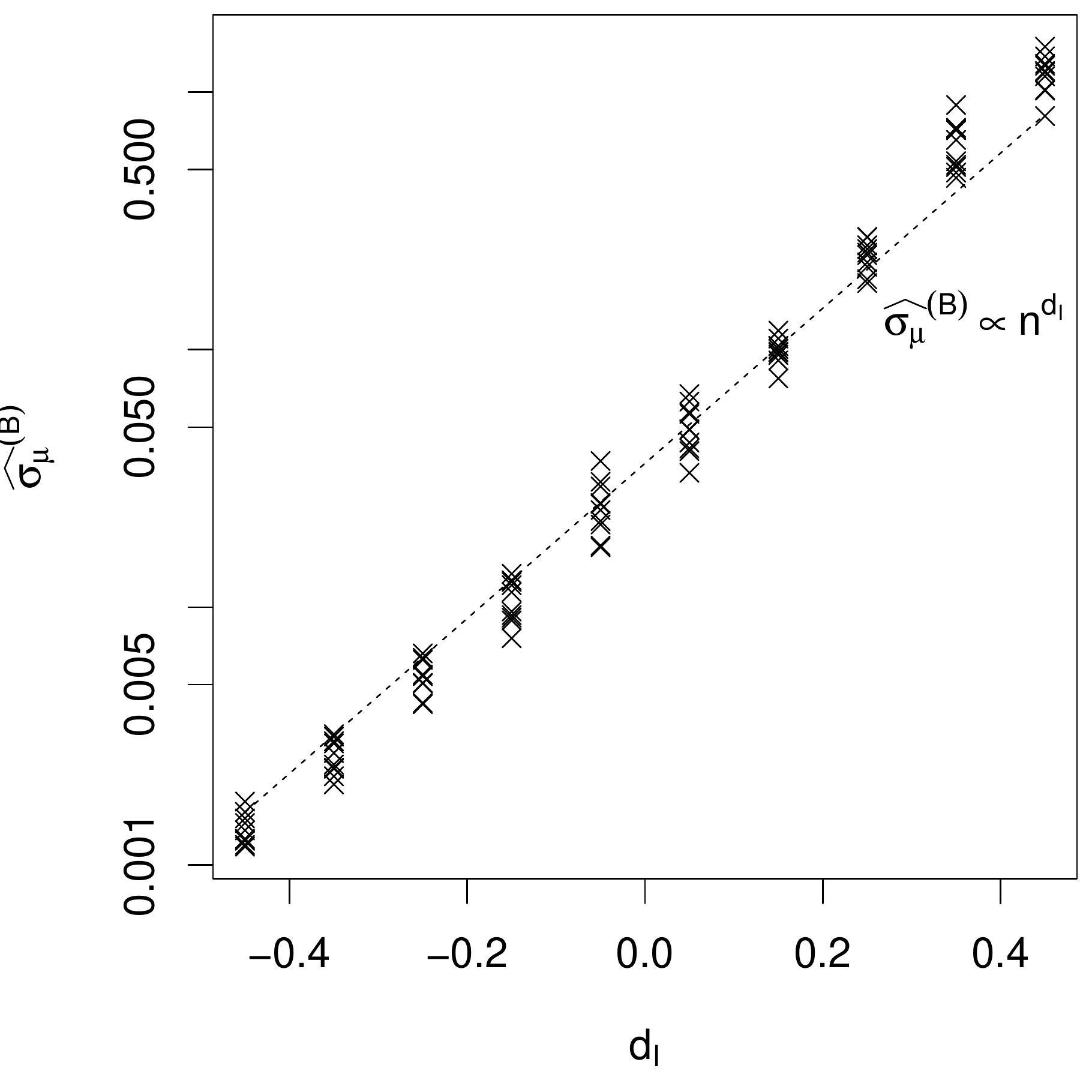} \\
$\quad\quad$ (a) & $\quad\quad$ (b) & $\quad\quad$ (c) 
\end{tabular}
\caption{Posterior outputs; (a) $\widehat{\sigma}^{(B)}$ against $d_I$, (b)
$\widehat{\mu}^{(B)}$ against $d_I$, and (c) $\widehat{\sigma_{\mu}}^{(B)}$ against $d_I$ (semi-log scale).}
\label{fig:1_b_gamma_1and1_b_mu_1}
\end{center}
\end{figure}

Next, the corresponding plots for the parameters $\sigma$ and $\mu$ are shown
in figure \ref{fig:1_b_gamma_1and1_b_mu_1}.
We see from plot (a) that the estimate of $\sigma$ also appears to be
unaffected by the input value $d_I$. 
% This is also further confirmed by directly studying the posterior variances (not shown). 
The situation is
different however in plot (b) for the location parameter $\mu$. Although the
bias appears to be roughly zero for all $d_I$, the posterior variance clearly
\emph{is} affected by $d_I$. To ascertain the precise functional dependence,
consider plot (c) which shows, on a semi-log scale, the
marginal posterior standard deviation of $\mu$,
$\widehat{\sigma_{\mu}}^{(B)}$, against $d_I$.

It appears that the marginal posterior standard deviation
$\widehat{\sigma_{\mu}}^{(B)}$ is a function of $d_I$; specifically:
$\widehat{\sigma_{\mu}}^{(B)} \propto A^{d_I}$,
for some $A$. The constant $A$ could be estimated via least-squares
regression. Instead however, inspired by asymptotic results in literature
concerning classical estimation of long memory processes \citep{Beran_1994} we
set $A=n$ and plotted the best fitting such line (shown in plot (c)). Observe that, although not fitting exactly, the relation
$\widehat{\sigma_{\mu}}^{(B)} \propto n^{d_I}$ holds reasonably well for $d_I
\in \dint{}$.  Indeed, \citeauthor{Beran_1994} motivated long memory in this way,
and derived asymptotic consistency results for optimum (likelihood-based)
estimators and found indeed that the standard error for $\mu$ is proportional
to $n^{d-1/2}$ (theorem 8.2) but the standard errors of all other parameters
are proportional to $n^{-1/2}$ (theorem 5.1).

%% Moreover, \cite{Dahlhaus_1989} % showed that these achieve the
%Cram\'{e}r--Rao bound, i.e.\ one is unable to do % better than this. The
%dependency on $n$ will be treated shortly.

% We conclude, then, that the methods of section \ref{sec:alik} are
% effective. The posterior variances are low and behave in a predictable manner,
% and the posterior means correspond closely to the `true values'. Once suitably
% centred and/or normalised by some appropriate function of $d_I$, the marginal
% posterior for each of $d$, $\sigma$ and $\mu$ is independent of $d_I$. This
% allows us to simply focus our Monte Carlo experiments on one particular value
% of $d_I$, say $0$, for brevity.

To study the impact of improper priors for $\mu$ and $\sigma$ 
we twice repeated the analysis of the 100 \FID{0} above, changing
$p_\mu(\cdot)$ in the first instance and $p_\sigma(\cdot)$ in the second. When
using $p_\mu(\cdot) \sim
\mathcal{N}(0,100^2)$, the maximum difference between estimates was 0.0012 for
$d$, 0.0017 for $\sigma$ and 0.0019 for $\mu$. When using $p_\sigma(\cdot)
\sim \mathcal{R}(0.01,0.01)$, the corresponding values were 0.0020 for $d$,
0.0017 for $\sigma$ and 0.0027 for $\mu$. Since these maximum differences are
well below the Monte Carlo error, %---i.e., changing only the initial random seed
%could produce results that are more different---we can conclude 
we conclude that there is
practically no difference between using an improper prior and a vague proper
prior.

\subsubsection*{Comparison of likelihood methods}

\begin{figure}[ht!]
\begin{center}
\small
\begin{minipage}{6cm}
\begin{tabular}{rrc|rr} 
 & mean & std & \multicolumn{2}{c}{95\% CI} \\ \hline 
approx $d$        & $0.006$  & $0.025$ & $-0.043$ & $0.056$  \\
exact $d$         & $0.014$  & $0.026$ & $-0.035$ & $0.065$ \\ \hline
approx $\mu$      & $-0.005$ & $0.034$ & $-0.073$ & $0.063$  \\
exact $\mu$       & $-0.001$ & $0.036$ & $-0.072$ & $0.069$  \\ \hline
approx $\sigma$ & $1.002$  & $0.022$ & $0.959$  & $1.046$  \\
exact $\sigma$  & $1.001$  & $0.022$ & $0.958$  & $1.044$  
\end{tabular}
\end{minipage}
\hfill
\begin{minipage}{8.35cm}
\includegraphics[width=20pc]{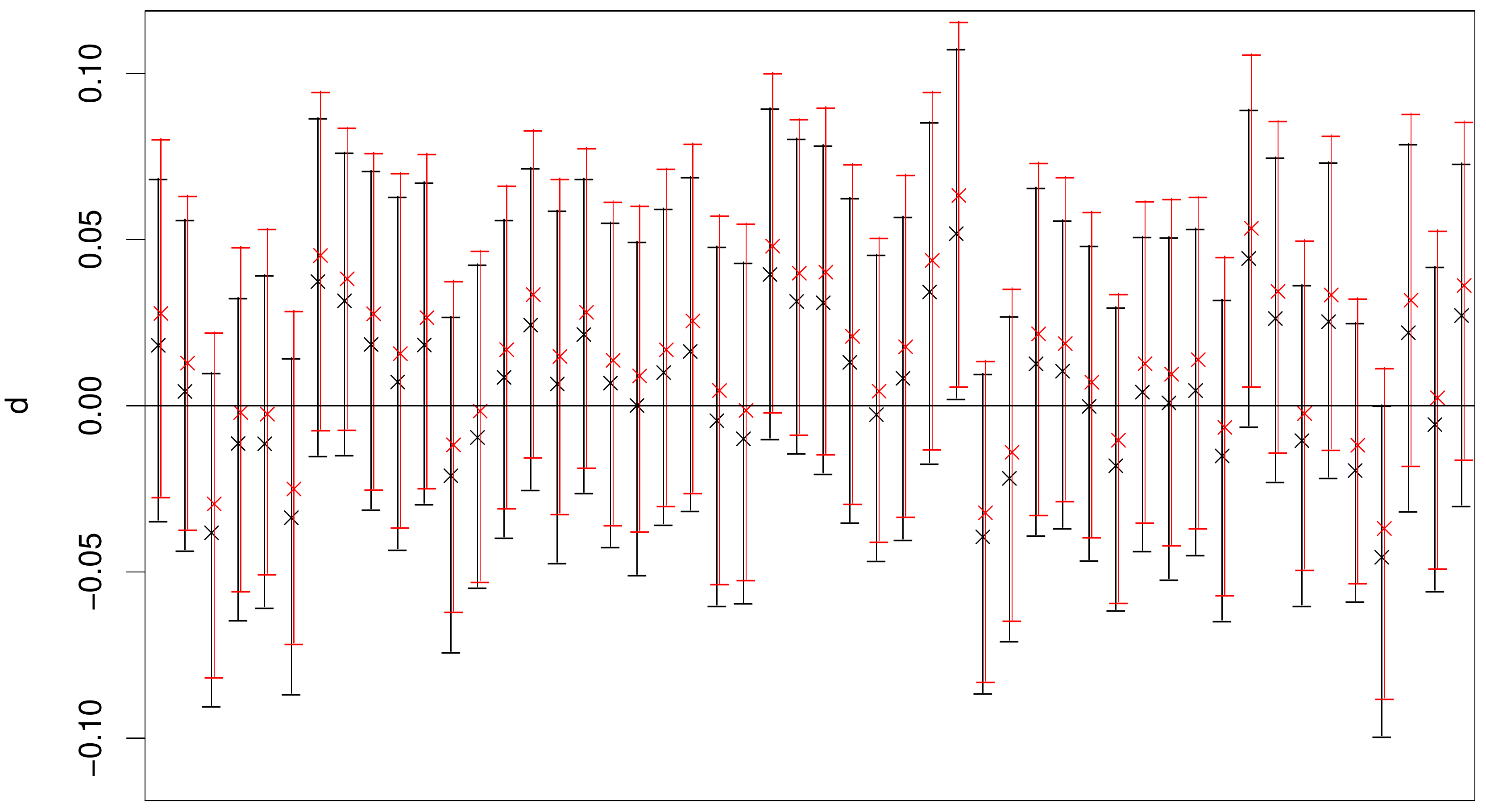}
\end{minipage}
\end{center}
\vspace{-0.25cm}
\caption{
{\em Table:} Comparison of posterior summary statistics for \FID{0} process
obtained via approximate and exact likelihood methods. Average of 50 runs.
{\em Plot:} Comparison of 95\% credibility intervals for $d$ from \FID{0}
processes obtained via approximate and exact likelihood methods. Black is used
for approximate, red for exact. The crosses are the respective point
estimates.}
\label{fig:1_a_2_a}
\end{figure}

To compare the methods based on approximate and exact likelihoods we consider
fifty \FID{0} generated as described above.  The output summary statistics are
presented in a table within figure \ref{fig:1_a_2_a}, which is accompanied
by a useful visualisation.
Observe that both methods produce highly correlated point estimates and
credibility interval endpoints. The plot shows that the posteriors of $d$ are
similar; there is only a very slight `shift' between the approximate and
exact results, implying that the approximate method is consistently
underestimating $d$ (by less than 0.01). The credibility
intervals have excellent frequentist coverage properties; 48 (96\%) of the
95\%-level intervals contain the input $d_I$. We also found (not shown)
that $\mu$ and $\sigma$ exhibit the same pattern of behaviour.

\begin{figure}[ht!]
\begin{center}
\begin{minipage}{6cm}
\small
\begin{tabular}{rrc|rr} 
 & mean & std & \multicolumn{2}{c}{95\% CI} \\ \hline 
approx $d_R$        & $0.005$  & $0.025$ & $-0.042$ & $0.048$  \\
exact $d_R$         & $0.014$  & $0.025$ & $-0.034$ & $0.057$
\end{tabular}
\end{minipage}
\hfill
\begin{minipage}{8.35cm}
\includegraphics[width=20pc]{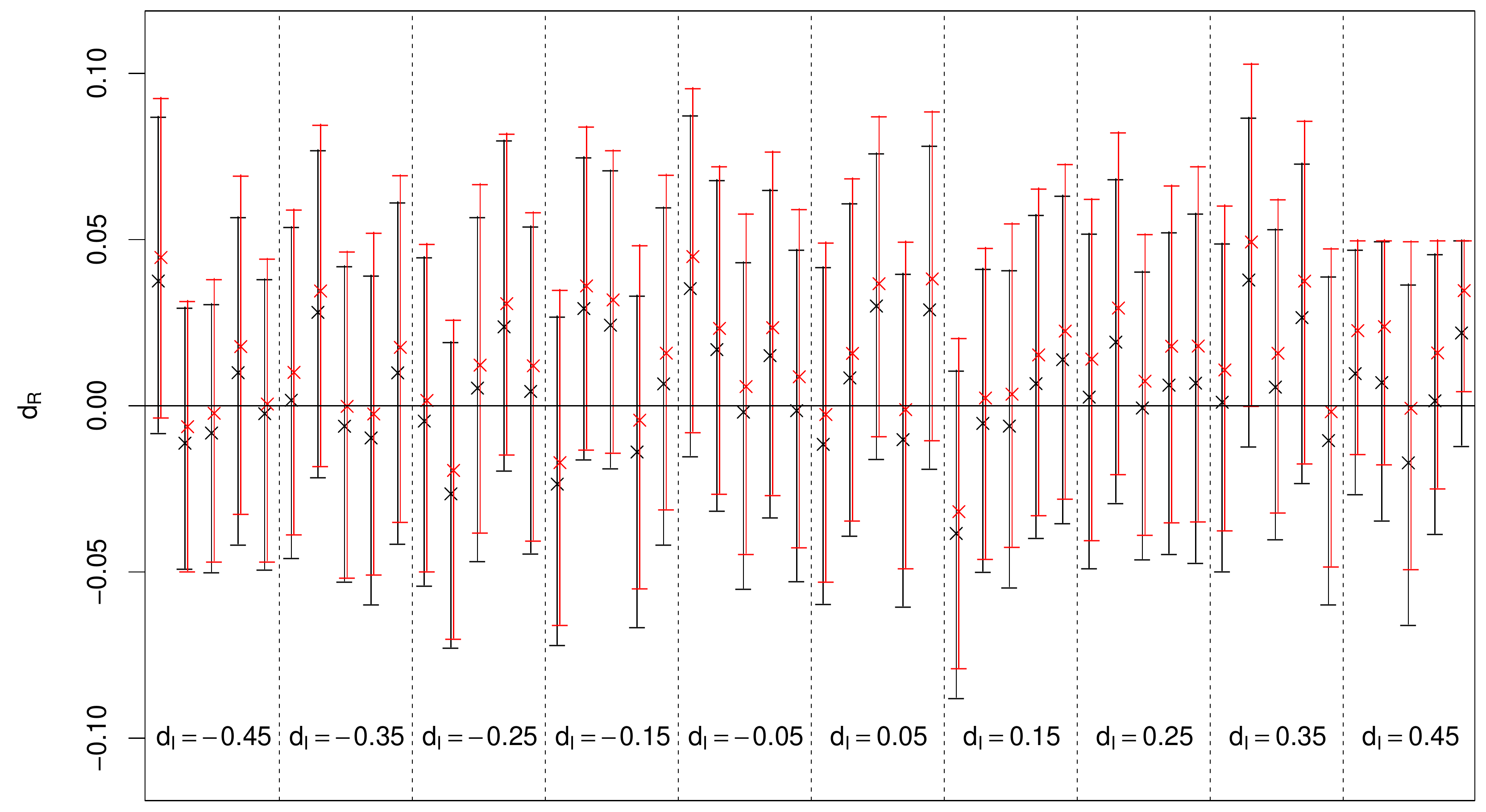}
\end{minipage}
\caption{Analog of Figure \ref{fig:1_a_2_a} for \FID{d}.}
\label{fig:1_b_2_b}
\end{center}
\end{figure}

To check for similarity between the approximate and exact likelihood methods
across the entire parameter space we repeated the \FID{d} simulations with
varying $d$ (analyzed in figure \ref{fig:1_b_dd_1and1_b_dd_3}). The results
for the residuals, $d_R$, are presented in figure \ref{fig:1_b_2_b}.
We observe the same pattern here as with the $d_I=0$ case. 
Note that the estimates of $d$ obtained using the two
methods appear to be slightly closer for negative $d_I$ than for positive
$d_I$, with very low discrepancies $(0.01)$ even for the largest $d_I=0.45$.
This suggests that the approximate method generally performs better for smaller
$d_I$.

\subsubsection*{Effect of varying time series length}

\begin{figure}[ht!]
\begin{center}
\begin{tabular}{ccc}
 \includegraphics[width=12pc,trim=25 0 0 0]{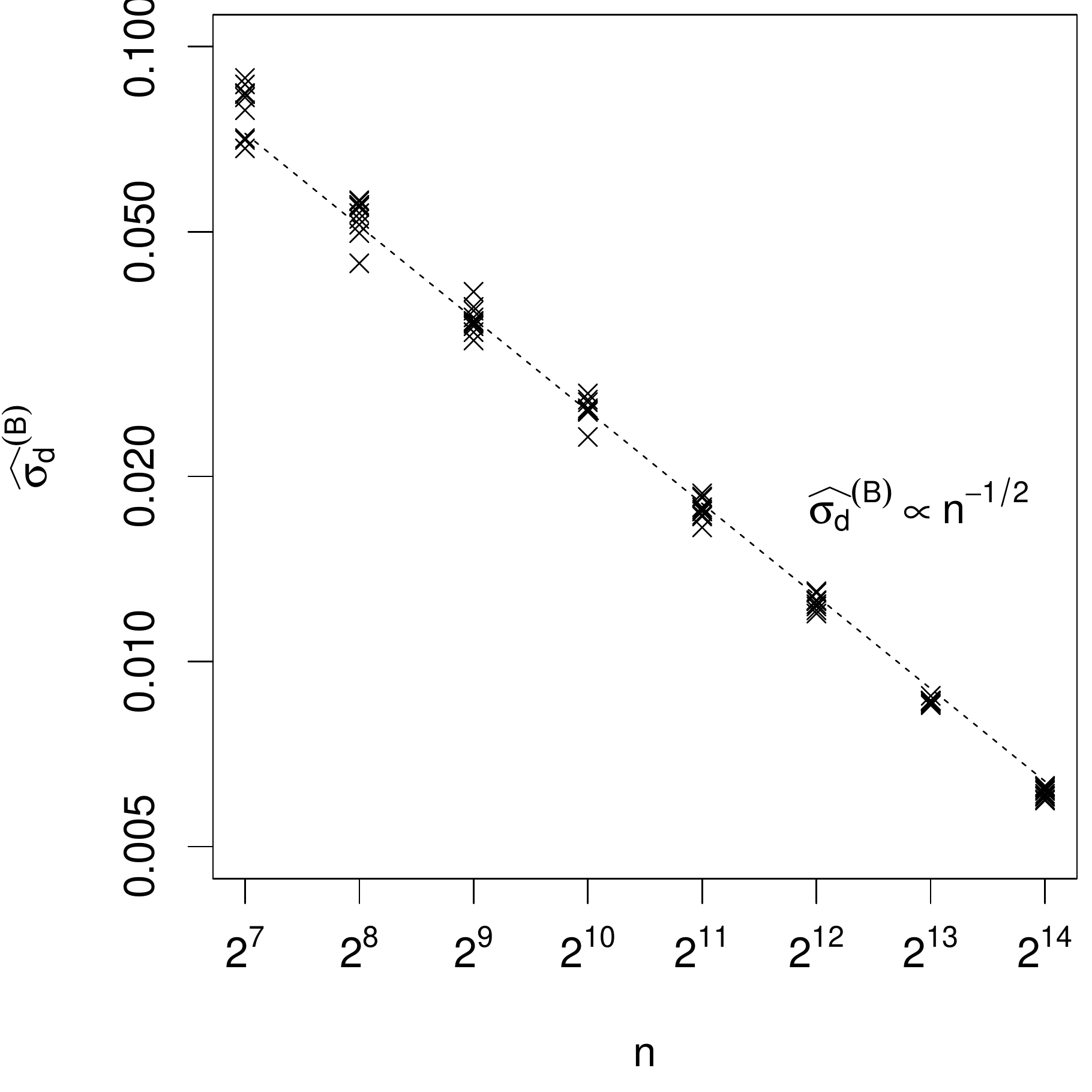} & \ 
 \includegraphics[width=12pc,trim=25 0 0 0]{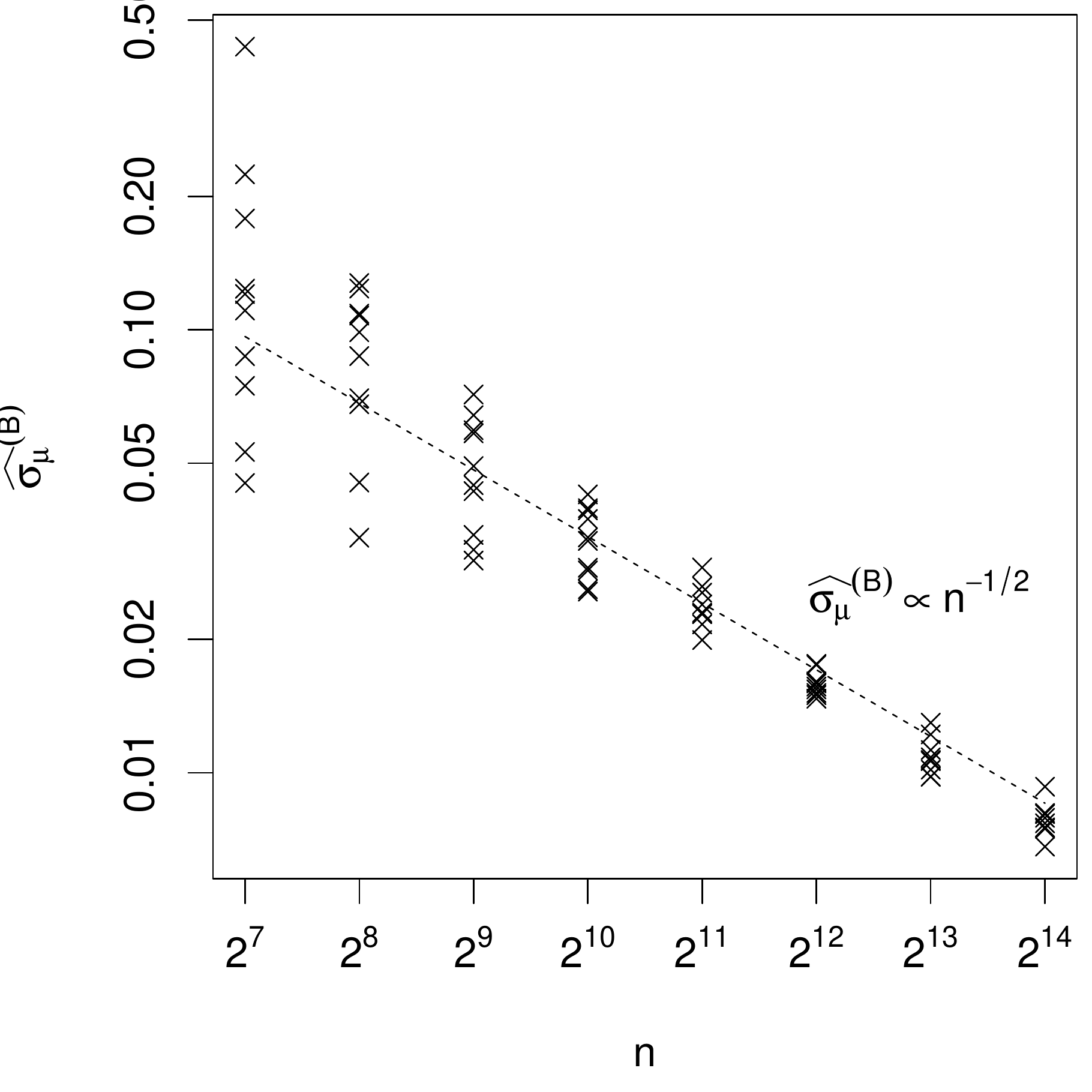} 
& \ \includegraphics[width=12pc,trim=25 0 0 0]{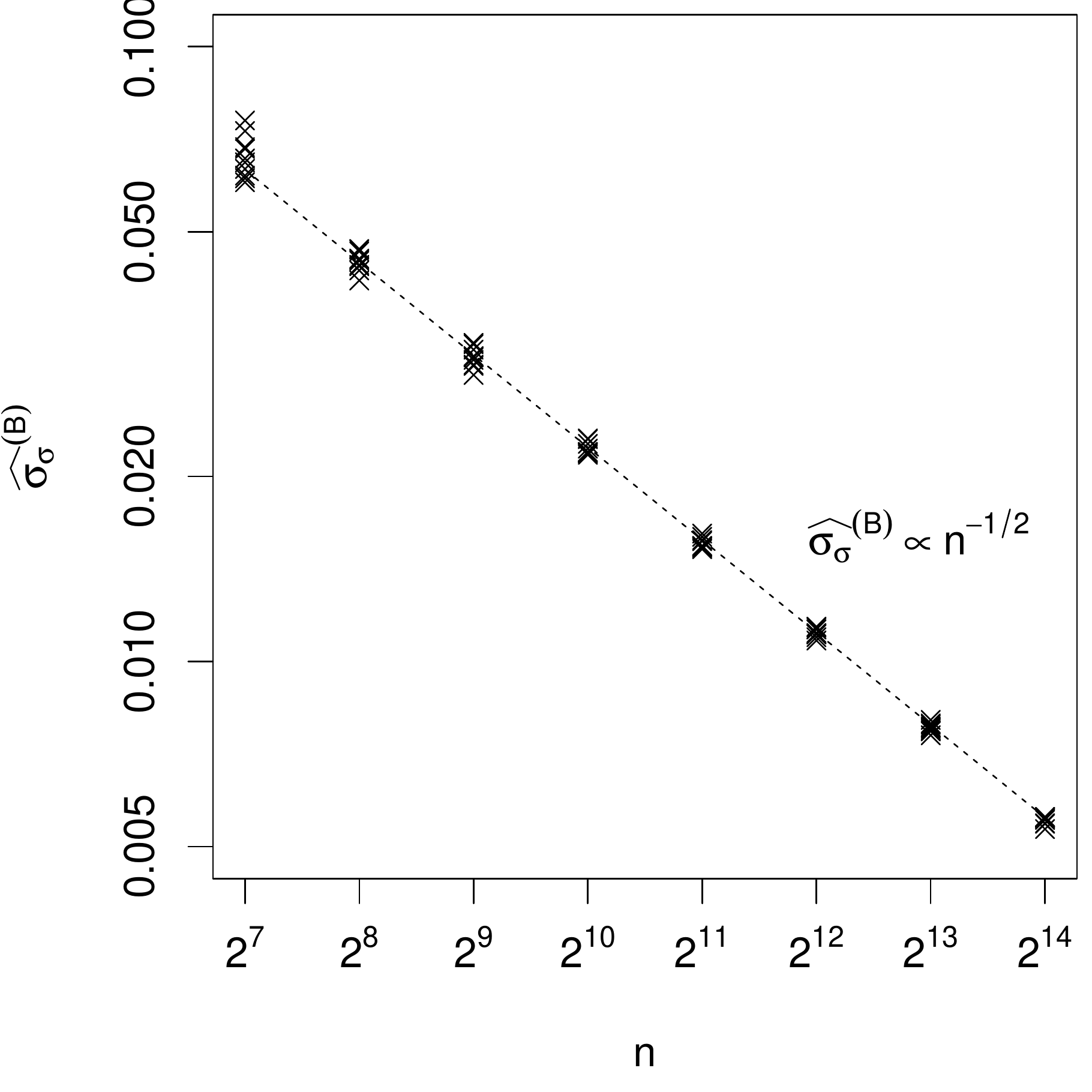} \\
 $\quad\quad$ (a) & $\quad\quad$ (b) & $\quad\quad$ (c) 
 \end{tabular}
\caption[Marginal posterior standard deviations as a function of
$n$]{Posterior outputs from \FID{0} series; (a) $\widehat{\sigma_d}^{(B)}$
against $n$, (b) $\widehat{\sigma_\mu}^{(B)}$ against $n$, (c)
$\widehat{\sigma_\sigma}^{(B)}$ against $n$ (log-log scale).}
\label{fig:3_a_dd_3and3_a_gamma_3and3_a_mu_3}
\end{center}
\end{figure}

We now analyse the effect of changing the time series length. For this 
we conduct a similar experiment but fix
$d_I=0$ and vary $n$. The posterior statistics of interest are the
posterior standard deviations $\widehat{\sigma_d}^{(B)}$,
$\widehat{\sigma_\mu}^{(B)}$ and $\widehat{\sigma_\sigma}^{(B)}$. For each $n
\in \{128 = 2^7,2^8,\ldots,2^{14} = 16,384\}$, 10 independent \FID{0} time
series are generated. The resulting posterior standard deviations are plotted
against $n$ (on log-log scale) in figure
\ref{fig:3_a_dd_3and3_a_gamma_3and3_a_mu_3}.

Observe that all three marginal posterior standard deviations are
proportional to $\frac{1}{\sqrt{n}}$, although the posterior of $\mu$ is less
`reliable'. Combining these observations with our earlier deduction that
$\sigma_\mu^{(B)} \propto n^{d_I}$, we conclude that for an \FID{d_I} process
of length $n$, the marginal posterior standard deviations
% $\pi(d,\mu,\sigma|\mathbf{x})$ follow 
% \begin{align*}
% \sigma_d^{(B)} & \propto n^{-1/2}, &
% \sigma_\mu^{(B)} & \propto n^{d-1/2}, &
% \sigma_{\sigma}^{(B)} & \propto n^{-1/2}.
% \end{align*}
follow those of \citeauthor{Beran_1994} given previously. 
% Although
% \citeauthor{Beran_1994} gave the above formulae as the theoretical asymptotic
% standard errors for likelihood-based estimators, it is unsurprising that the
% same relationships hold for the Bayesian posterior standard deviations since
% our approach is heavily likelihood-based. It is both interesting and
% reassuring that this asymptotic behaviour manifests itself for such small $n$,
% in particular it allows us to develop rules-of-thumb regarding appropriate
% lengths of time series.

\subsubsection*{Influence of prior}

Throughout, uninformative priors are used by default. However, one of the
principal advantages of the Bayesian approach is the ability to include
genuine prior information. To explore the effect prior variations we performed
a series of tests in which  $p_d(d) \propto \mathcal{N}(0,0.15^2)$ in an
analysis of \FID{0.25} processes. Note that the true value
$d=0.25$ is outside of the central 90\% of the prior distribution, with the
density there being less than a quarter of that at the maximum (i.e.\
when $d=0$). For each length of $n=2^7,2^8,2^9,2^{10}$, ten series were
analysed and compared with the equivalent analyses with the flat prior.

\begin{figure}[ht!]
\begin{center}
\begin{tabular}{c|c} 
$n$ & mean difference   \\ \hline
128   & 0.057   \\
256   & 0.029  \\
512   & 0.015  \\
1024  & 0.007   
\end{tabular}
\begin{tabular}{cc}
\includegraphics[width=12pc]{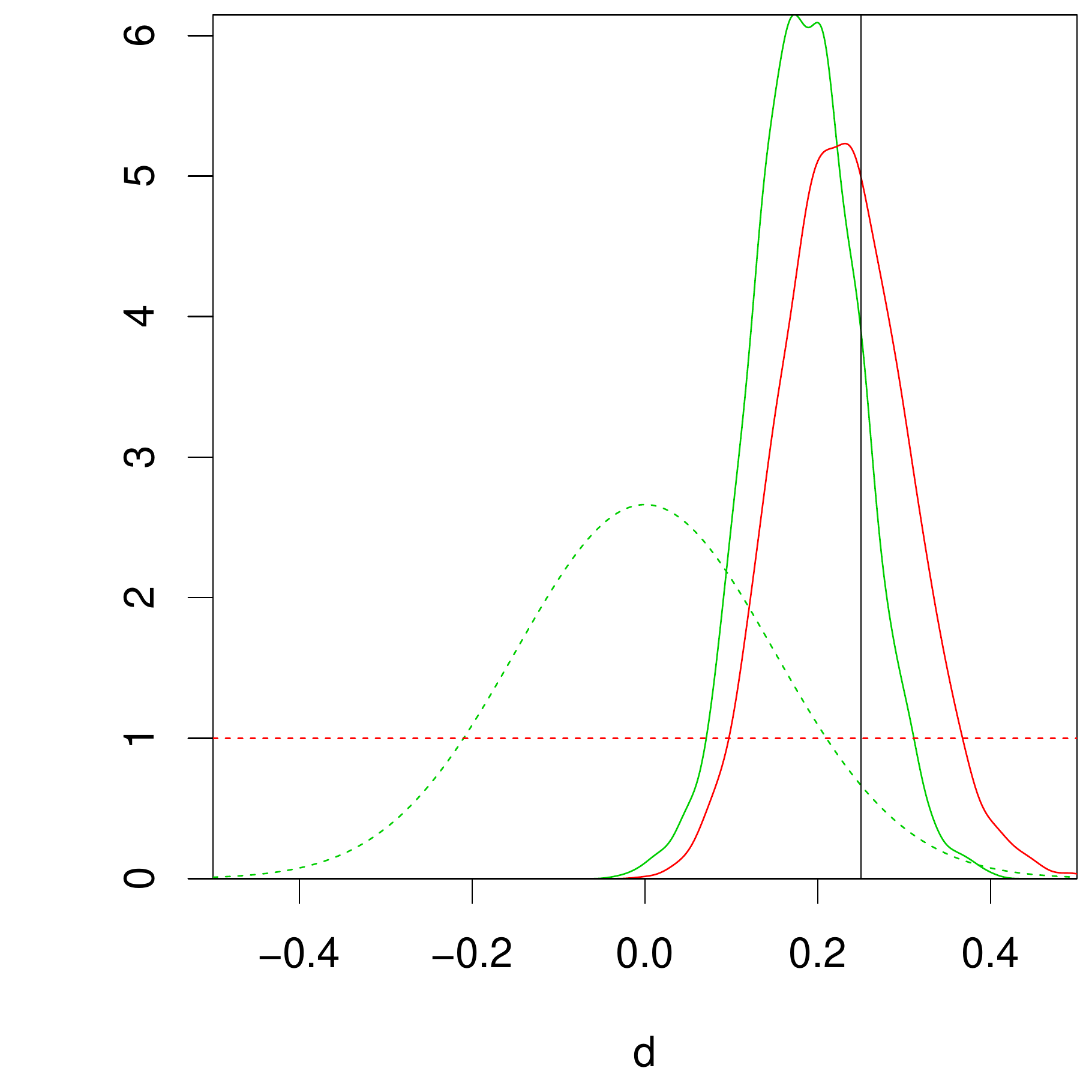} & \includegraphics[width=12pc]{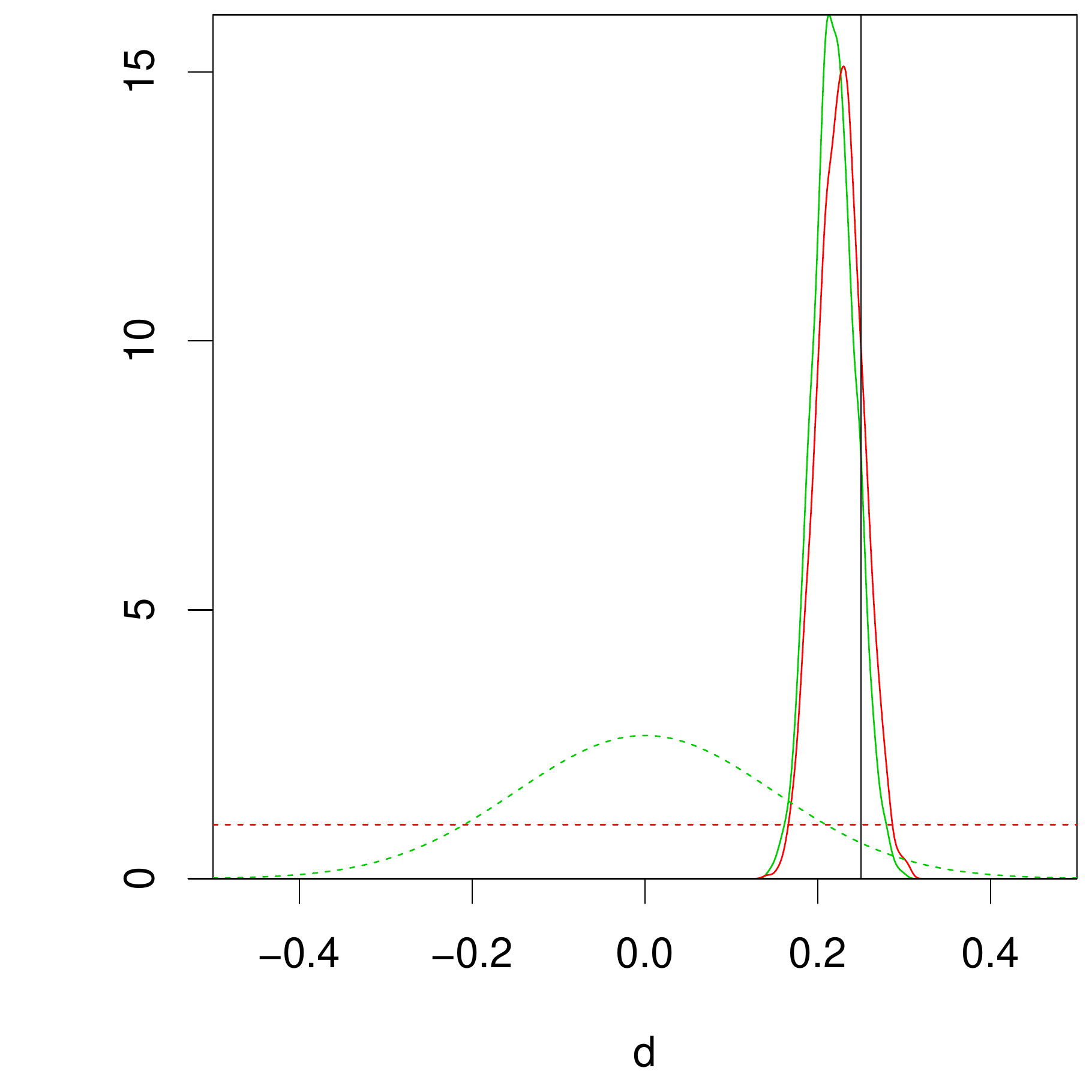} \\
$\quad\quad\quad$ (a) & $\quad\quad\quad$ (b)
\end{tabular}
\end{center}
\vspace{-0.5cm}
\caption{{\em Table:} Mean difference of estimates $\hat{d}^{(B)}$ under
alternative prior assumption.  {\em Plots:} Comparison of posteriors (solid
lines) obtained under different priors (dotted lines). Time series used:
\FID{0.25}; (a) $n=2^7 = 128$, (b) $n=2^{10}=1024$.}
\label{fig:10_1and10_2}
\end{figure}

As expected, in each case the Bayesian estimate $\hat{d}^{(B)}$ is always
lower when the Gaussian prior is used, compared to the flat prior. However the
average \emph{difference} between the two estimates shows a clear inverse
relationship with $n$; see the table in figure \ref{fig:10_1and10_2}.
For small $n$, the average difference is fairly significant, but for
$n=1024$, the difference is below the Monte Carlo error, as the likelihood has
overwhelmed the prior in this case. For a graphical demonstration the
convergence of the priors, refer to the plots in figure \ref{fig:10_1and10_2}.
Plot (a) shows that for $n=128$, the two posteriors are quite clearly
distinct. However plot (b) shows that for $n=1024$, the posteriors have
essentially coincided.

\subsubsection*{Comparison with common estimators}

In many practical applications, the long memory parameter is estimated using
non/semi-parametric methods. These may be
appropriate in many situations, where the exact form of the underlying process
is unknown. However when a specific model form is known (or at least assumed)
they tend to perform poorly compared with fully parametric alternatives 
\citep{Franzke_2012}.
Our aim here
is to demonstrate, via a short Monte Carlo study involving \FID{d} data, that
our Bayesian likelihood-based method significantly outperforms other common
methods in that case.  We consider the following comparators:
% There have been many previous Monte Carlo studies of long memory
% estimators:
% \citet{Taqqu_1995},
% \cite{Bardet_2003}, \citet{Mielniczuk_2007},
% \citet{Kristoufek_2010} and \citet{Franzke_2012}. 
%
% Other LRD estimators include 
(i) rescaled adjusted range, or $R/S$
\citet{Hurst_1951,graves:phd}---we use the \textsf{R}
implementation in the \texttt{FGN} \citep{R_FGN} package; (ii)
Semi-parametric Geweke--Porter-Hudak (GPH) method
\citep{Geweke_1983}---implemented in \textsf{R} package
\texttt{fracdiff} \citep{R_fracdiff}; (iii) detrended fluctuation analysis (DFA),
originally devised by \citet{Peng_1994}---in the
\textsf{R} package \texttt{PowerSpectrum}
\citep{R_PowerSpectrum}. (iv) wavelet-based semi-parametric estimators
\citet{Abry_2003} available in \textsf{R} package
\texttt{fARMA} \citep{R_fArma}. 

Each of these four methods will be applied to the same 100 time series with
varying $d_I$ as were used earlier experiments above. We extend the idea of a
residual, $\widehat{d_R}^{(R)}$, $\widehat{d_R}^{(G)}$, $\widehat{d_R}^{(D)}$,
and $\widehat{d_R}^{(W)}$, to accomodate the new comparators, respectively, and
plot them against $\widehat{d_R}^{(B)}$ in figure \ref{fig:
classical_1}.
\begin{figure}[ht!]
\begin{center}
\begin{tabular}{cc}
\includegraphics[width=15pc]{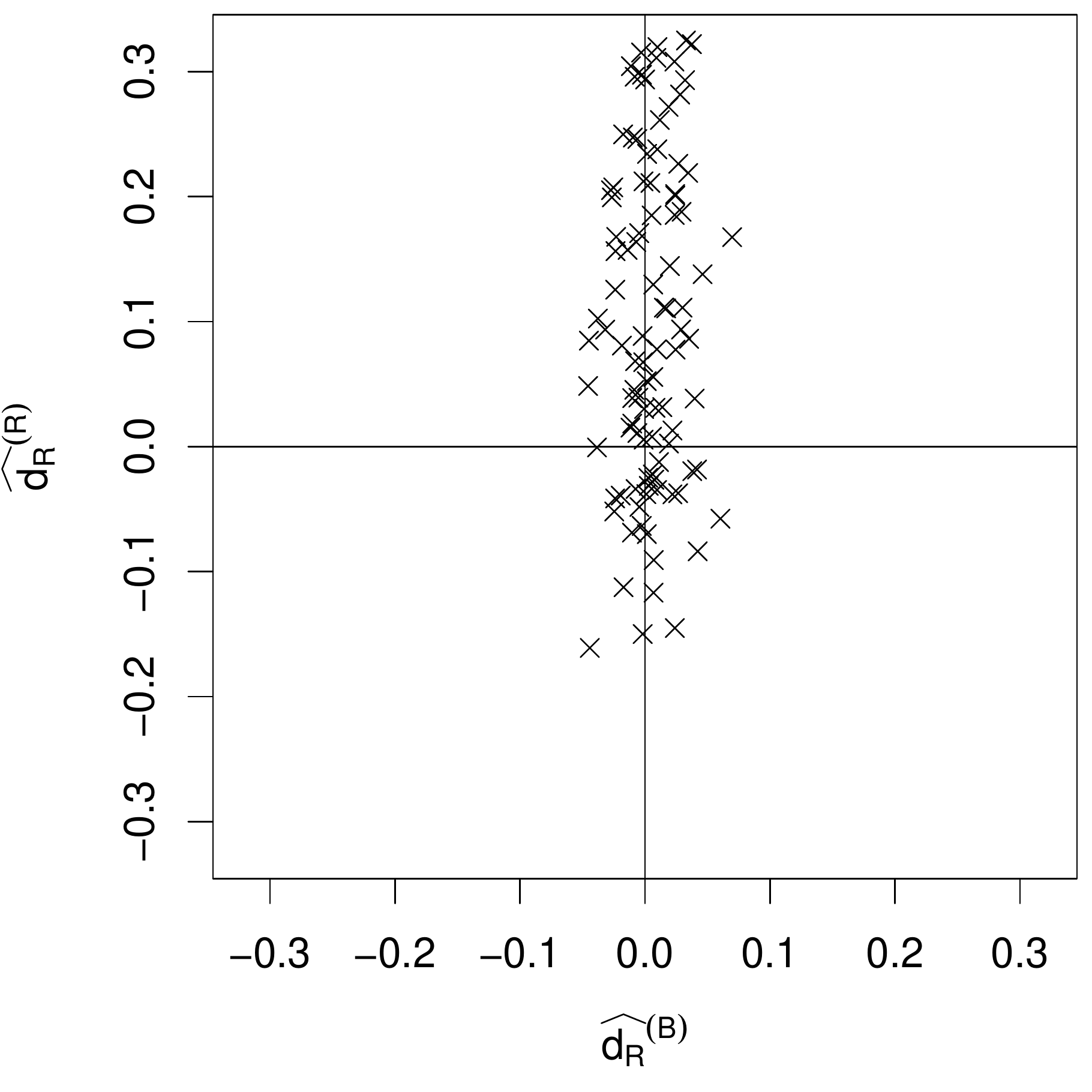} & \includegraphics[width=15pc]{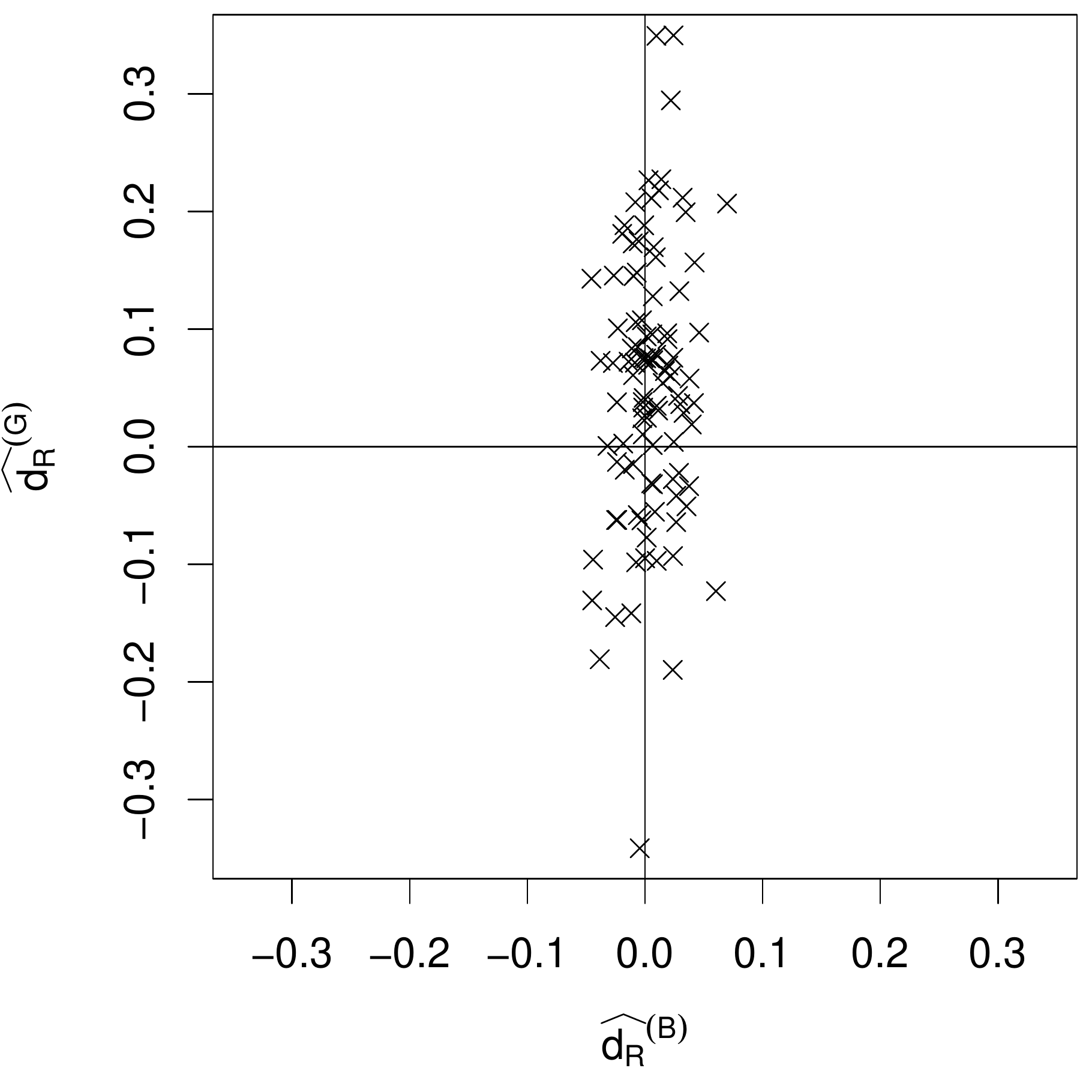} \\
$\quad\quad\quad$ (a) & $\quad\quad\quad$ (b) \\
 \\
\includegraphics[width=15pc]{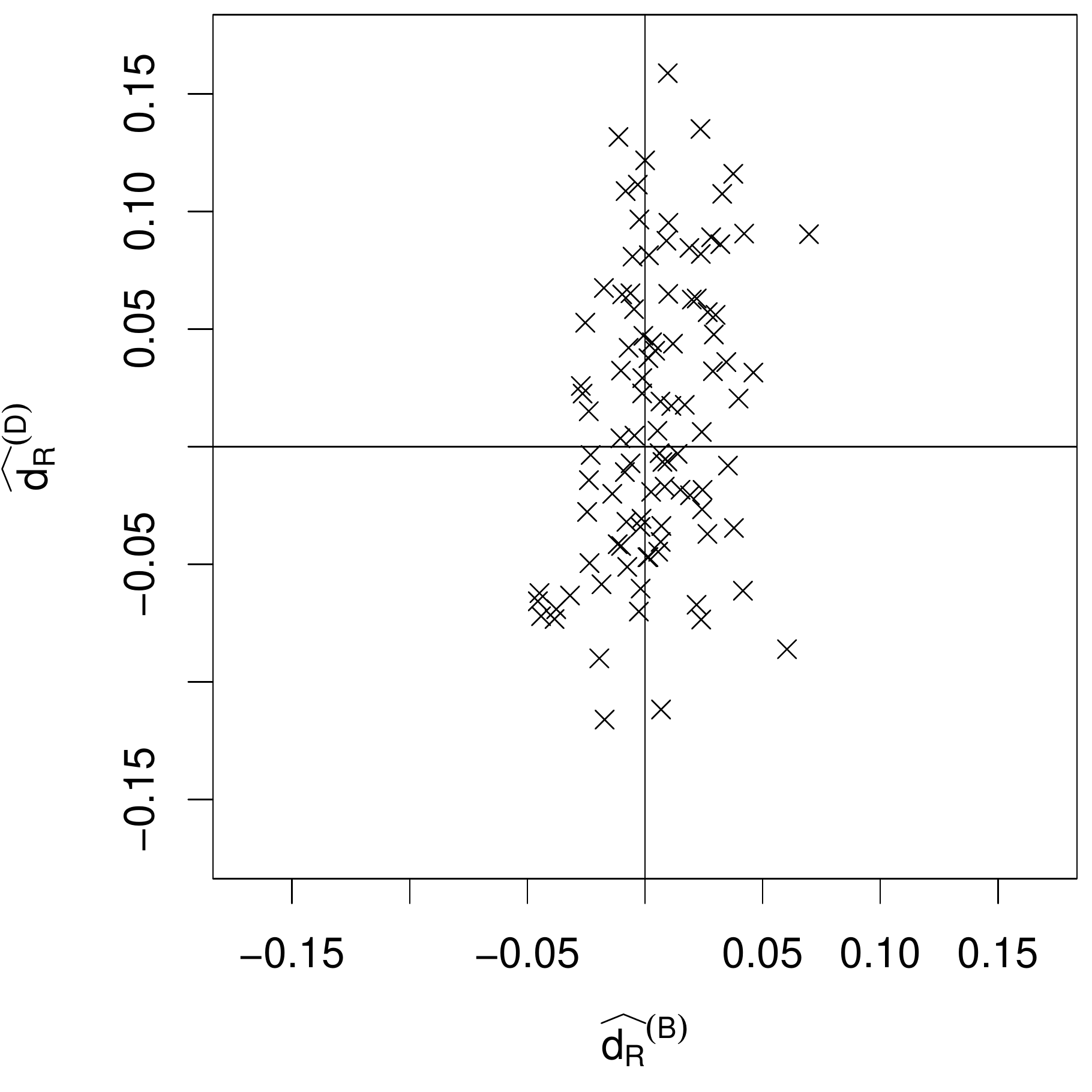} & \includegraphics[width=15pc]{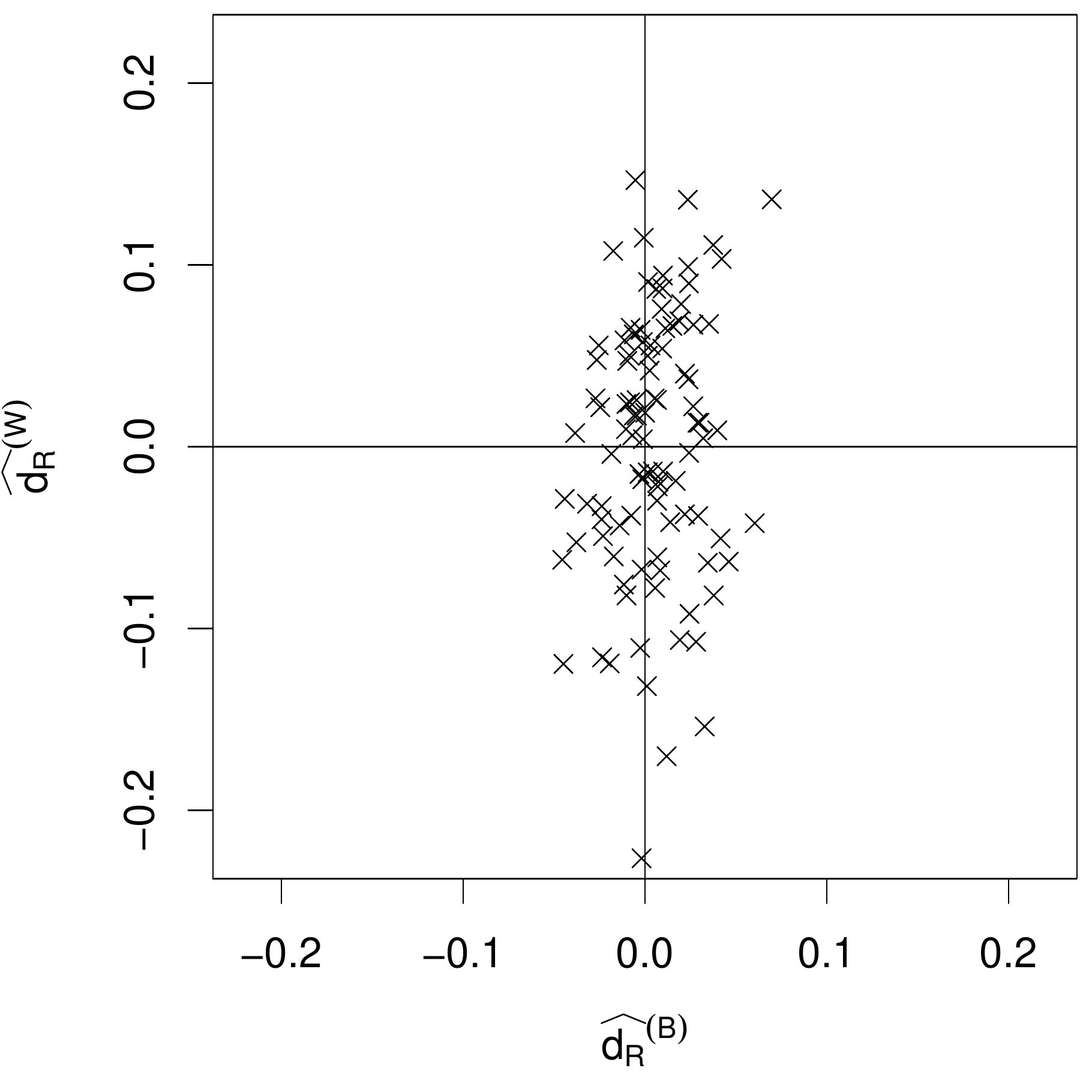} \\
$\quad\quad\quad$ (c) & $\quad\quad\quad$ (d)
\end{tabular}
\end{center}
\caption[Comparison of Bayesian estimator with classical estimators]{Comparison of Bayesian estimator with common classical estimators; (a) $R/S$, (b) GPH, (c) DFA, (d) Wavelet.}
\label{fig: classical_1}
\end{figure}
Observe that all four methods have a much larger variance than our Bayesian
method, and moreover the $R/S$ is positively biased. Actually, the
bias in some cases would seem to depend on $d_I$: $R/S$  is
significantly (i.e.\ $>0.25$) biased for $d_I<-0.3$ but slightly
negatively biased for $d>0.3$ (not shown); DFA is only unbiased for
$d_I>0$; both the GPH and wavelet methods are unbiased for all $d\in
\dint{}$.

\subsection{Extensions for short memory}
\label{sec:empsm}

{\bf Known form:} We first consider the MCMC
algorithm from section \ref{sec:pdqlik} for sampling under an
\FAR{1,d,0} model where the full memory parameter is
$\bm{\omega}=(d,\phi_1)$. Recall that that method involved proposals
from a hypercuboid MVN using a pilot-tuned covariance matrix. Also
recall that it is a special case of the re-parametrised method from
section \ref{sec:rj}.

\begin{figure}[ht!]
\begin{center}
\begin{tabular}{cc}
\includegraphics[width=15pc]{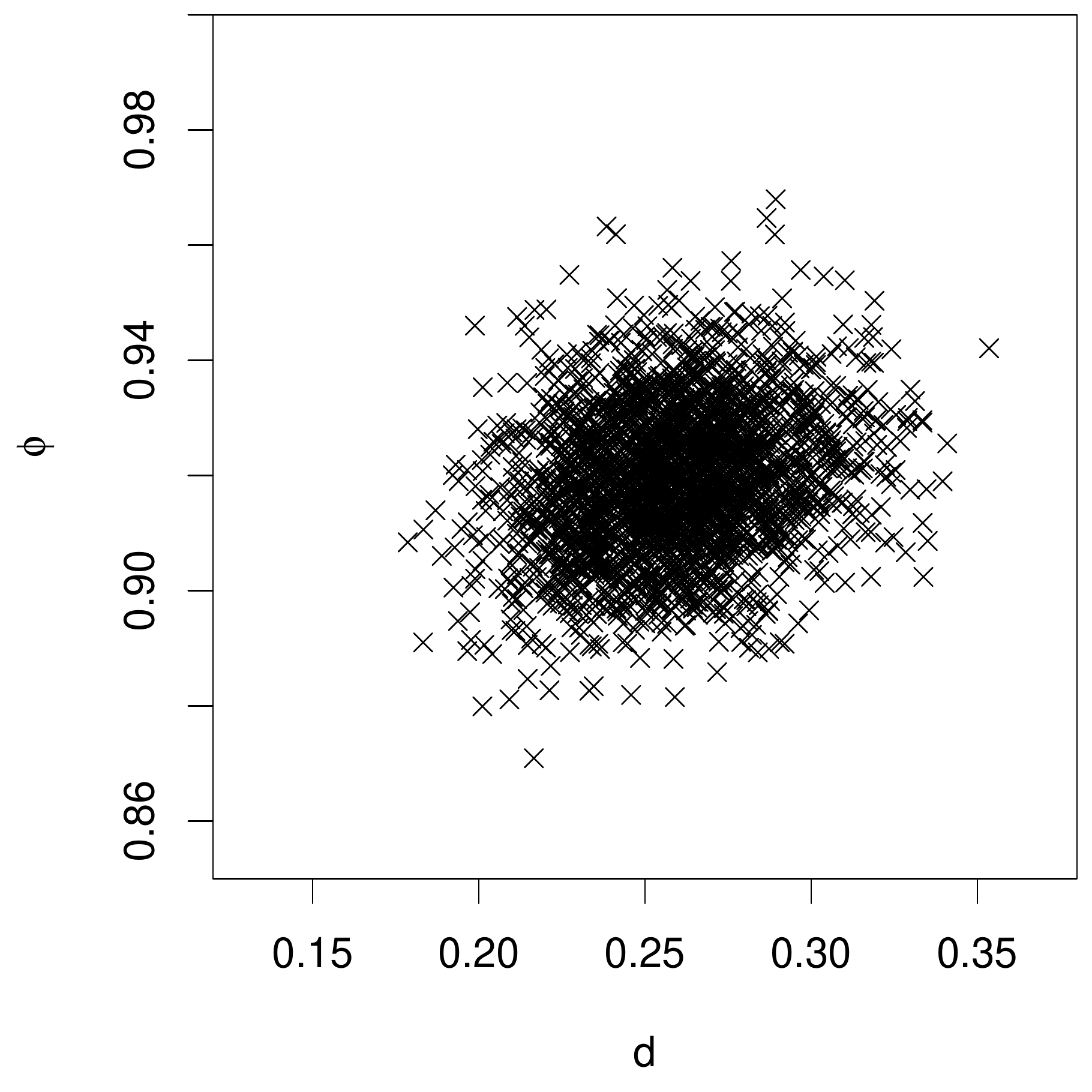} & 
\includegraphics[width=15pc]{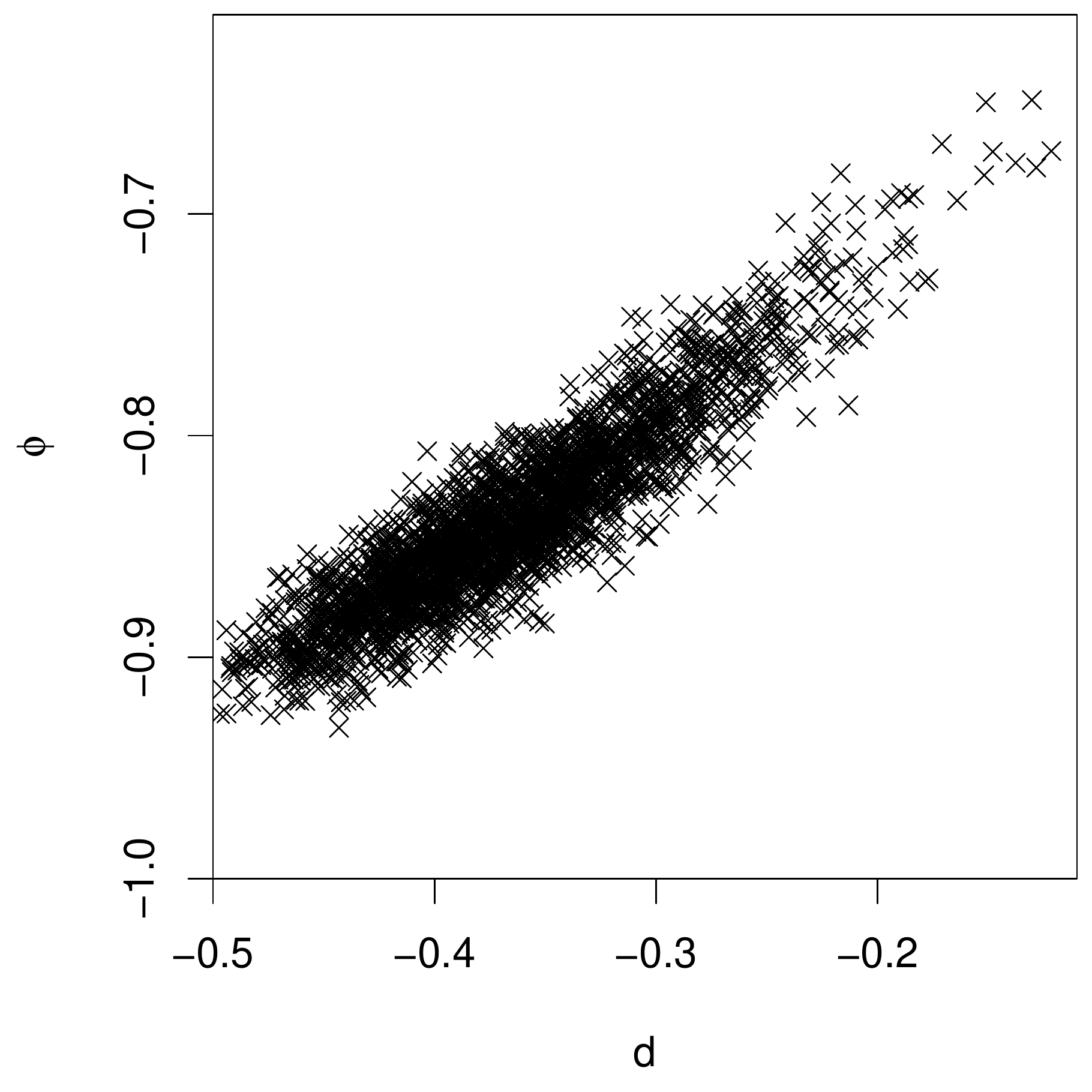} \\
$\quad\quad\quad$ (a) & $\quad\quad\quad$ (b)
\end{tabular}
\end{center}
\caption[Joint posterior samples of $(d,\phi)$ from \FAR{1,d,0}
processes]{Posterior samples of $(d,\phi)$; input time series (a)
$(1+0.92\mathcal{B})(1-\mathcal{B})^{0.25}X_t=\innov{}$, (b)
$(1-0.83\mathcal{B})(1-\mathcal{B})^{-0.35}X_t=\innov{}$.}
\label{fig:5_a_i_dd_phi_038and5_a_i_dd_phi_002}
\end{figure}

\begin{figure}[ht!]
\begin{center}
\begin{tabular}{ccc}
\includegraphics[width=12.25pc,trim=20 0 0 0]{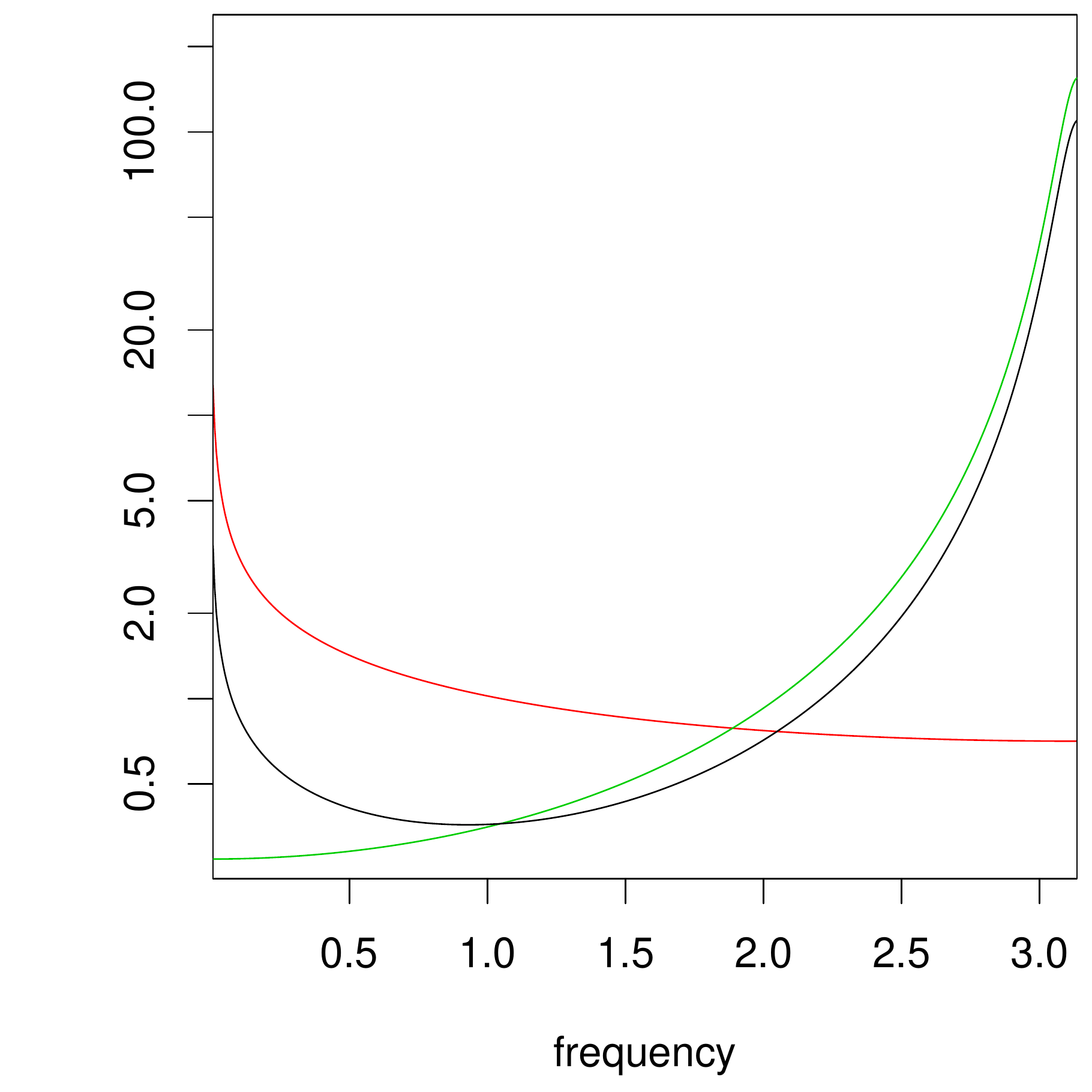} & 
\includegraphics[width=12.25pc,trim=20 0 0 0]{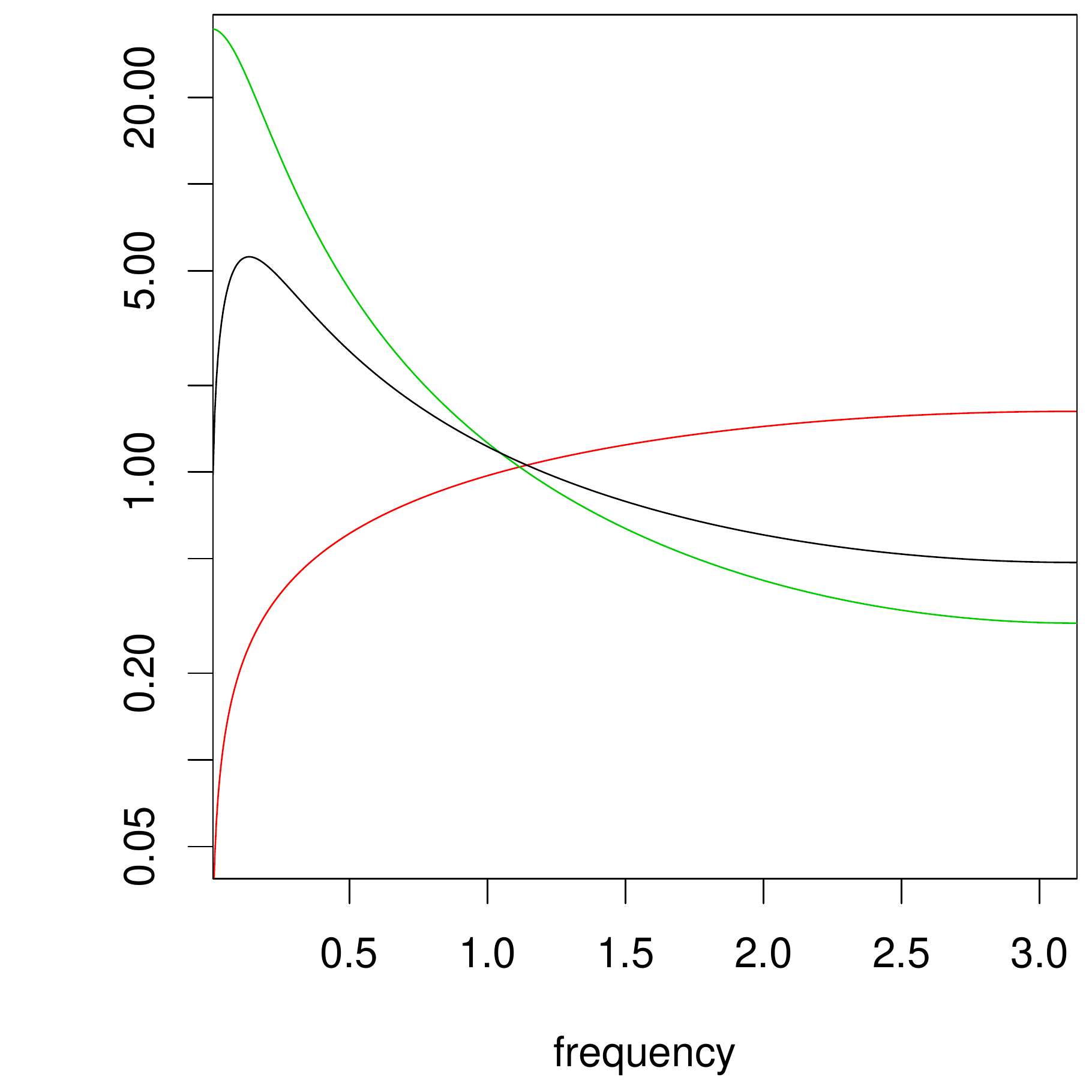} &
\includegraphics[width=12.25pc,trim=20 0 0 0]{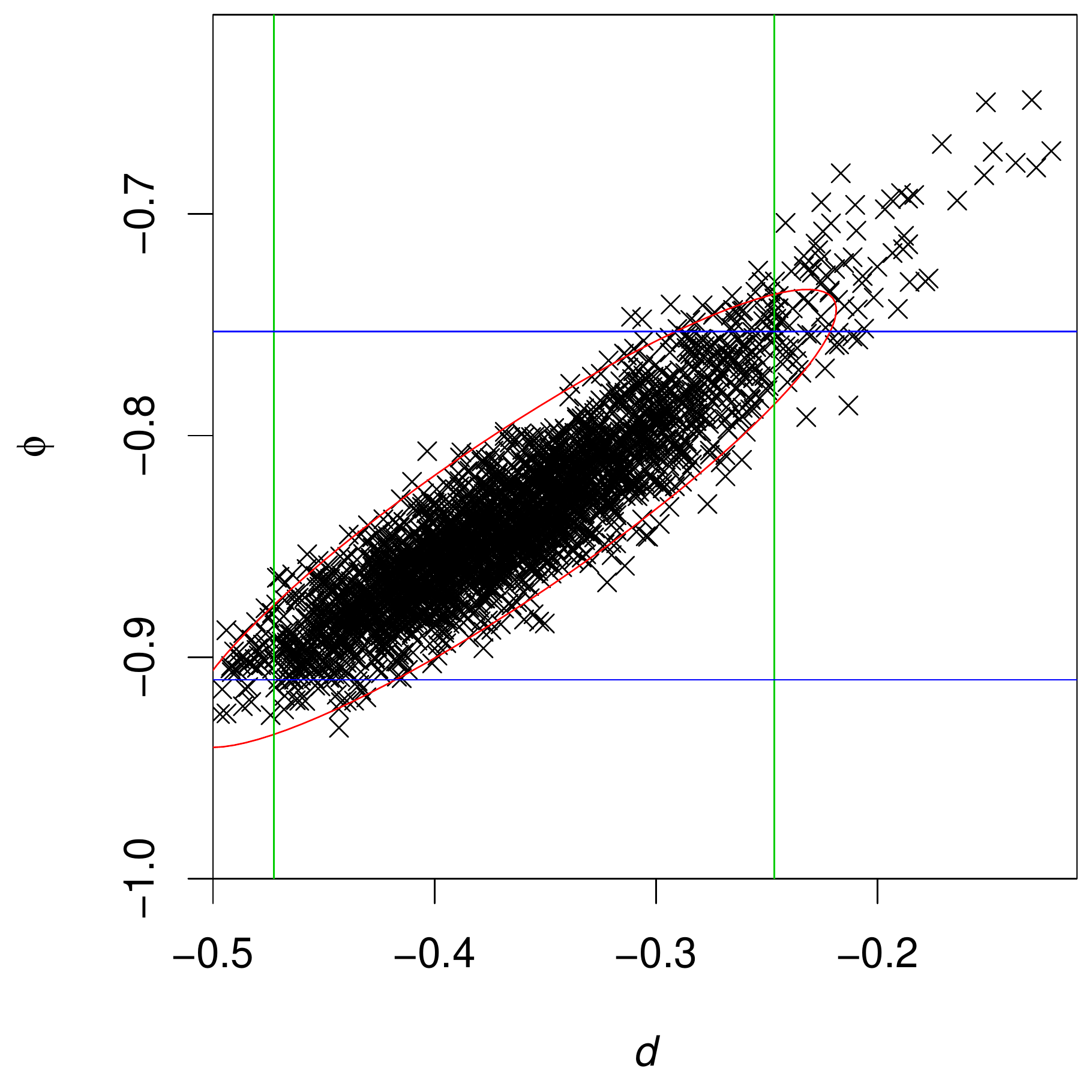} \\ 
$\quad\quad$ (a) & $\quad\quad$ (b) & $\quad\quad$ (c)
\end{tabular}
\end{center}
\vspace{-0.25cm}
\caption{Spectra for processes in figure
\ref{fig:5_a_i_dd_phi_038and5_a_i_dd_phi_002}. Green line is relevant
\ARMA{1,0} process, red line is relevant \FID{d} process, black line is
\FAR{1,d,0} process; (a)
$(1+0.92\mathcal{B})(1-\mathcal{B})^{0.25}X_t=\innov{}$; (b)
$(1-0.83\mathcal{B})(1-\mathcal{B})^{-0.35}X_t=\innov{}$. Pane (c)
shows posterior samples of $(d,\phi)$ from series considered in pane (b) with
credibility sets: red is 95\% credibility set for $(d,\phi)$, green is 95\%
credibility interval for $d$, blue is 95\% credibility interval for $\phi$. }
\label{fig:C_1_38_spectrumandC_1_2_spectrum}
\end{figure}

In general, this method works very well; two example outputs are presented in
figure \ref{fig:5_a_i_dd_phi_038and5_a_i_dd_phi_002}, under two similar
data generating mechanisms.
Plot (a) shows relatively mild correlation ($\rho=0.21$) compared with (b)
which shows strong correlation ($\rho=0.91$). This differential behaviour can
be explained heuristically by considering the differing data-generating
values. For the process in plot (a) the short memory and long memory
components exhibit their effects at opposite ends of the spectrum; see figure
\ref{fig:C_1_38_spectrumandC_1_2_spectrum}(a). The resulting ARFIMA
spectrum, with peaks at either end, makes it easy to distinguish between short
and long memory effects, and consequently the posteriors of $d$ and $\phi$ are
largely uncorrelated. In contrast, the parameters of the process in plot (b)
express their behaviour at the same end of the spectrum. With negative $d$
these effects partially cancel each other out, except very near the origin
where the negative memory effect dominates; see figure
\ref{fig:C_1_38_spectrumandC_1_2_spectrum}(b). Distinguishing between the
effects of $\phi$ and $d$  is much more difficult in this case, consequently
the posteriors are much more dependent.
In cases where there is significant correlation between $d$ and $\phi$, it
arguably makes little sense to consider only the marginal posterior
distribution of $d$. For example the 95\% credibility interval for $d$ from
plots (b) is $(-0.473,-0.247)$, and the corresponding interval for $\phi$ is
$(-0.910,-0.753)$, yet these clearly give a rather pessimistic view of our
joint knowledge about $d$ and $\phi$---see figure
\ref{fig:C_1_38_spectrumandC_1_2_spectrum}(c).
In theory an ellipsoidal credibility set could be constructed,
although this is clearly less practical when $\dim \bm{\omega} > 2$.

{\bf Unknown form:} The RJ scheme outlined in
section \ref{sec:rj} works well for data simulated with $p$ and $q$ up
to 3. The marginal posteriors for $d$ are generally roughly centred
around $d_I$ (the data generating value) and the modal posterior model
probability is usually the `correct' one.  To illustrate, consider
again the two example data generating contexts used above.

\begin{figure}[ht!]
\begin{center}
\begin{tabular}{cccc}
\includegraphics[width=8.9pc,trim=85 0 0 0]{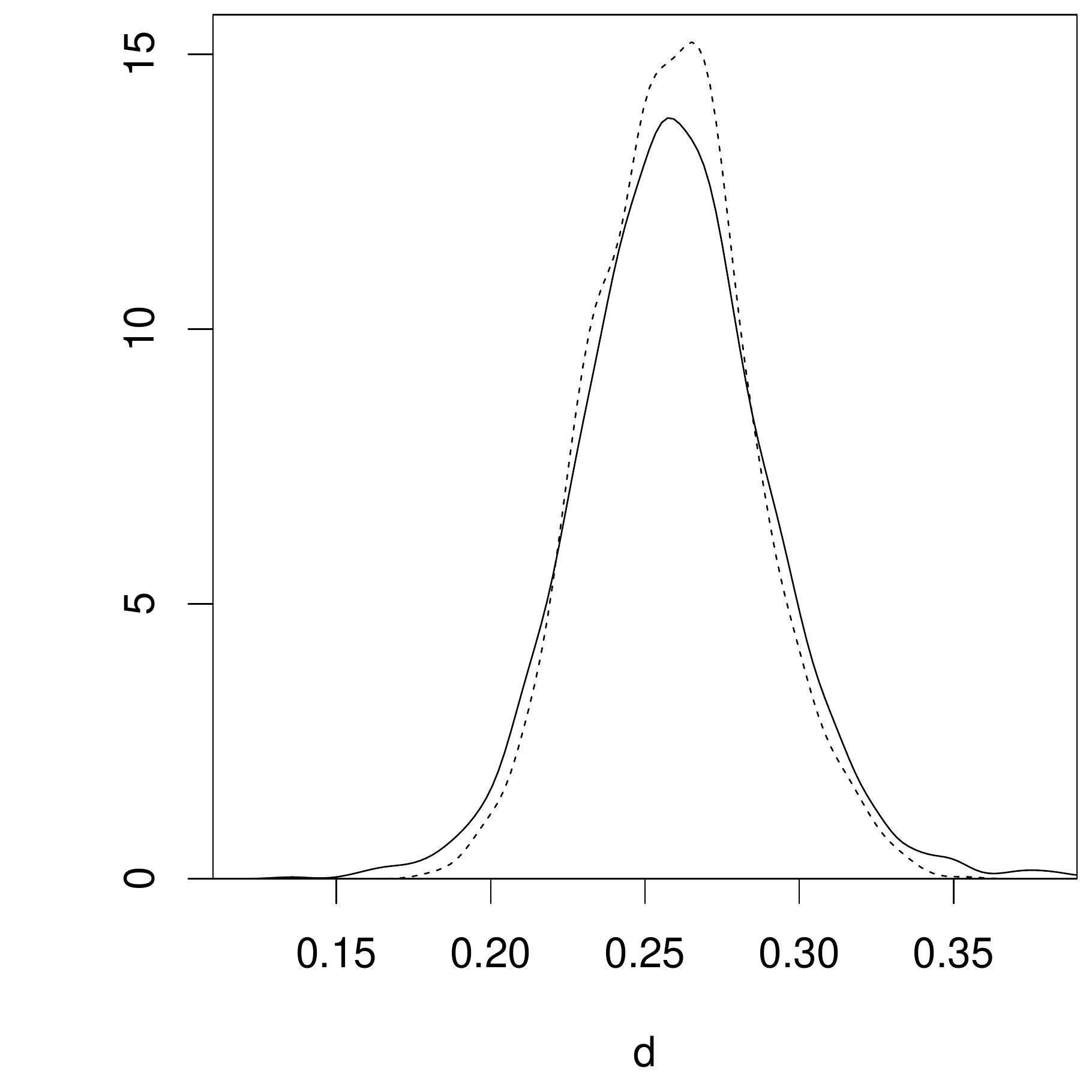} & 
\includegraphics[width=8.9pc,trim=85 0 0 0]{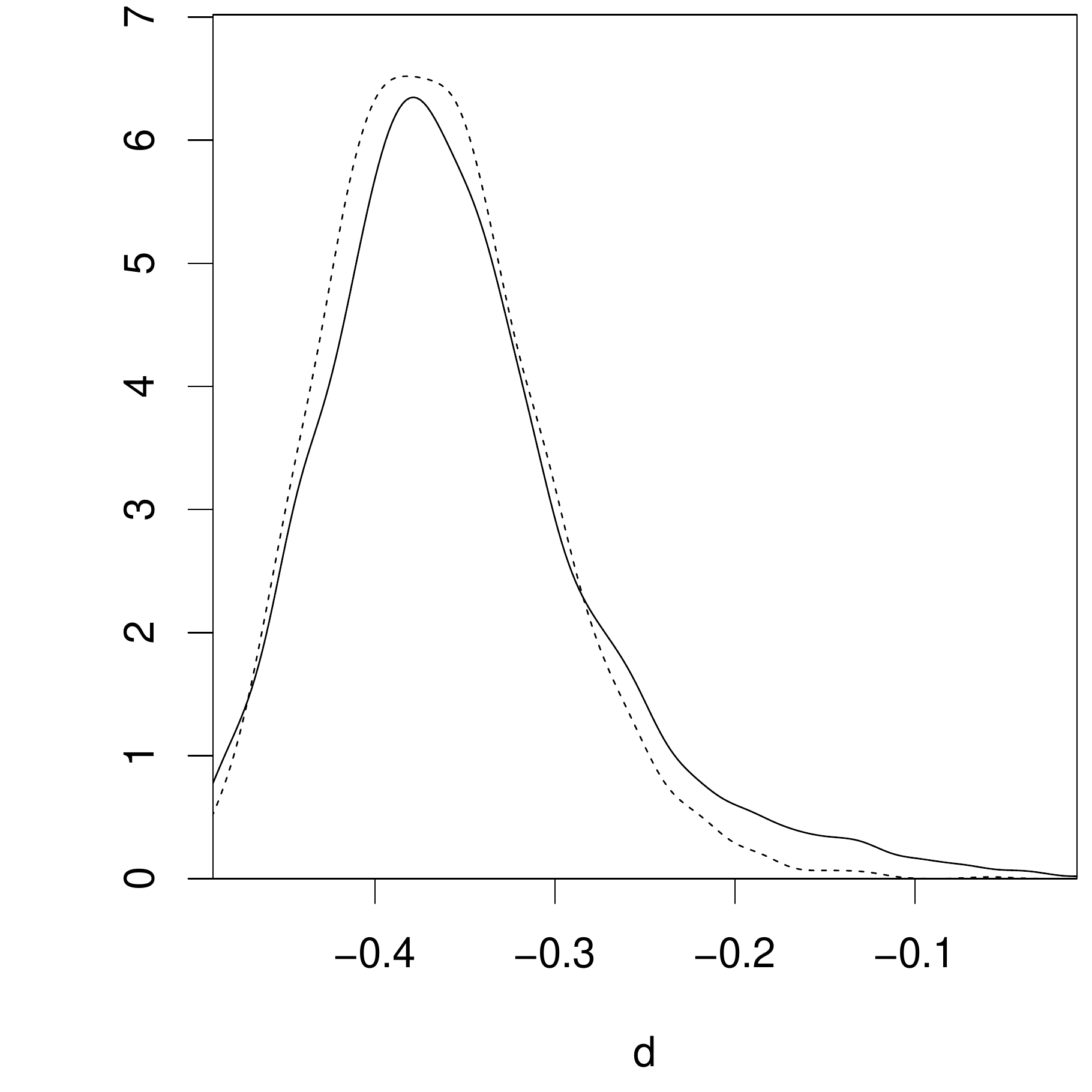} &
\includegraphics[width=8.9pc,trim=85 0 0 0]{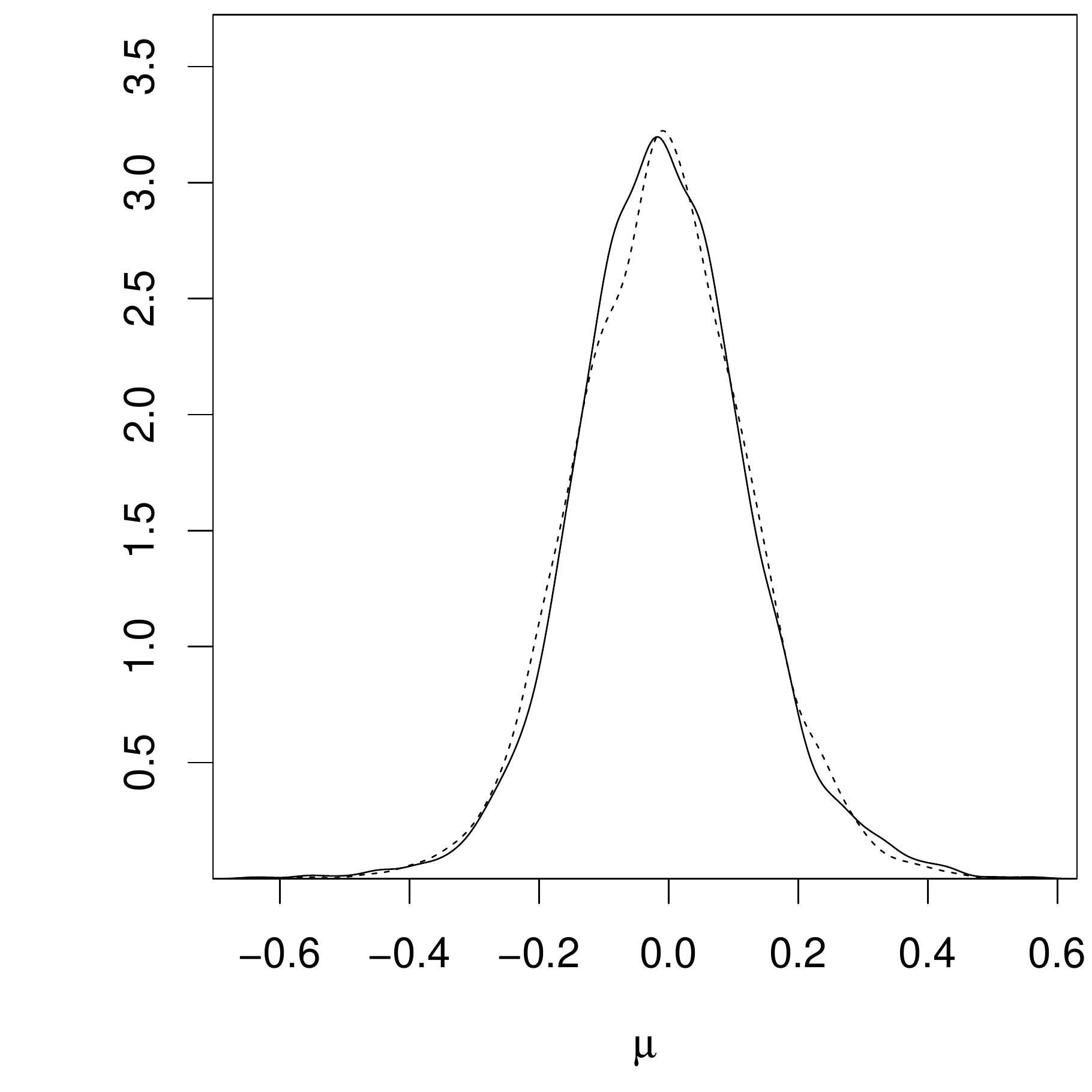} & 
\includegraphics[width=8.9pc,trim=85 0 0 0]{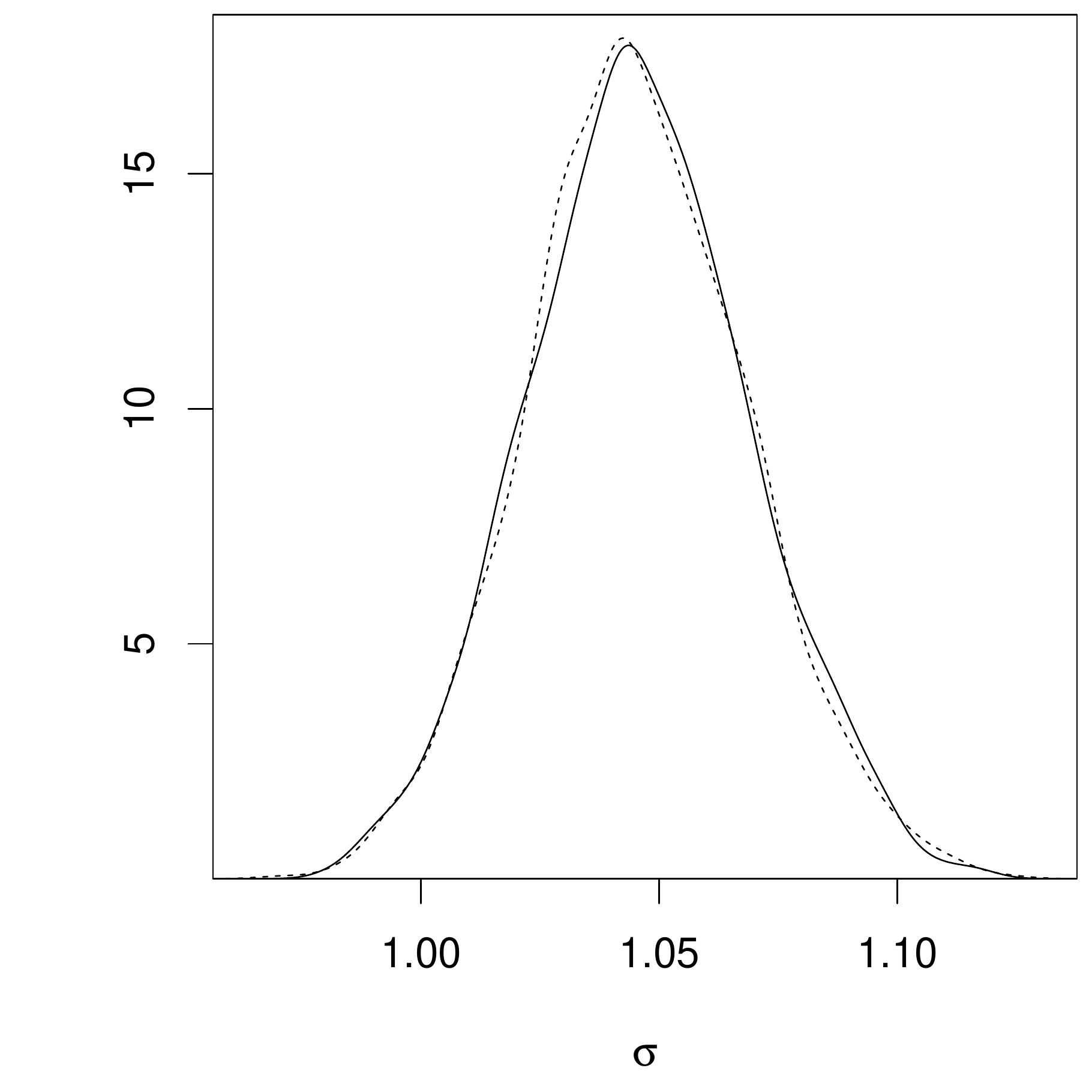} \\
\  (a) & \ (b) & \ (c) & \ (d)
\end{tabular}
\caption{Marginal posterior density of $d$ from series in figure 
\ref{fig:5_a_i_dd_phi_038and5_a_i_dd_phi_002}, (a)--(b) respectively. Solid
line is density obtained using reversible-jump algorithm. Dotted line is
density obtained using fixed $p=1$ and $q=0$. Panels (c)--(d) shows
the posterior densities for $\mu$ and $\sigma$, respectively, corresponding
to the series in \ref{fig:5_a_i_dd_phi_038and5_a_i_dd_phi_002}(a); those
for \ref{fig:5_a_i_dd_phi_038and5_a_i_dd_phi_002}(b) look similar.}
\label{fig:5_b_i_dd_038and5_b_i_dd_002}
\end{center}
\end{figure}

For both series, kernel density for the marginal posterior for
$d$ are plotted in figure \ref{fig:5_b_i_dd_038and5_b_i_dd_002}(a)--(b), together with
the equivalent density estimated assuming unknown model orders.
Notice how the densities obtained via the RJ method are very close to those
obtained assuming $p=1$ and $q=0$. The former are slightly more heavy-tailed,
reflecting a greater level of uncertainty about $d$.
Interestingly, the corresponding plots for the posteriors of $\mu$ and
$\sigma$ do not appear to exhibit this effect---see figure
\ref{fig:5_b_i_dd_038and5_b_i_dd_002}(c)--(d).
The posterior model probabilities are presented in table
\ref{table:postmodprobsfor38}, showing that the `correct' modes are
being picked up consistently.
\begin{table}[ht!]
\centering
\caption{Posterior model probabilities for time series from figures
\ref{fig:5_a_i_dd_phi_038and5_a_i_dd_phi_002}(a)--(b) and
\ref{fig:5_b_i_dd_038and5_b_i_dd_002}(a)--(b).}
\label{table:postmodprobsfor38}
\small
\vspace{3 mm}
(a) \begin{tabular}{r|rrrrrr|r} 
$p\backslash q$ & 0 & 1 & 2 & 3 & 4 & 5 & marginal \\ \hline
0 & 0.000 & 0.000 & 0.000 & 0.000 & 0.000 & 0.000 & 0.000 \\
1 & {\bf 0.805} & 0.101 & 0.003 & 0.000 & 0.000 & 0.000 & 0.908 \\
2 & 0.038 & 0.043 & 0.001 & 0.000 & 0.000 & 0.000 & 0.082 \\
3 & 0.005 & 0.004 & 0.000 & 0.000 & 0.000 & 0.000 & 0.009 \\
4 & 0.000 & 0.001 & 0.000 & 0.000 & 0.000 & 0.000 & 0.001 \\
5 & 0.000 & 0.000 & 0.000 & 0.000 & 0.000 & 0.000 & 0.000  \\ \hline
marginal & 0.848 & 0.148 & 0.004 & 0.000 & 0.000 & 0.000 & 
\end{tabular}

\vspace{0.25cm}
(b) \begin{tabular}{r|rrrrrr|r} 
$p\backslash q$ & 0 & 1 & 2 & 3 & 4 & 5 & marginal \\ \hline
0 & 0.000 & 0.000 & 0.000 & 0.000 & 0.000 & 0.000 & 0.000 \\
1 & {\bf 0.829} & 0.125 & 0.002 & 0.000 & 0.000 & 0.000 & 0.956 \\
2 & 0.031 & 0.013 & 0.000 & 0.000 & 0.000 & 0.000 & 0.044 \\
3 & 0.000 & 0.000 & 0.000 & 0.000 & 0.000 & 0.000 & 0.000 \\
4 & 0.000 & 0.000 & 0.000 & 0.000 & 0.000 & 0.000 & 0.000 \\
5 & 0.000 & 0.000 & 0.000 & 0.000 & 0.000 & 0.000 & 0.000  \\ \hline
marginal & 0.860 & 0.138 & 0.002 & 0.000 & 0.000 & 0.000 & 
\end{tabular}
\end{table}

\begin{figure}[ht!]
\begin{center}
\begin{tabular}{cc}
\includegraphics[width=12pc]{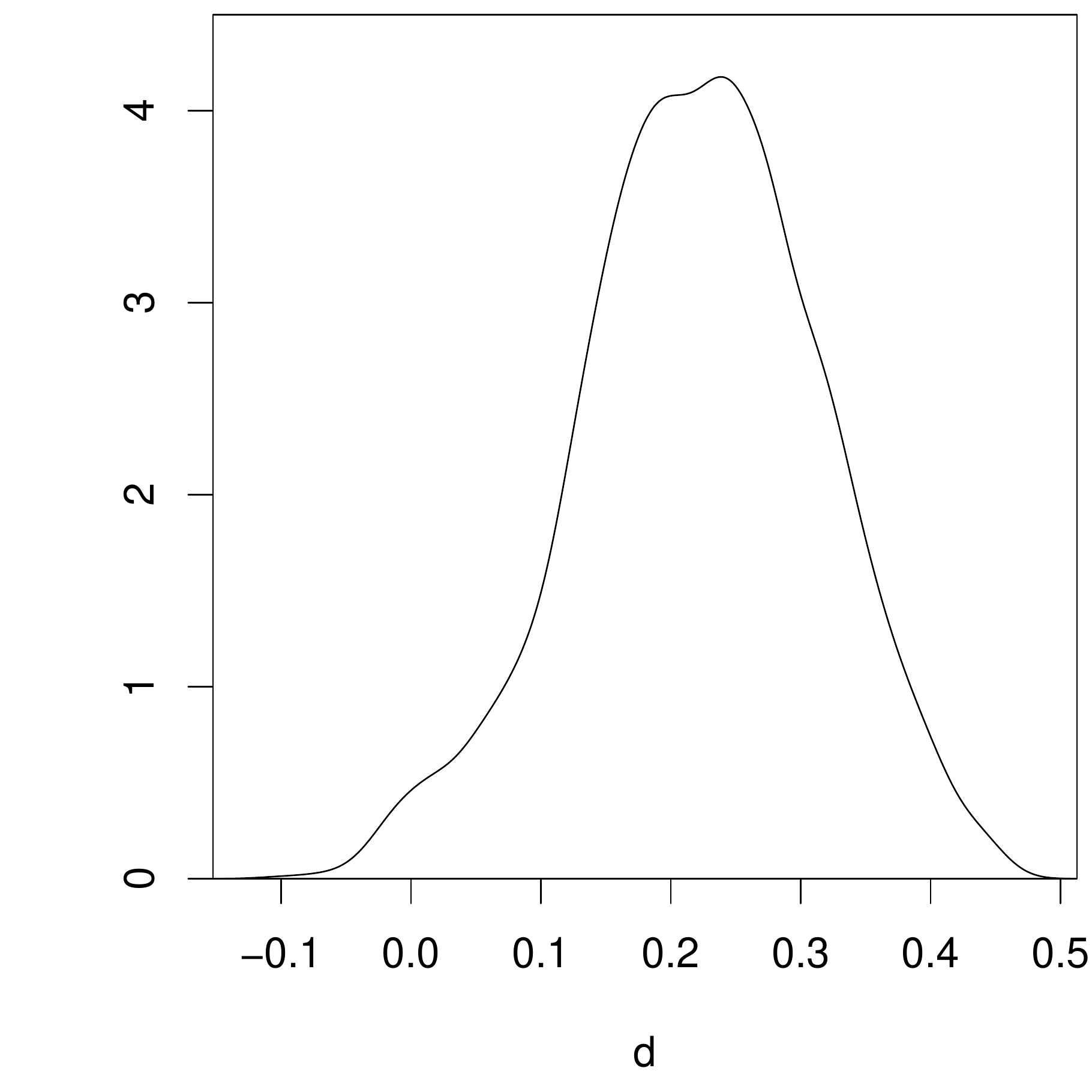} & \includegraphics[width=12pc]{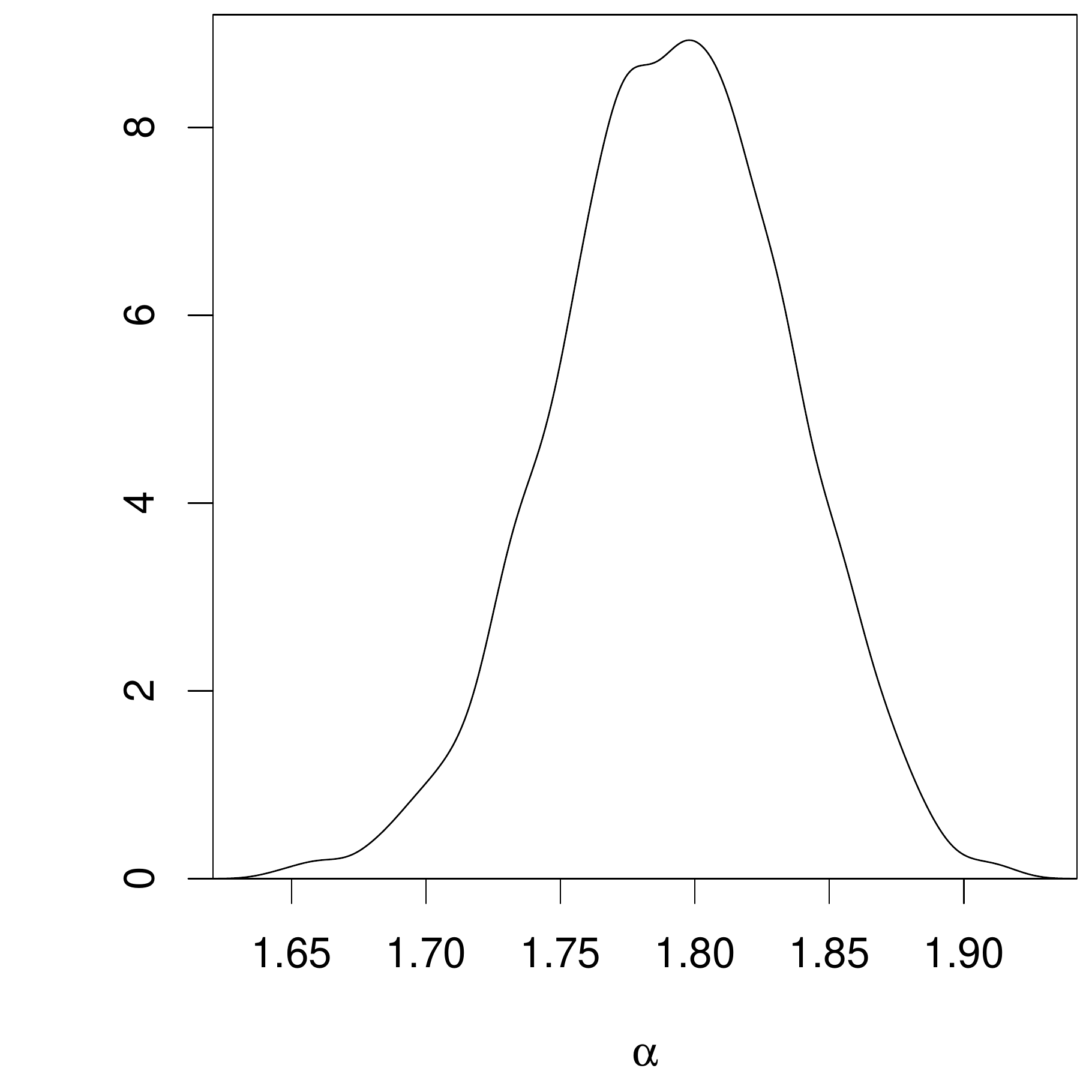} \\
$\quad\quad$ (a) & $\quad\quad$ (b)
\end{tabular}
\caption{Marginal posterior densities (a) $d$, (b) $\alpha$.}
\label{fig:5_e_dd_1and5_e_alpha_1}
\end{center}
\end{figure}

As a test of the robustness of the method, consider a complicated short memory
input combined with a heavy tailed $\alpha$-stable innovations distribution.
Specifically, the time series that will be used is the following \FAR{2,d,1}
process
\begin{equation}
\label{eqn:defnoftrickyalpha2,d,1timeseries}
\left(1-\frac{9}{16}\mathcal{B}^2\right)\left(1-\mathcal{B}\right)^{0.25}X_t =
\left(1 + \frac{1}{3}\mathcal{B}\right)\innov{},\qquad \hbox{where } \innov
\sim \Sstandard{\alpha=1.75}{0}.
\end{equation}
For more details, see \citep[][\S7.1]{graves:phd}.
The marginal posterior densities of $d$ and $\alpha$ are presented in figure
\ref{fig:5_e_dd_1and5_e_alpha_1}.
Performance looks good despite the complicated structure. The posterior
estimate for $d$ is $\widehat{d}^{(B)}= 0.22$, with 95\% CI
$(0.04,0.41)$. Although this interval is admittedly rather wide, it is
reasonably clear that long memory is present in the signal. The corresponding
interval for $\alpha$ is $(1.71,1.88)$ with estimate
$\widehat{\alpha}^{(B)}=1.79$. Finally, we see from table
\ref{table:postmodprobsfor5e} that the algorithm is very rarely in the `wrong'
model.
\begin{table}[ht!]
\centering
\caption{Posterior model probabilities.}
\label{table:postmodprobsfor5e}
\vspace{3 mm}
\begin{tabular}{c|cccccc|c} 
$p\backslash q$ & 0 & 1 & 2 & 3 & 4 & 5 & marginal \\ \hline
0 & 0.000 & 0.000 & 0.000 & 0.000 & 0.000 & 0.000 & 0.000 \\
1 & 0.000 & 0.000 & 0.000 & 0.000 & 0.000 & 0.000 & 0.000 \\
2 & 0.000 & {\bf 0.822} & 0.098 & 0.001 & 0.000 & 0.000 & 0.921 \\
3 & 0.014 & 0.056 & 0.004 & 0.000 & 0.000 & 0.000 & 0.075 \\
4 & 0.003 & 0.001 & 0.000 & 0.000 & 0.000 & 0.000 & 0.004 \\
5 & 0.000 & 0.000 & 0.000 & 0.000 & 0.000 & 0.000 & 0.000 \\ \hline
marginal & 0.017 & 0.880 & 0.102 & 0.002 & 0.000 & 0.000 & 
\end{tabular}
\end{table}

\begin{figure}[ht!]
\begin{minipage}{5.5cm}
\small
\begin{tabular}{rr|rr} 
 & mean & \multicolumn{2}{c}{95\% CI} \\ \hline
$d$       & $0.039$ & $0.336$ & $0.482$ \\
$\mu$      & $62$  & $1037$ & $1284$ \\
$\sigma$ & $1.91$ & $66.46$  & $73.97$
\end{tabular}
\end{minipage} \hfill
\begin{minipage}{11cm}
\begin{tabular}{cc}
\includegraphics[width=12pc]{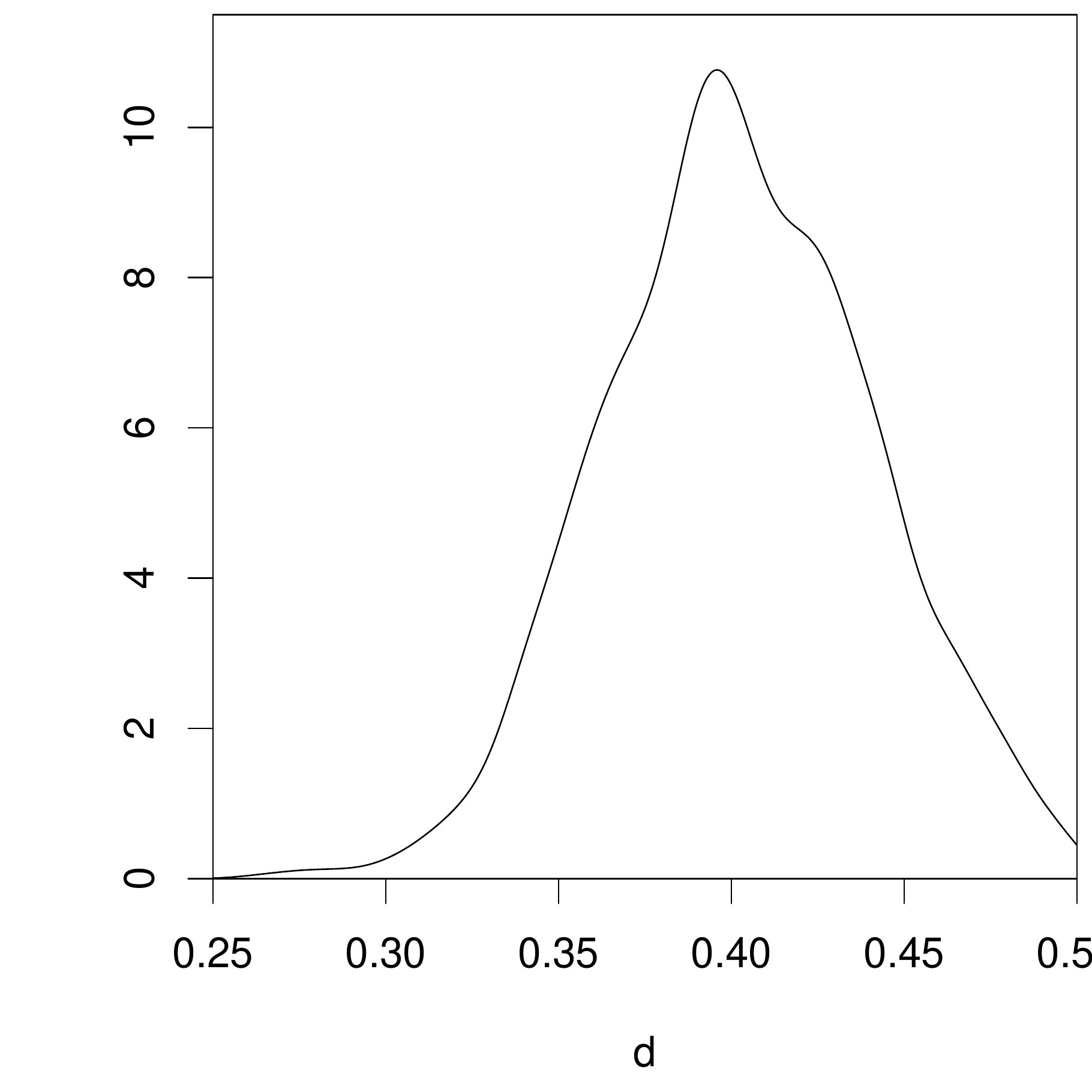} & \includegraphics[width=12pc]{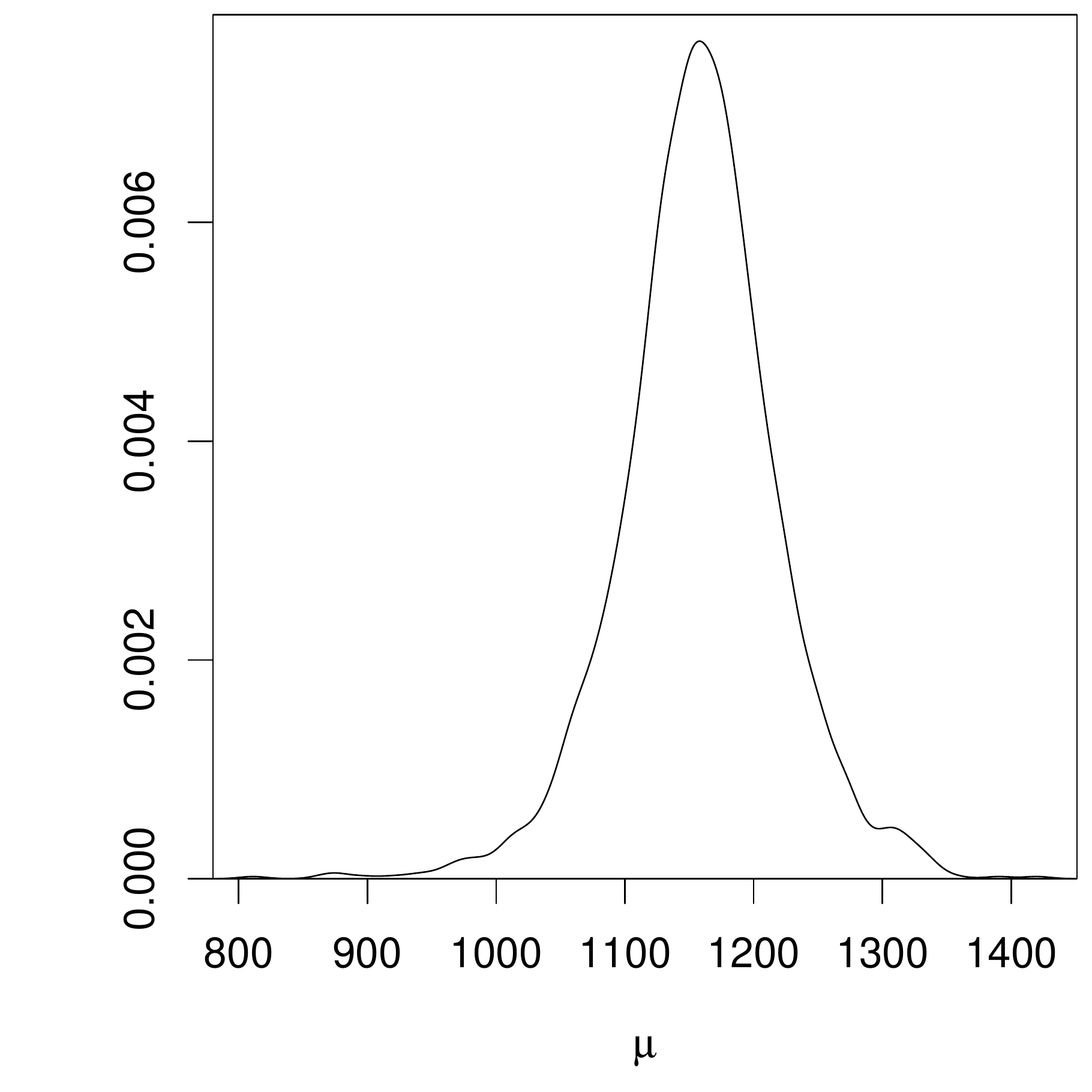} \\
$\quad\quad$ (a) & $\quad\quad$ (b)
\end{tabular}
\end{minipage}
\caption{{\em Table:} Summary posterior statistics for Nile minima. {\em
Plots:} Marginal posterior densities for Nile minima; (a) $d$, (b) $\mu$.}
\label{fig:nile_dd_1andnile_mu_1}
\end{figure}

{\bf The Nile Data:} We conclude with an application of our % suite of
methods to the famous annual Nile minima data. Because of the
fundamental importance of the river to the civilisations it has
supported, local rulers kept measurements of the annual maximal and
minimal heights obtained by the river at certain points (called
gauges). The longest uninterrupted sequence of recordings is from the
Roda gauge (near Cairo), between 622 and 1284 AD ($n=663$).\footnote{There is evidence
\citep[e.g.][]{Ko_2006b} that the sequence is not actually homogeneous.}
% estimation for the location of a change point in the data coincides with
% historical evidence of a new construction at the site.} 
% These data are plotted
% in figure
% \ref{fig:nile_data}.
% \begin{figure}[ht!]
% \begin{center}
% \includegraphics[width=25pc]{nile_data}
% \caption[Nile minima]{Annual Nile minima.}
% \label{fig:nile_data}
% \end{center}
% \end{figure}
%
The posterior summary
statistics and marginal densities of $d$ and $\mu$ for the Nile data
are presented in figure
\ref{fig:nile_dd_1andnile_mu_1}. Posterior model probabilities are presented
in table \ref{table:postmodprobsfornile}.
\begin{table}[ht!]
\centering
\caption{Posterior model probabilities for Nile minima.}
\label{table:postmodprobsfornile}
\vspace{3 mm}
\begin{tabular}{r|rrrrrr|r} 
$p\backslash q$ & 0 & 1 & 2 & 3 & 4 & 5 & marginal \\ \hline
0 & {\bf 0.638} & 0.101 & 0.010 & 0.000 & 0.000 & 0.000 & 0.750 \\
1 & 0.097 & 0.124 & 0.011 & 0.000 & 0.000 & 0.000 & 0.232 \\
2 & 0.007 & 0.010 & 0.000 & 0.000 & 0.000 & 0.000 & 0.018 \\
3 & 0.000 & 0.000 & 0.000 & 0.000 & 0.000 & 0.000 & 0.000 \\
4 & 0.000 & 0.000 & 0.000 & 0.000 & 0.000 & 0.000 & 0.000 \\
5 & 0.000 & 0.000 & 0.000 & 0.000 & 0.000 & 0.000 & 0.000 \\ \hline
marginal & 0.742 & 0.236 & 0.022 & 0.000 & 0.000 & 0.000 & 
\end{tabular}
\end{table}
We see that the model with the highest posterior probability is the \FID{d}
model with $d\approx 0.4$. This suggests a strong, `pure', long memory
feature. Our results compare favourably with other studies
\citep{Liseo_2001,Hsu_2003,Ko_2006a}.

\section{Discussion}
\label{sec:discuss}
We have provided a systematic treatment of efficient Bayesian inference for
ARFIMA models, the most popular parametric model combining long and
short memory effects.  Through a mixture of theoretical and empirical work we
have demonstrated that the methods can handle the sorts of time series
data that are typically confronted with possible long memory in mind.  

Many of the choices made throughout, but in particular those leading to our
likelihood approximation stem from a
need to accommodate further extension.  For example, in future work we intend
to extend them to cope with a heavy-tailed
innovations distribution.  For more evidence of potential in this context, see
\citet[][\S7]{graves:phd}. Along similar lines, there is scope for further
generalisation to incorporate seasonal (long memory) effects.
Finally, an advantage of the Bayesian approach is that it provides a
natural mechanism for dealing with missing data, via data augmentation. This
is particularly relevant for long historical time series which may, for a
myriad of reasons, have recording gaps. For example, some of the data
recorded at other gauges along the river Nile have missing observations
although otherwise span a similarly long time frame.  For a demonstration of
how this might fit within our framework, see \S5.6 of \citeauthor{graves:phd}
dissertation.
 
\appendix

\section{Gibbs sampling of $\mu$ and $\sigma$}
\label{sec:gibbs}

Assuming Gaussianity, and the
 Gaussian and root-inverse-gamma independent priors (deliberately chosen to
 ensure prior-posterior conjugacy), it is possible to use Gibbs updating for
 the parameters $\mu$ and $\sigma$. Demonstrating this requires the following
 result: If $g(x) \propto \exp \left[-\frac{1}{2}(\alpha x^2 - 2\beta x
 )\right]$, and $\int_\mathbb{R}g(x)\,dx = 1$, then $g \sim
 \mathcal{N}(\frac{\beta}{\alpha} ,\frac{1}{\alpha} )$. 

Now, let $\bar{c} =: \frac{1}{n}\sum_{t=1}^{n}c_t$. Then,  updating $\mu$
by combining its prior with the approximate (log) likelihood
\eqref{eqn:formofapproxloglikelihood}, yields the following conditional
posterior:
\begin{align*}
\pi_{\mu|\psi_{-\mu}}(\mu|\bm{\psi}_{-\mu},\mathbf{x}) & \propto  \frac{1}{\sqrt{2\pi\sigma_0^2}} \exp\left\{-\frac{(\mu-\mu_0)^2}{2\sigma_0^2} \right\}  \sigma^{-n}\prod_{t=1}^{n}  \left\{\frac{1}{\sqrt{2\pi}}\exp\left[-\frac{(c_t-\Pi_P\mu)^2}{2\sigma^2} \right]\right\} \\
& \propto  \exp\left\{-\frac{(\mu-\mu_0)^2}{2\sigma_0^2} \right\} \exp  \left\{-\frac{1}{2\sigma^2}\sum_{t=1}^n(c_t-\Pi_P\mu)^2\right\} \\
& \propto \exp \left\{-\frac{1}{2\sigma^2_0}(\mu - \mu_0)^2 - \frac{1}{2\sigma^2}\left(n\Pi_P^2\mu^2-2\mu\Pi_Pn\bar{c}\right) \right\} \\
%& = \exp \left\{-\frac{1}{2}\left[\frac{1}{\sigma^2_0}(\mu^2 - 2\mu\mu_0) + \frac{1}{\sigma^2}\left(n\Pi_P^2\mu^2-2\mu\Pi_Pn\bar{c}\right) \right]\right\} \\
& \propto \exp \left\{-\frac{1}{2}\left[\mu^2\left(\frac{1}{\sigma^2_0}\ + \frac{n\Pi_P^2}{\sigma^2}\right) - 2\mu\left(\frac{\mu_0}{\sigma^2_0} + \frac{\Pi_Pn\bar{c}}{\sigma^2}\right) \right]\right\},
\end{align*}
Then, our result reveals
\begin{equation}
\label{eqn:Gibbsmuapprox}
\mu|\bm{\psi}_{-\mu},\mathbf{x} \sim \mathcal{N}\left(\left[\frac{1}{\sigma^2_0} + \frac{n\Pi_P^2}{\sigma^2}\right]^{-1}\left[\frac{\mu_0}{\sigma^2_0} + \frac{\Pi_Pn\bar{c}}{\sigma^2}\right],\left[\frac{1}{\sigma^2_0} + \frac{n\Pi_P^2}{\sigma^2}\right]^{-1}\right).
\end{equation}

When using the exact likelihood method, 
% from \eqref{eqn: likelihood form 2}:
% \begin{equation*}
% \pi_{\mu|\psi_{-\mu}}(\mu,|\bm{\psi}_{-\mu},\mathbf{x}) \propto \exp \left\{-\frac{1}{2\sigma^2_0}(\mu - \mu_0)^2\right\} \exp \left\{  -\frac{1}{2\sigma^2} \left[\mu^2 \mathbf{1}_n^t\Sigma_d^{-1}\mathbf{1}_n - 2\mu\mathbf{x}^t\Sigma_d^{-1} \mathbf{1}_n   \right] \right\},
% \end{equation*}
% and by using lemma \ref{lemma: conditional Gaussian} again:
we obtain
\begin{align}
\label{eqn:Gibbsmuexact}
\mu|\bm{\psi}_{-\mu},\mathbf{x} &\sim  \mathcal{N}\left(\left[\frac{1}{\sigma_0^2} + \frac{\mathbf{1}_n^t\Sigma_d^{-1}\mathbf{1}_n}{\sigma^2}\right]^{-1} \left[\frac{\mu_0}{\sigma_0^2} + \frac{\mathbf{x}^t\Sigma_d^{-1} \mathbf{1}_n }{\sigma^2}\right],\left[\frac{1}{\sigma_0^2} + \frac{\mathbf{1}_n^t\Sigma_d^{-1}\mathbf{1}_n}{\sigma^2}\right]^{-1} \right).
\intertext{In the limiting case of the flat prior, the conditional density 
\eqref{eqn:Gibbsmuapprox} becomes:}
\mu|\bm{\psi}_{-\mu},\mathbf{x} %= \left(\frac{n\Pi_P^2}{\sigma^2}\right)^{-1}\mathcal{N}\left(\left(\frac{\Pi_Pn\bar{c}}{\sigma^2}\right),\left(\frac{n\Pi_P^2}{\sigma^2}\right)\right) 
& \sim \mathcal{N}\left(\frac{\bar{c}}{\Pi_P},\frac{\sigma^2}{n\Pi_P}\right),
\quad \hbox{and \eqref{eqn:Gibbsmuexact} is} \quad
\mu|\bm{\psi}_{-\mu},\mathbf{x} \sim \mathcal{N}\left(\frac{\mathbf{x}^t\Sigma_d^{-1} \mathbf{1}_n}{\mathbf{1}_n^t\Sigma_d^{-1}\mathbf{1}_n} ,\frac{\sigma^2}{\mathbf{1}_n^t\Sigma_d^{-1}\mathbf{1}_n} \right). \nonumber
\end{align}

Likewise, when updating $\sigma$  we
obtain the conditional posterior:
\begin{align}
\pi_{\sigma|\psi_{-\sigma}}(\sigma|\bm{\psi}_{-\sigma},\mathbf{x}) 
&\propto  \frac{2}{\Gamma(\alpha_0)}\beta_0^{\alpha_0} \sigma^{-(2\alpha_0 +1)} \exp\left(-\frac{\beta_0}{\sigma^2}\right) \sigma^{-n}\prod_{t=1}^{n}  \left\{\frac{1}{\sqrt{2\pi}}\exp\left[-\frac{(c_t-\Pi_P\mu)^2}{2\sigma^2} \right]\right\} \nonumber \\
& \propto  \sigma^{-(2\alpha_0 + n +1)} \exp\left\{-\frac{1}{\sigma^2}\left[\beta_0 + \frac{1}{2}\sum_{t=1}^n(c_t-\Pi_P\mu)^2\right]\right\}, \quad \hbox{giving} \nonumber \\
\sigma|\bm{\psi}_{-\sigma},\mathbf{x} &
\sim \mathcal{R}\left(\alpha_0 + \frac{n}{2},\beta_0 + 
\frac{1}{2}\sum_{t=1}^n(c_t-\Pi_P\mu)^2\right).
\label{eq:Gibbssigmaapprox}
\intertext{Similarly, when using the exact likelihood \eqref{eqn:log-likelihoodwithQ}:}
% \pi_{\sigma|\psi_{-\sigma}}(\sigma,|\bm{\psi}_{-\sigma},\mathbf{x}) & 
% \propto \frac{2}{\Gamma(\alpha_0)}\beta_0^{\alpha_0} \sigma^{-(2\alpha_0 +1)} \exp\left(-\frac{\beta_0}{\sigma^2}\right) \sigma^{-n} [\det(\Sigma_d)]^{-1/2} \exp\left\{-\frac{1}{2\sigma^2} Q(\mathbf{x}|\mu,d) \right\} \\
% & \propto \sigma^{-(2\alpha_0 + n +1)}  \exp\left\{-\frac{1}{\sigma^2} 
% \left[ \beta_0 + \frac{1}{2}Q(\mathbf{x}|\mu,d) \right] \right\}, \quad \hbox{giving} \\
\sigma|\bm{\psi}_{-\sigma},\mathbf{x} &\sim \mathcal{R}\left(\alpha_0 + \frac{n}{2},\beta_0 + \frac{1}{2}Q(\mathbf{x}|\mu,d)\right),
\label{eq:Gibbssigmaexact}
\intertext{Finally, the limiting case of the diffuse prior yields we
obtain the following for \eqref{eq:Gibbssigmaapprox} and \eqref{eq:Gibbssigmaexact} }
\sigma|\bm{\psi}_{-\sigma},\mathbf{x} &\sim \mathcal{R}\left(\frac{n}{2},\frac{1}{2}\sum_{t=1}^n(c_t-\Pi_P\mu)^2\right)
\quad \hbox{and} \quad 
\sigma|\bm{\psi}_{-\sigma},\mathbf{x} \sim \mathcal{R}\left(\frac{n}{2},\frac{1}{2}Q(\mathbf{x}|\mu,d)\right). \nonumber
\end{align}

\section{Preservation of ACF under bijection}
\label{sec:bijection}

Consider the bijection from \citet{Monahan_1984} in
\eqref{eqn:reparameterisationforwardrecursion}.  Now, suppose we have the
vector $\bm{\varphi}^{(p)} =
(\varphi_1^{(p)},\ldots,\varphi_{p-1}^{(p)},\varphi_p^{(p)})$, and its
truncated version:
\begin{equation*}
\bm{\varphi}^{(p-1)} = (\varphi_1^{(p)},\ldots,\varphi_{p-1}^{(p)}) =: (\varphi_1^{(p-1)},\ldots,\varphi_{p-1}^{(p-1)}).
\end{equation*}
Let the `classical' representations of these vectors be denoted
$\bm{\phi}^{(p)}$ and $\bm{\phi}^{(p-1)}$ respectively, and note that
$(\phi_1^{(p-1)},\ldots,\phi_{p-1}^{(p-1)}) \neq
(\phi_1^{(p)},\ldots,\phi_{p-1}^{(p)})$. These define two polynomials
$\Phi^{(p)}$ and $\Phi^{(p-1)}$, so a natural question to ask is: What is the
relationship between $\Phi^{(p)}$ and $\Phi^{(p-1)}$?
\begin{Theorem}
\label{theorem:firstmatching}
Consider the pair of \AR{p} and \AR{p-1} models, with polynomials $\Phi^{(p)}$ and $\Phi^{(p-1)}$ as defined above. Denote the ACV of the \AR{p} model by $\gamma(\cdot)$, and that of the \AR{p-1} model by $\gamma'(\cdot)$. Then for $0\leq k <p$:
$\gamma'(k) = \gamma(k) \left[1-\left(\phi^{(p)}_p\right)^2\right]$.
\begin{proof}
Our strategy is to show that $\gamma'(\cdot)$ satisfies the relevant Yule--Walker equations, 
and then appeal to uniqueness. 
Firstly, from \eqref{eqn:reparameterisationforwardrecursion} we have that
$\left[1-\left(\phi_p^{(p)}\right)^2\right] \phi_i^{(p-1)} =
\phi_i^{(p)}-\phi_p^{(p)}\phi_{p-i}^{(p)}$, $i=1,\ldots,p-1$. (Note that this
also is valid for special case of $i=0$.) We are given
\citep[\S3.3]{Brockwell_1991} that the Yule--Walker equations for
$\bm{\phi}^{(p)}$ are
$%\begin{equation}
%\label{eqn:YWforphi^p}
\sum_{i=0}^{p}\phi^{(p)}_i\gamma(k-i) = \sigma^2 \mathbb{I}_{\{k=0\}}i%
% \left\{\begin{array}{lcl} \sigma^2 & \mathrm{for} & k=0 \\ 0 & \mathrm{for} & k=1,\ldots,p \end{array} \right. .
$. %\end{equation}
Now consider the following for $k=0,\ldots,p-1$:
\begin{align*}
\sum_{i=0}^{p-1}\phi^{(p-1)}_i\gamma'(k-i) &= \sum_{i=0}^{p-1}\phi^{(p-1)}_i \gamma(k-i)\left[1-\left(\phi_p^{(p)}\right)^2\right] 
= \sum_{i=0}^{p-1}\gamma(k-i)\left(\phi^{(p)}_i-\phi^{(p)}_p\phi^{(p)}_{p-i}\right) \\
&= \sum_{i=0}^{p}\gamma(k-i)\left(\phi^{(p)}_i-\phi^{(p)}_p\phi^{(p)}_{p-i}\right) - \gamma(k-p)\left(\phi^{(p)}_p-\phi^{(p)}_p\phi^{(p)}_{p-p}\right) \\
%&= \sum_{i=0}^{p}\gamma(k-i)\phi^{(p)}_i - \phi^{(p)}_p\sum_{i=0}^{p}\gamma(i-k)\phi^{(p)}_{p-i}  \\
&= \sum_{i=0}^{p}\phi^{(p)}_i\gamma(k-i) - \phi^{(p)}_p\sum_{i=0}^{p}\phi^{(p)}_i\gamma((p-k)-i) .
\end{align*}
Using Yule--Walker when $k=0$, the last is $\sigma^2 -
\phi_p^{(p)} 0 = \sigma^2$. And for $k=1,\ldots,p-1$ it equals $0-\phi_p^{(p)}
0=0$ (because $k=1,\ldots,p-1$ corresponds to $p-k =p-1,\ldots,1$). So we have
that:
\begin{equation*}
\sum_{i=0}^{p-1}\phi^{(p-1)}_i\gamma'(k-i) = \left\{\begin{array}{lcl}
\sigma^2 & \mathrm{for} & k=0 \\
0 & \mathrm{for} & k=1,\ldots,p-1
\end{array} \right. .
\end{equation*}
These equations are the Yule--Walker equations for the \AR{p-1} process so we are done.
\end{proof}
\end{Theorem}
\begin{Corollary}
\label{corollary: matching acfs!}
Under Thm.~\ref{theorem:firstmatching} assumptions, ACFs of
$\bm{\phi}^{(p)}\!$ and  $\bm{\phi}^{(p-1)}\!$ are equal for $0\leq k <p$.
\begin{proof}
\begin{equation*}
\rho'(k) = \frac{\gamma'(k)}{\gamma'(0)} = \frac{\gamma(k) \left[1-\left(\phi^{(p)}_p\right)^2\right]}{\gamma(0) \left[1-\left(\phi^{(p)}_p\right)^2\right]} = \frac{\gamma(k)}{\gamma(0)} = \rho(k).
\end{equation*}
\end{proof}
\end{Corollary}

\section{Positive definiteness of $\Sigma_{\bm{\varpi}}$}
\label{sec:posdef}

\begin{Lemma}
Suppose $A$ is an $M\times M$ matrix which can be block divided as:
\begin{equation*}
A \quad=\quad \left(
\begin{MAT}(@,40pt,20pt){c.c}
\aligntop
B_{m,m} & B_{m,m'}  \\.
\alignbottom
B_{m',m} & B_{m',m'}   \\
\end{MAT}
\right),
\end{equation*}
where $M=m+m'$, and $B_{m,m'}=B_{m',m}^t$. Furthermore, let $\mathbf{x} \in
\mathbb{R}^M$ be block divided as $\mathbf{x}^t =
\left(\begin{MAT}(@,20pt,10pt){c.c}
\mathbf{y}^t & \mathbf{z}^t \\\end{MAT}\right)$, where $\mathbf{y} \in
\mathbb{R}^m$ and $\mathbf{z} \in \mathbb{R}^{m'}$. Then:
\begin{equation}
\label{eqn:usefulstatementaboutblockmatrix/vectorproducts}
\mathbf{x}^tA\mathbf{x} = \mathbf{y}^tB_{m,m}\mathbf{y} + 2\mathbf{y}^tB_{m,m'}\mathbf{z} + \mathbf{z}^tB_{m',m'}\mathbf{z}.
\end{equation}
\end{Lemma}
\begin{Theorem}
\label{theorem:positivedefinitenessofSigmaoneway}
Let $\Sigma_M$ be an $M\times M$ positive-definite matrix, and let $\Sigma_m$
be the leading $m$-submatrix of $\Sigma_M$. Then $\Sigma_m$ is also positive
definite.
\begin{proof}
$\Sigma_M$ is positive definite so $\mathbf{x}^t\Sigma_M\mathbf{x} > 0$, for all $\mathbf{x} \in \mathbb{R}^M$. In particular, this is true for all block-divided $\mathbf{x}^t = \left(\begin{MAT}(@,20pt,10pt){c.c}
\mathbf{y}^t & \mathbf{0}^t \\\end{MAT}\right)$, where $\mathbf{y} \in \mathbb{R}^m$. Then from \eqref{eqn:usefulstatementaboutblockmatrix/vectorproducts}:
\begin{equation*}
0 < \mathbf{x}^t\Sigma_M\mathbf{x} = \mathbf{y}^t\Sigma_m\mathbf{y} + 2\mathbf{y}^tB_{m,m'}\mathbf{0} + \mathbf{0}^tB_{m',m'}\mathbf{0} = \mathbf{y}^t\Sigma_m\mathbf{y}.
\end{equation*} 
Clearly therefore, $\mathbf{y}^t\Sigma_m\mathbf{y}>0$ for all $\mathbf{y} \in
\mathbb{R}^m$, i.e.\ $\Sigma_m$ is positive-definite. \end{proof}
\end{Theorem}
\begin{Theorem}
\label{theorem:positivedefinitenessofSigmaotherway}
Suppose the $m\times m$ matrix $\Sigma_m$ and the $n\times n$ matrix
$\Sigma_n$ are positive definite. Then the following $(m+n)\times(m+n)$ matrix
is also positive definite:
\[\Sigma_{m,n} \quad=\quad \left(
\begin{MAT}(@,40pt,20pt){c.c}
\aligntop
\Sigma_m & \mathbf{O}  \\.
\alignbottom
\mathbf{O} & \Sigma_n   \\
\end{MAT}
\right).\]
\begin{proof}
Let $\mathbf{x} \in \mathbb{R}^{m+n}$ be partitioned as $\mathbf{x}^t =
\left(\begin{MAT}(@,20pt,10pt){c.c}
\mathbf{y}^t & \mathbf{z}^t \\\end{MAT}\right)$ where $\mathbf{y} \in
\mathbb{R}^m$ and $\mathbf{z} \in \mathbb{R}^n$. Then from
\eqref{eqn:usefulstatementaboutblockmatrix/vectorproducts},
$
\mathbf{x}^t\Sigma_{m,n}\mathbf{x} = \mathbf{y}^t\Sigma_m\mathbf{y} + 2\mathbf{y}^t\mathbf{O}\mathbf{z} + \mathbf{z}^t\Sigma_n\mathbf{z} 
= \mathbf{y}^t\Sigma_m\mathbf{y} + \mathbf{z}^t\Sigma_n\mathbf{z}$.
And because these two terms are positive, we have that
$\mathbf{x}^t\Sigma_{m,n}\mathbf{x}>0$, i.e.\ $\Sigma_{m,n}$ is
positive-definite. \end{proof}
\end{Theorem}

\bibliography{LRD_database}
\bibliographystyle{jasa}

\end{document}